\documentclass[Afour, sageh, times]{sagej}

\usepackage[utf8]{inputenc}
\usepackage{amsthm}
\usepackage{amsmath}

\usepackage{floatrow}
\usepackage{graphicx}
\usepackage{calc}
\usepackage[label font=bf,labelformat=simple]{subfig}
\usepackage{caption}

\usepackage{standalone}
\usepackage{tikz,pgf}
\usepackage{mathtools}
\usepackage{framed}
\usepackage{diagbox}
\newcommand{\defeq}{\vcentcolon=}

\usetikzlibrary{shapes,backgrounds}
\usepackage{verbatim}
\usepackage{wasysym}
\pgfdeclarelayer{back}
\pgfsetlayers{back,main}
\usepackage{appendix}
\usepackage{subfig}
\usepackage[bottom]{footmisc}
\usepackage{stackengine}
\makeatletter
\pgfkeys{%
  /tikz/on layer/.code={
    \def\tikz@path@do@at@end{\endpgfonlayer\endgroup\tikz@path@do@at@end}%
    \pgfonlayer{#1}\begingroup%
  }%
}
\makeatother

\usepackage{geometry}

\definecolor{wikigreen}{HTML}{B3DDD5}
\definecolor{wikiblue}{HTML}{D5D6FA}
\definecolor{wikipink}{HTML}{DBBAD2}
\usetikzlibrary{fit}
\usepackage{dsfont}
\usetikzlibrary{shapes.geometric}
\usepackage[font=small,labelfont=sf]{caption}
\usepackage{amssymb}
\DeclareMathOperator*{\argmax}{arg\,max}
\DeclareMathOperator*{\argmin}{arg\,min}

\usetikzlibrary{decorations.text}
\usepackage{algorithmic,float}
\usepackage[linesnumbered,ruled]{algorithm2e}
\long\def\symbolfootnote[#1]#2{\begingroup%
	\def\thefootnote{\fnsymbol{footnote}}\footnote[#1]{#2}\endgroup}

\newcommand{\ccnodes}{V_{t}^{cc}(\mathcal{S})}

\newcommand{\wpc}{WPC}
\newcommand{\rulesep}{\unskip\ \hrule\ }

\usepackage[utf8]{inputenc}
\usepackage[linesnumbered,ruled]{algorithm2e}
\usepackage{amsthm}
\usepackage{amsfonts}
\usepackage{amssymb}
\usepackage{standalone}
\usepackage[super]{nth}
\usepackage{amsmath}
\usepackage{tikz,pgf}
\usepackage{enumitem}
\usepackage{tkz-euclide}
\usepackage[font=small,labelfont=sf]{caption}

\newtheorem{theorem}{Theorem}
\newtheorem{corollary}{Corollary}
\newtheorem{lemma}[theorem]{Lemma}
\theoremstyle{definition}
\newtheorem{defn}{Definition}

\usepackage{array,xcolor,colortbl}

\usepackage{mathtools}
\usetikzlibrary{quotes,angles}

\DeclarePairedDelimiter\floor{\lfloor}{\rfloor}

\usepackage{moreverb,url}

\usepackage[colorlinks,bookmarksopen,bookmarksnumbered,citecolor=red,urlcolor=red]{hyperref}

\newcommand\BibTeX{{\rmfamily B\kern-.05em \textsc{i\kern-.025em b}\kern-.08em
T\kern-.1667em\lower.7ex\hbox{E}\kern-.125emX}}

\begin{document}


\title{Too Global To Be Local:\\Swarm Consensus in Adversarial Settings}

\author{Lior Moshe\affilnum{1} and Noa Agmon\affilnum{1}}

\affiliation{\affilnum{1}Computer Science Department, Bar-Ilan University, Israel}




\begin{abstract}
Reaching a consensus in a swarm of robots is one of the fundamental problems in swarm robotics, examining the possibility of reaching an agreement within the swarm members. The recently-introduced contamination problem offers  a new perspective of the problem, in which swarm members should reach a consensus in spite of the existence of adversarial members that intentionally act to divert the swarm members towards a different consensus.
In this paper, we search for a consensus-reaching algorithm under the contamination problem setting by taking a top-down approach: We transform the problem to a
centralized two-player game in which each player controls the behavior of a subset of the swarm, trying to force the entire swarm to converge to an agreement on its own value. We define a performance metric for each player’s performance, proving a correlation between this metric and the chances of the player to win the game. We then present the globally optimal solution to the game and prove that unfortunately it is unattainable in a distributed setting, due to the challenging characteristics of the swarm members. We therefore examine the problem on a simplified swarm model, and compare the performance of the globally optimal strategy with locally optimal strategies, demonstrating its superiority in rigorous simulation experiments.
\end{abstract}


\maketitle

\section{Introduction}
\label{Introduction}
Robot swarms use simple local behavioral rules to achieve an emergent behavior over time. The study of swarms gained considerable interest in the scientific community due to its applicability in a variety of areas, such as search and rescue~(\citet*{chen2009survey,skinner2018uav,leon2016robot}), space exploration~(\citet*{nguyen2019swarmathon,sabatini2009collective,huang2014collective}) and disaster relief~(\citet*{schurr2005future,siddiqui2017development,ganesan2011small}).
As a part of these various applications, swarm members may need to achieve an agreement over a set of variables using local interactions among themselves. This problem is referred to as the {\em consensus} problem. Developing consensus-reaching algorithms for swarms has proven to be quite challenging due to the swarm members' limitations in sensing and computation capabilities.

Recently, there has been an interest in the contamination problem (\citet*{contaminationProblem}) which acts as an extension to the consensus problem. In the contamination problem, swarm members must reach a consensus in spite of the existence of adversarial swarm members that may act to divert the swarm from reaching the {\em desired} consensus. The swarm members are divided such that each one can adapt to one of two different states: \textit{healthy} or \textit{contaminated}. Each swarm member can change its state, and thus also its behavior, based on \textit{external factors} by other swarm members and its own \textit{internal state}. The goal of a consensus protocol in the contamination problem is to guide the swarm towards a desired state.

Past work showed that we can utilize the process of formation creation as a mean to reach consensus (\citet*{contaminationProblem}). It was shown that gathering swarm members in a geometrical structure named \textit{maximal stable cycle} (MSC), a clique, can maintain their initial state under certain conditions.

In this paper we examine the contamination problem using a top-down approach, in which we first define the problem as a two player game where each player has full information regarding the swarm members in the game, and has full control of a given group of swarm members that share the same state. We take advantage of the simplified settings of the game to define a measure for the performance of players in the game. We then use this measure to convert the game to an optimization problem. We show that although there is a defined globally optimal solution to the game, it is unattainable in a distributed setting. To overcome this, we perform several relaxations over the initial distributed setting of the problem. Finally, we present simulation results showing that the globally optimal strategy outperforms previously proposed locally optimal strategies for various numbers of swarm members.

The unattainability of the globally optimal solution under the standard distributed setting forces us to focus on the larger space of locally optimal solutions to devise an effective distributed solution to the problem. Since the size of the space of locally optimal solutions is directly correlated to the swarm's population, we can conclude that an efficient method to identify the subset of efficient locally optimal solutions is required to construct scalable distributed solutions to problems in the domain of swarm robotics.

\section{Contributions}
In this work, we show a thorough analysis for a problem which had very limited theoretical coverage in the past. Several dynamical aspects of the problem are overlooked in our analysis, as the movement of agents can lead to a continuously changing network of interactions that can impact the ability to reach consensus. Accordingly, previous works chose to tackle this kind of problem  under a stochastic perspective (\citet*{castellano2009statistical,valentini2017best}).
Since this is the first known theoretical analysis of the problem of reaching swarm consensus in adversarial settings we chose to take a different approach and focus on the inherent properties of the problem in hopes that it would lead us to a series of discoveries which would simplify the initially complicated problem. Indeed, we discovered that the initial problem could be simplified to the problem of forming dense circles while facing adversaries. A problem which was shown to be insurmountable under the current swarm settings. This result means that even while ignoring the dynamical aspects of the problem altogether, a globally optimal solution is unattainable under the current swarm settings.
The main contribution of this work is the discovery that there is no attainable globally optimal solution for the problem of swarm consensus under adversarial settings, this puts the spotlight on the problem of finding an effective local optimum solution for the problem. Constructing mechanisms that identify subsets of effective local optimums will be of utmost importance in the future since the space of local optimums grows exponentially according to the number of agents in the swarm, making most of the local optimums result in an overall underperformance of the whole swarm.

\section{Related Work}
\label{sec:2relatedwork}
The contamination problem was motivated from the set of popular problems in distributed computing of reaching a  {\em consensus}  (\citet*{olfati2007consensus}).
In networks of agents, a consensus is an agreement regarding a chosen value which depends on the states of all the agents.
A consensus protocol is a proposed rule that guides the interaction between an agent and all of its neighbors in the network towards reaching consensus. The instrumental work in paving the way for the development of self organizing swarms is surveyed in \citet*{olfati2007consensus}. A popular approach to solving consensus problems is a graph-based approach. In this approach a consensus reaching algorithm is expressed as a nth-order linear  system on  a given graph. These algorithms use the \textit{algebraic connectivity} (\citet*{fiedler1973algebraic}) as a measure that quantifies the speed of convergence of consensus algorithms.

The contamination problem is related to the Byzantine-consensus problem (\citet*{feldman1988optimal}), in which a network of distributed processors should reach an agreement on a value despite byzantine faults which might transpire in the system. In the contamination problem each agent resembles a processor in the Byzantine-consensus problem, though in our problem the processors are mobile, work asynchronously and use no explicit communication mechanisms. To the best of our knowledge, these properties were not considered in the research of the Byzantine-consensus problem.

In another line of research, \citet*{valentini2016collective} examined the speed of reaching an agreement in a swarm of robots with limited sensing capabilities which do not use explicit communication while having major uncertainties in their actuators and sensors. In this work, the majority rule was used for reaching a decision by each individual robot. It was shown that the main deciding factor for the speed of reaching an agreement is the size of the external group that each individual robot can observe. In this research, there are no external robots which act intentionally against the group of robots that has to reach an agreement, as in our case.

The problem of resilient asymptotic consensus (\citet*{leblanc2013resilient}) addresses reaching a consensus in a large-scale distributed system while facing misbehaving nodes in the form of adversaries. They defined the problem of reaching asymptotic consensus in the presence of misbehaving nodes given a particular threat model of those presented in \citet*{agmon2006fault} and a scope of threat. A local consensus protocol that is resilient to F adversarial nodes was  proposed. While the results of this research have a strong theoretic basis, it assumes that nodes of the network have full knowledge of the network and intentions of the other nodes whereas the agents in the contamination problem do not have full knowledge of the network's structure.

\citet*{saldana2018triangular} presented a formation topology that can be constructed in a fully distributed manner in static networks which guarantees resilient asymptotic consensus facing an unknown malicious agent in the network. The presented topology which is termed \textit{triangular robust networks} is based on the notion of network robustness presented in \citet*{leblanc2013resilient} and has a variety of appealing theoretical properties. While the verification of network robustness was shown to be an NP-hard problem (\citet*{leblanc2013algorithms}), triangular network robustness can be verified in polynomial time. Moreover, a triangular robust network can be incrementally expanded in a distributed manner which is competent with distributed robotic systems. The proposed formation can achieve resilient asymptotic consensus in a static networks facing a lone malicious agent, whereas the contamination game requires convergence to consensus in dynamic networks facing a group of malicious agents.

Continuing this line of work, \citet*{saldana2017resilient} proposed an approach that provides resilience for networks of agents which are time varying while diverting from the notion of high connectivity rates which are required in the topological measure of network robustness presented in \citet*{leblanc2013resilient}. The resilient consensus algorithm provided in this work relies on specific topological properties of the communication graph of agents in the network while in the contamination problem the agents do not have explicit communication capabilities.

Recently, some attention has been given to analysis of the strategies of robotic swarms from a game-theoretic perspective (\citet*{givigi2006game,givigi2007swarm,douchan2019swarms}).
Game theory was discovered to be a valuable tool for controlling behavior in distributed systems, as 
there are various parallels between the decision making architectures of societal systems which are common in the game theory literature and distributed systems. Particularly, both are comprised of a collection of connected decision making components whose collective behavior depends on the local decisions which are made by the components based on partial information about each other.
Finally, a learning process can be used to guide agents towards a solution. This learning process also constitutes a design choice as it is beneficial using a learning process which ensures convergence to a chosen game theoretic solution concept.
Game theoretic methods have been used for the purpose of distributed control~(\citet*{marden2015game}). In particular, there have been efforts to use game theory as a tool for modelling cooperative behavior in swarms~(\citet*{givigi2006game,givigi2007swarm}) which is the opposite case of the non-cooperative behavior of swarm members in adversarial settings. \citet*{douchan2019swarms} proposed a solution method to the problem of spatial coordination by forming a connection between the global utility theoretically reached using the extensive-form game which describes the environment of the robotic swarm and the utility of a single agent in the swarm. This connection is formed using \textit{potential games} (\citet*{monderer1996potential}) as a tool to aid agents in the swarm in learning optimal actions. It is shown that if a problem in swarm domains can be formalized as a potential game, then agents in the swarm can choose to maximize their own individual payoffs and the system will converge to pure-strategy Nash Equilibrium. This result means that there are cases in which agents in the swarm can be rational and achieve a local optimum of the problem by reaching their own local optimum by choosing an action which gives them maximal expected utility. The question of what should the agents do when the game cannot be represented as a potential game was not answered yet. 
The contamination problem cannot be represented as a potential game since each agent can only observe a limited area surrounding it which makes it impossible to craft a local utility function that is directly correlated to the global utility of all the members of the swarm.
\citet*{vamvoudakis2018game} developed a game theoretical solution method to the consensus problem for networked systems with the presence of adversaries. The proposed algorithm enabled the agents to reject adversarial input, thereby leading to consensus. Even though they solve the consensus problem in a distributed environment, they do so by rejecting adversarial input completely. In our problem we wish to reach a consensus in spite of the collected ${\text{adversarial input}}$.
\citet*{neto2005minimax} developed a dynamic programming algorithm to solve a class of stochastic games called two-person zero-sum games and evaluated its performance in the game of robotic soccer. The proposed work intends to model situations of teams with opposing objectives by approaching each team as an augmented agent such that the overall problem reduces to two-person zero-sum games. The appeal in this kind of games is that each equilibria has a similar reward structure which means that all equilibria are interchangeable. The proposed algorithm uses linear programming in order to find the Nash equilibrium in each state of the game which ensures optimal behavior in the worst-case scenario.
Despite the effective performance of the algorithm while using a minified model of robotic soccer, the algorithm requires a transition function which cannot be defined in complex swarm environments. Furthermore, the system of equations which are solved in order to find Nash equilibria for each state of the game scales up in the domain of robotic swarms as the approach of regarding a homogeneous swarm as an augmented agent does not perform well in practice.

\section{Contamination Problem as a Two-Player Game}
\label{sec:3contamgame}
In this section, we formally present the contamination problem and take our first step in our top-down approach by providing a simplified representation for the problem in the form of a two-player game where each player controls a swarm that can be represented by a graph of connected components. We then use this representation to show that the performance of a swarm in the contamination problem is directly correlated to the strength of its constructed connected components. Finally, we present the problem of constructing effective components as an optimization problem.
\subsection{Preliminaries}
The contamination problem can be represented as a time continuous game between two groups of agents that move simultaneously at each time step. We refer to this game as the {\em contamination game}. Each player in the contamination game controls a swarm of agents that share the same state. Let $\mathcal{S}$ be the group of all the members of the swarm. We assume that all the members of the swarm face the same physical limitations. Each agent has a physical diameter of length $D_{r}$, and can observe a limited area surrounding it. Let $S_{min}$ and $S_{max}$ denote the minimal and maximal observation radii of each swarm member, respectively.
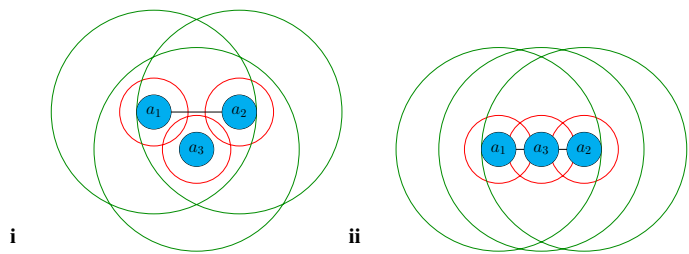
\begin{figure}[!b]
    \centering
     \sidesubfloat[]{
     \scalebox{0.45}{ 
\def \agentone {(-1.25,0) circle (0.5) node {$a_1$}}
\def \agenttwo {(1.25,0) circle (0.5) node {$a_2$}}
\def \agentthree {(0,-1.1) circle (0.5) node {$a_3$}}

\def \smin {[red] (-1.25,0) circle (1) (1.25,0) circle (1) (0,-1.1) circle (1)
}
\def \smax {[black!45!green] 
(-1.25,0) circle (3) 
(1.25,0) circle (3)
(0,-1.1) circle (3)
}

\begin{tikzpicture}
\tikzstyle{every node}=[font=\Large]

    \draw \agentone;
    \fill[fill=cyan] \agentone;
    
    \draw \agenttwo;
    \fill[fill=cyan] \agenttwo;
     
    \draw \agentthree;
    \fill[fill=cyan] \agentthree;

    \begin{pgfonlayer}{back}
    \draw \smin;
    \draw \smax;
    \draw[-] (-1.25,0)--(1.25,0);

    \end{pgfonlayer}

\end{tikzpicture}}
     }
    \sidesubfloat[]{
    \scalebox{0.45}{ 
\def \agentone {(-1.25,0) circle (0.5) node {$a_1$}}
\def \agenttwo {(1.25,0) circle (0.5) node {$a_2$}}
\def \agentthree {(0,0) circle (0.5) node {$a_3$}}

\def \smin {[red] (-1.25,0) circle (1) (1.25,0) circle (1) (0,0) circle (1)
}
\def \smax {[black!45!green] 
(-1.25,0) circle (3) 
(1.25,0) circle (3)
(0,0) circle (3)
}

\begin{tikzpicture}
\tikzstyle{every node}=[font=\Large]


    \draw \agentone;
    \fill[fill=cyan] \agentone;
    
    \draw \agenttwo;
    \fill[fill=cyan] \agenttwo;
     
    \draw \agentthree;
    \fill[fill=cyan] \agentthree;

    \begin{pgfonlayer}{back}
    \draw \smin;
    \draw \smax;
    \draw[-] (-1.25,0)--(1.25,0);
    \end{pgfonlayer}

\end{tikzpicture}}
    }
   
    \caption{The possible relationships between a set of three agents in the contamination game. Each agent is represented in cyan filled circles and the $S_{min}$ and $S_{max}$ circles of each agent are colored in red and green, respectively. (i) Agent $a_{1}$ can clearly see agent $a_{2}$. (ii) $a_{3}$ obstructs a part of $a_{2}$ from $a_{1}$ and vice versa.}
    \label{fig:concealment}
    
\end{figure}
In other words, each agent $a_j \in \mathcal{S}$ can be seen by an agent $a_i \in \mathcal{S}$ if it lies within a distance greater than $S_{min}$ and smaller than $S_{max}$ from it. Furthermore, we assume that two agents can observe one another only if we can connect a line between both of their centers without intersecting any other agent (concealing the view) as displayed in Figure \ref{fig:concealment}. Let $O(S_{min}, S_{max}, a_i)$ be the observation area of agent $a_i \in \mathcal{S}$ based on radii $S_{min}$ and $S_{max}$ and physical concealments that may be caused by other agents. In short, we use the notation $O(a_i)$ as we assume that the $S_{min}$ and $S_{max}$ have fixed values. 
We say that an agent $a_i \in \mathcal{S}$ can observe another agent $a_j \in \mathcal{S}$ if $a_j \in O(a_i)$. Furthermore, each agent is included in its own observation area, i.e., ${a_i \in O(a_i)~~\forall a_{i} \in \mathcal{S}}$.

As part of the contamination game, each agent $a_i \in \mathcal{S}$ is in either one of two different states: \textit{healthy} or  \textit{contaminated}, denoted by ${s(a_i) \in \{s_{H}, s_{C}\}}$, respectively.
The state of an agent $a_i \in \mathcal{S}$ is decided by using a majority rule over the agents that $a_i$ observes. Formally, if we denote the number of agents in an area $A$ by $h(A)$ and similarly the number of contaminated agents in $A$ by $c(A)$, then the state of agent $a_i$ is decided by the following update rule:
$$s(a_i) = 
\left\{\begin{array}{lr}
        s_{H}, & h(O(a_i)) \geq c(O(a_i)) \\
        s_{C}, & c(O(a_i)) > h(O(a_i))
        \end{array}\right\}$$
The observation of each agent in $\mathcal{S}$ changes through time. Consequently, the state of each agent in $\mathcal{S}$ might change as well. Denote the observation (resp. state) of agent $a_i \in \mathcal{S}$ at time $t$ by $O_{t}(a_i)$ (resp. $s_{t}(a_i)$).  We say that an agent $a_i \in \mathcal{S}$ is \textit{conquered} at time $t+1$ if ${s_{t+1}(a_i) = \{s_{H},s_{C}\} \setminus \{s_{t}(a_i)\}}$.
The goal of each agent in the game is to conquer agents of the opposing state, i.e., the goal of healthy agents is to conquer contaminated agents and vice versa.
{\small
\begin{quote}
{\bf The Contamination Game}\\
The contamination game is a time-continuous two-player game of length $T$ ($T$ is unknown to the players). Each player controls swarm members that share the same state and has full knowledge of the state of all the members of the swarm.
In each time-step each agent decides its state based on a majority rule which is applied on its surroundings.
The game ends when either one player controls all the swarm members or $T$ time-steps had passed.
The goal of each player in the game is to control the majority of swarm members at the end of the game.
\end{quote}
}
Note that since $T$ is unknown to the players, they constantly
strive to maximize the number of agents in their swarm.
\subsubsection{Graph Representations\\}

The observations of agents can be depicted by an observation graph. 

\begin{defn}
Let ${G_{t}(\mathcal{S})=(V_{t}(\mathcal{S}),E_{t}(\mathcal{S}))}$ be the undirected observation graph of the contamination game of the agents in $\mathcal{S}$ where
\begin{align*}
    V_{t}(\mathcal{S}) &= \mathcal{S}\\
    E_{t}(\mathcal{S}) &= \{(a_i, a_j)|~~a_i \in O(a_j)\}
\end{align*}

Simply put, each agent in the contamination game is represented as a node in the observation graph and each edge between two nodes represents two agents that can observe one another.
\end{defn}
Furthermore, we can simplify the representation of the contamination game by interpreting it as a group of connected components of agents which share the same state. Each pair of agents that belong to the same connected component necessarily have a path between their nodes in the observation graph.

\begin{defn}
Let $G_{t}^{cc}(\mathcal{S}) = (V_{t}^{cc}(\mathcal{S}),E_{t}^{cc}(\mathcal{S}))$ denote the undirected graph of connected components in the contamination game of the agents in $\mathcal{S}$ at time $t$. Each node $v \in V_{t}^{cc}(\mathcal{S})$ represents a connected component of agents in the observation graph $G_{t}(\mathcal{S})$ which share the same state. For any pair of components ${v_i, v_j \in V_{t}^{cc}(\mathcal{S})}$ with opposing states, there will be an edge in $E_{t}^{cc}(\mathcal{S})$ if there is any agent in $v_i$ that can observe another agent in $v_j$. 
\end{defn}

Let $h(\ccnodes)$ (resp. $c(\ccnodes)$) be the set of healthy (resp. contaminated) components in the contamination game of the agents in $\mathcal{S}$ at time $t$.

\begin{figure}[!t]
    \centering
    \scalebox{10}{\input{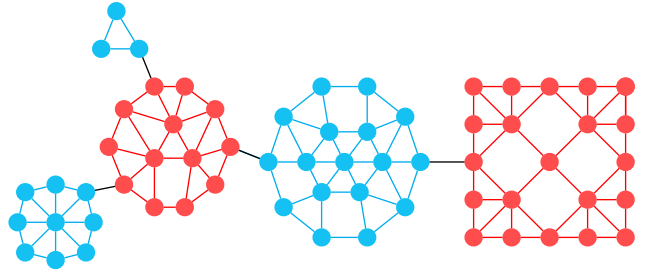}}
    \caption{An example of a connected components graph containing five connected components of agents.}
    \label{fig:connected_graph}
\end{figure}

Any pair of connected components that are connected by an edge in the connected components graph must be of opposing states since if they share the same state they will be a part of the same connected component. Figure \ref{fig:connected_graph} illustrates an example of a connected components graph describing an instance of the contamination game. The healthy and contaminated agents are colored in cyan and red, respectively. The cyan and red edges represent the connections between healthy and contaminated agents, respectively. The black edges are the edges of the connected components graph.

\def \tkzscl{6}

\subsection{The WPC Algorithm}
Each swarm in the contamination problem can be described as a set of connected components of agents. Therefore, a swarm that acts optimally in the contamination problem must have strong connected components according to the rules of the problem, that is, connected components that are hard to conquer.
Hence, a metric that evaluates the strength of connected components of agents in the contamination problem  can be utilized to develop an optimal strategy for swarm members. An intuitive measure for the strength of a connected component would be the minimal number of required agents to conquer it.
We start by defining the strength of an individual agent, which is the number of agents it observes that share its own state.
\begin{defn}
Given a connected component of agents ${C \in \ccnodes}$, the \textit{connectivity factor} of an agent $a_i \in C$ at time $t$, denoted by $cf_{t}(a_i)$, is defined to be the number of agents of $C$ that $a_i$ can observe at time $t$, that is
\begin{eqnarray*}
cf_{t}(a_i) = \left\{\begin{array}{lr}
        \lvert h(O(a_{i}) \rvert, & s(a_{i}) = s_{H} \\
      \lvert c(O(a_{i}) \rvert, & s(a_{i}) = s_{C}
        \end{array}\right\}
\end{eqnarray*}
\end{defn}

\begin{figure}[!b]
    \centering
    \scalebox{0.9}{ 
\def \agentone {(0,0) circle (0.25) node {$a_1$}}
\def \agenttwo {(-0.5,0.5) circle (0.25) node {$a_2$}}
\def \agentthree {(0.5,0.5) circle (0.25) node {$a_3$}}
\def \agentfour {(0.75, 1.25) circle (0.25) node {$a_4$}}
\def \agentfive {(-0.75, 1.25) circle (0.25) node {$a_5$}}

\def \agentsix {(0, 2) circle (0.25) node {$a_6$}}

\def \agentseven {(-1.25, 2) circle (0.25) node {$a_7$}}

\def \agenteight {(-1.5, 0.75) circle (0.25) node {$a_8$}}

\def \agentnine {(-1.1, -0.4) circle (0.25) node {$a_9$}}

\def \agentten {(-0.2, -1.2) circle (0.25) node {$a_{10}$}}

\def \agenteleven {(0.9, -0.7) circle (0.25) node {$a_{11}$}}

\def \agentwelve {(1.7, 0.3) circle (0.25) node {$a_{12}$}}

\def \agentthirteen {(1.95, 1.45) circle (0.25) node {$a_{13}$}}

\def \smin {[red] (0,0) circle (0.5) (-0.5,0.5) circle (0.5) (0.5,0.5) circle (0.5) (0.75,1.25) circle (0.5) (-0.75,1.25) circle (0.5) (0, 2) circle (0.5) (-1.25, 2) circle (0.5) (-1.5, 0.75) circle (0.5) (-1.1, -0.4) circle (0.5) (-0.2, -1.2) circle (0.5) (0.9, -0.7) circle (0.5) (1.7, 0.3) circle (0.5) (1.95, 1.45) circle (0.5)}
\def \smax {[black!45!green] (0,0) circle (1.5) (-0.5,0.5) circle (1.5) (0.5,0.5) circle (1.5) (0.75,1.25) circle (1.5) (-0.75,1.25) circle (1.5)
(0, 2) circle (1.5) (-1.25, 2) circle (1.5) (-1.5, 0.75) circle (1.5) (-1.1, -0.4) circle (1.5) (-0.2, -1.2) circle (1.5) (0.9, -0.7) circle (1.5) (1.7, 0.3) circle (1.5) (1.95, 1.45) circle (1.5)}

\begin{tikzpicture}
\tikzstyle{every node}=[font=\small]
    \draw (-0.75,1.25) -- (-0.5,0.5) -- (0,0) -- (0.5,0.5) -- (0.75,1.25) -- (-0.75, 1.25)--(0,2)--(0.75,1.25);
    
    \draw (-1.25,2)--(-0.75,1.25);
    \draw (-1.25,2)--(0,2);
    \draw (-1.5, 0.75)--(-0.5,0.5);
    \draw (-1.5, 0.75)--(-0.75, 1.25);
    \draw (-1.5, 0.75)--(-1.25, 2);
    
    \draw (-1.1, -0.4)--(-1.5, 0.75);
    \draw (-1.1, -0.4)--(-0.5,0.5);
    \draw (-1.1, -0.4)--(0,0);
    
    \draw (-0.2, -1.2)--(-1.1, -0.4);
    \draw (-0.2, -1.2)--(0,0);
    
    \draw (0.9, -0.7)--(-0.2, -1.2);
    \draw (0.9, -0.7)--(0,0);
    \draw (0.9, -0.7)--(0.5,0.5);
    
    \draw (1.7, 0.3)--(0.9, -0.7);
    \draw (1.7, 0.3)--(0.5,0.5);
    \draw (1.7, 0.3)--(0.75, 1.25);
    
    \draw (1.95, 1.45)--(0.75, 1.25);
    \draw (1.95, 1.45)--(1.7, 0.3);
    
    \draw (0.5,0.5)--(-0.75, 1.25);
    \draw (0.5,0.5)--(-0.5,0.5);
    \draw (-0.5,0.5)--(0.75, 1.25);
    

    \draw \agentone;
    \fill[fill=cyan] \agentone;
    
    \draw \agenttwo;
    \fill[fill=cyan] \agenttwo;
    
    \draw \agentthree;
    \fill[fill=cyan] \agentthree;
    
    \draw \agentfour;
    \fill[fill=cyan] \agentfour;
    
    \draw \agentfive;
    \fill[fill=cyan] \agentfive;
    
    \draw \agentsix;
    \fill[fill=yellow] \agentsix;
    
    \draw \agentseven;
    \fill[fill=yellow] \agentseven;
    
    \draw \agenteight;
    \fill[fill=yellow] \agenteight;
    
    \draw \agentnine;
    \fill[fill=yellow] \agentnine;
    
    \draw \agentten;
    \fill[fill=yellow] \agentten;
    
    \draw \agenteleven;
    \fill[fill=yellow] \agenteleven;
    
    \draw \agentthirteen;
    \fill[fill=yellow] \agentthirteen;
    
    \draw \agentwelve;
    \fill[fill=yellow] \agentwelve;
    

    \begin{pgfonlayer}{back}
    \draw \smin;
    \draw \smax;
    \end{pgfonlayer}

\end{tikzpicture}}
     \caption{Example of a connected component of agents ${C = \{a_1, a_2, \dots, a_{13}\}}$, $S_{min}$ and $S_{max}$ circles are drawn around each agent colored red and green, respectively. A black line connecting two agents means they can observe each other. The agents which are on the fence of $C$ are colored in yellow.}
     \label{fig:barenessExample}
\end{figure}
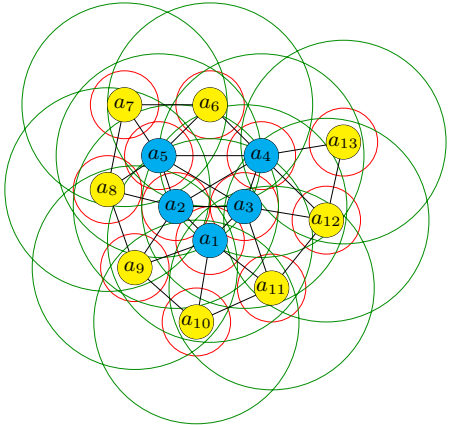

In order to define the strength of a connected component of agents we first need to inspect the first line of defence which guards the agents in the interior of the structure, which are the agents on the exterior of the structure. We refer to this group of agents as the \textit{fence} of the component, while each individual agent is referred to as a \textit{bare} agent. Formally, we define it as follows:
\begin{defn}
Given a connected component of agents ${C \in \ccnodes}$, an agent $a_i \in C$ is said to be \textit{bare} at time $t$ if there is some part of its observation area that can be intruded by other agents outside of $C$. Moreover, the \textit{fence} of $C$ at time $t$, denoted by $F_{t}(C)$, is the group of bare agents in $C$, or formally
\begin{eqnarray*}
 F_{t}(C) = \{a_i \in C~| ~~O_{t}(a_i) \setminus (\bigcup_{a_{j} \in C \setminus \{a_{i}\}} O_{t}(a_j)) \neq \varnothing\}
\end{eqnarray*}
\end{defn}

Figure \ref{fig:barenessExample} demonstrates the notion of a bare agent in a connected component of agents. It depicts a connected component of thirteen healthy agents where the bare agents are represented by yellow filled circles whereas the other agents are represented by cyan filled circles.

We wish to be able to measure the strength of each agent on the fence of the connected component, which is measured by the minimal number of required agents to conquer it.
\begin{defn}
The \textit{bareness factor} of a bare agent $a_i \in \mathcal{S}$ at time $t$, denoted by $b_{t}(a_i)$, is the number of required agents to force $a_i$ to switch its state.
\end{defn}

The connectivity factor can be used as a lower bound for the bareness factor. In other words, for each agent $a_i \in \mathcal{S}$ we can say that
$$ cf_{t}(a_i) + 1 \leq b_{t}(a_i)$$

\begin{figure}[!t]
    \centering
    \scalebox{0.8}{ 
\def \agentone {(0,0) circle (0.25) node {$a_1$}}
\def \agenttwo {(-0.4,0.4) circle (0.25) node {$a_2$}}
\def \agentthree {(-0.4,-0.4) circle (0.25) node {$a_3$}}
\def \agentfour {(-1.7, 0) circle (0.25) node {$a_4$}}

\def \five {(1.91, -2.31) circle (0.25) node {$a_5$}}
\def \six {(2.57, 0.00) circle (0.25) node {$a_6$}}
\def \seven {(1.91, 2.31) circle (0.25) node {$a_7$}}
\def \eight  {(-0.85, -2.88) circle (0.25) node {$a_8$}}
\def \nine  {(-0.19, -2.59) circle (0.25) node {$a_9$}}

\def \ten  {(-0.19, 2.59) circle (0.25) node {$a_{10}$}}
\def \eleven  {(-0.85, 2.88) circle (0.25) node {$a_{11}$}}

\def \smin {[red] 
(0,0) circle (1)
(-0.4,0.4) circle (1)
(-0.4,-0.4) circle (1)
(-1.7, 0) circle (1)
}
\def \smax {[black!45!green] 
(0,0) circle (3) 
(-0.4,0.4) circle (3) 
(-0.4,-0.4) circle (3) 
(-1.7, 0) circle (3)
}

\begin{tikzpicture}
\tikzstyle{every node}=[font=\small]
    \draw \agentone;
    \fill[fill=cyan] \agentone;
    
    \draw \agenttwo;
    \fill[fill=cyan] \agenttwo;
    
    \draw \agentthree;
    \fill[fill=cyan] \agentthree;
    
    \draw \agentfour;
    \fill[fill=cyan] \agentfour;
    
    
    
    
    



    \begin{pgfonlayer}{back}
     
         \draw[fill=white,even odd rule]  (0,0) circle (0.25)
                                   (0,0) circle (1);
       
       \draw[fill=white,even odd rule]  (-1.7, 0) circle (0.25)
       (-1.7,0) circle (1);
       \draw[fill=white,even odd rule]  (-0.4, 0.4) circle (0.25)
       (-0.4,0.4) circle (1);
       \draw[fill=white,even odd rule]  (-0.4, -0.4) circle (0.25)
       (-0.4,-0.4) circle (1);

         \draw (0,0)--(-1.7, 0);
    \draw (-0.5,0.5)--(-1.7, 0);
    \draw (-0.5,-0.5)--(-1.7, 0);
    \draw \smin;
    \draw \smax;
     
    \end{pgfonlayer}

\end{tikzpicture}}
    \caption{An example of a connected component of healthy agents $C = \{a_1, a_2, a_3, a_4\}$ at time $t$. Similarly to the previous figure, the $S_{min}$ and $S_{max}$ circles are drawn in red and green, respectively.}
    \label{fig:enlarged_example}
\end{figure}
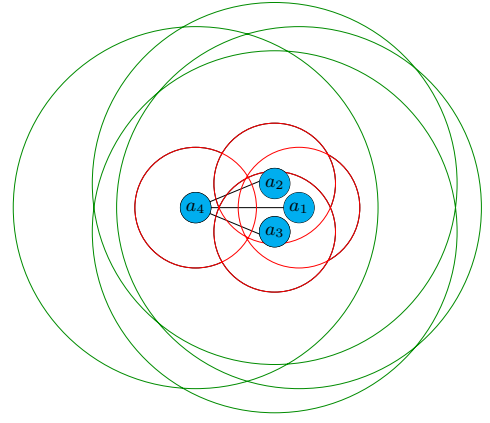
Initially, it may seem that this bound is the actual value of the bareness factor since if healthy agent $a_{i}$ observes $cf_{t}(a_i) + 1$ contaminated agents at time $t$ then it will necessarily change its state based on the applied majority rule. However, we must also take into account the fact that the $cf_{t}(a_i) + 1$ contaminated agents must preserve their own contaminated state. Therefore, we need to make sure that the number of healthy agents observed by each contaminated agent is at most the number of contaminated agents observed by it.
This can result in scenarios where an opponent must use more than $cf_{t}(a_i) + 1$ agents to conquer $a_{i}$. As an example, consider the connected component presented in Figure \ref{fig:enlarged_example}.
It can easily be seen that all the agents in the connected component are bare, meaning that an opponent can conquer each one of them. Assume the opponent chooses to conquer agent $a_{1}$.

\begin{figure}[!t]
    \centering
    \scalebox{0.8}{ 
\def \agentone {(0,0) circle (0.25) node {$a_1$}}
\def \agenttwo {(-0.4,0.4) circle (0.25) node {$a_2$}}
\def \agentthree {(-0.4,-0.4) circle (0.25) node {$a_3$}}
\def \agentfour {(-1.7, 0) circle (0.25) node {$a_4$}}

\def \five {(1.91, -2.31) circle (0.25) node {$a_5$}}
\def \six {(2.57, 0.00) circle (0.25) node {$a_6$}}
\def \seven {(1.91, 2.31) circle (0.25) node {$a_7$}}
\def \eight  {(-0.85, -2.88) circle (0.25) node {$a_8$}}
\def \nine  {(-0.19, -2.59) circle (0.25) node {$a_9$}}

\def \ten  {(-0.19, 2.59) circle (0.25) node {$a_{10}$}}
\def \eleven  {(-0.85, 2.88) circle (0.25) node {$a_{11}$}}

\def \smin {[red] 
(0,0) circle (1)
(-0.4,0.4) circle (1)
(-0.4,-0.4) circle (1)
(-1.7, 0) circle (1)
}
\def \smax {[black!45!green] 
(0,0) circle (3) 
}

\begin{tikzpicture}
\tikzstyle{every node}=[font=\small]
    \draw \agentone;
    \fill[fill=cyan] \agentone;
    
    \draw \agenttwo;
    \fill[fill=cyan] \agenttwo;
    
    \draw \agentthree;
    \fill[fill=cyan] \agentthree;
    
    \draw \agentfour;
    \fill[fill=cyan] \agentfour;
    
    
    
    
    



    \begin{pgfonlayer}{back}

         \filldraw[fill=cyan!10!white, draw=cyan!50!black]
     (1.91,-2.31) arc (320.4147020709557:367.670448104312:3)
     arc (352.329551895688:399.5852979290443:3)
     arc (50.414702070955684:-50.414702070955684:3);
     
     \filldraw[fill=cyan!30!white, draw=cyan!50!black]
     (1.91,-2.31) arc (-50.41470207095567:-106.44337961485542:3)
     arc (-73.55662038514458:-59.75731932588542:3)
     arc (-85.98247895634071:-7.6704481043120065:3)
     arc (7.6704481043120065:-39.58529792904434:3);
     
     \filldraw[fill=cyan!30!white, draw=cyan!50!black]
     (2.57,0.00) arc (7.6704481043120065:85.98247895634071:3)
        arc (59.75731932588542:73.55662038514458:3)
        arc(106.44337961485542:50.41470207095567:3)
        arc(39.58529792904434:-7.6704481043120065:3);
        
        \filldraw[fill=cyan!50!white, draw=cyan!50!black]
     (-0.19, -2.59) arc (-59.75731932588542:59.75731932588542:3)
        arc (85.98247895634071:7.6704481043120065:3)
        arc(-7.6704481043120065:-85.98247895634071:3);
        
          \filldraw[fill=cyan!70!white, draw=cyan!50!black]
     (-0.85, -2.88) arc (253.55662038514458:106.44337961485542:3)
     arc(73.55662038514458:-73.55662038514458:3);
     
         \draw[fill=white,even odd rule]  (0,0) circle (0.25)
                                   (0,0) circle (1);
       
       \draw[fill=white,even odd rule]  (-1.7, 0) circle (0.25)
       (-1.7,0) circle (1);
       \draw[fill=white,even odd rule]  (-0.4, 0.4) circle (0.25)
       (-0.4,0.4) circle (1);
       \draw[fill=white,even odd rule]  (-0.4, -0.4) circle (0.25)
       (-0.4,-0.4) circle (1);

         \draw (0,0)--(-1.7, 0);
    \draw (-0.5,0.5)--(-1.7, 0);
    \draw (-0.5,-0.5)--(-1.7, 0);
    \draw \smin;
    \draw \smax;
     
    \end{pgfonlayer}

\end{tikzpicture}}
    \caption{Partition of the observation area of $a_1$ into sub-areas based on the observations of the other agents.}
    \label{fig:enlarged_colored}
\end{figure}
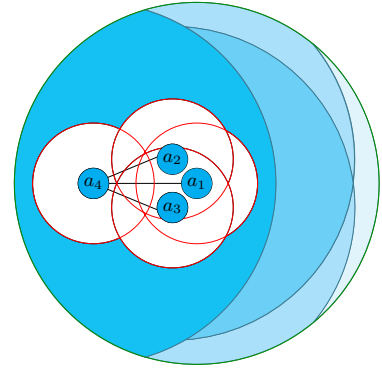

Consider the area observed by the agent $a_1$. This area can be described as a union of sub-areas such that each sub area is observed by a different subset of agents from $\{a_1, a_2, a_3, a_4\}$. Figure \ref{fig:enlarged_colored} shows the partition of the observation area of $a_1$ into several sub-areas based on the observations of the other agents. The strength of the cyan color of an area is proportional to the number of agents observing this area.
The connectivity factor of $a_1$ is $1$ since $a_1$ can only observe $a_4$. In order to find the bareness factor of $a_1$ we must devise a way to conquer $a_1$ using the minimal number of agents. We must take into account the partition of the observation area of $a_1$ presented in Figure \ref{fig:enlarged_colored} throughout this process since if an attacking agent is located in an area which is observed by $n > 0$ healthy agents than it must observe at least $n$ contaminated agents to preserve its state.

While attacking $a_1$ we must consider the physical dimensions of the agents. Given a sub area of the observation area of $a_1$, denoted by $O' \subseteq O(a_1)$, we can place a contaminated agent in $O'$ if it physically fits in $O'$.
In other words, each sub area of $O(a_1)$ can inhabit a limited number of contaminated agents in a way that they can observe one another.
To the purpose of our example, we assume that the the sub area of $O(a_1)$ which is observed by only $a_1$ can inhabit only a single contaminated agent based on the agent's diameter $D_{r}$.

Therefore, we must place an agent in an area of $O(a_1)$ that is observed by at least two healthy agents. Consequently, this will require placing another agent to preserve the contaminated state of the agents which were previously placed during the attack. This leads to the fact that each possible attack against $a_1$ will require more than ${cf_{t}(a_1) + 1 = 2}$ agents, i.e., ${b_{t}(a_1) > cf_{t}(a_1) + 1}$.

\begin{figure}[!t]
    \centering
\sidesubfloat[]{\scalebox{0.65}{ 
\def \agentone {(0,0) circle (0.2) node {$a_1$}}
\def \agenttwo {(-0.4,0.4) circle (0.2) node {$a_2$}}
\def \agentthree {(-0.4,-0.4) circle (0.2) node {$a_3$}}
\def \agentfour {(-1.7, 0) circle (0.2) node {$a_4$}}

\def \contone {(2.8, 0) circle (0.2) node {$c_1$}}

\def \smin {[red] 
(0,0) circle (1)
(-0.4,0.4) circle (1)
(-0.4,-0.4) circle (1)
(-1.7, 0) circle (1)

(2.8, 0) circle (1)
}
\def \smax {[black!45!green] 
(0,0) circle (3) 
}

\begin{tikzpicture}
\tikzstyle{every node}=[font=\small]
    \draw \agentone;
    \fill[fill=cyan] \agentone;
    
    \draw \agenttwo;
    \fill[fill=cyan] \agenttwo;
    
    \draw \agentthree;
    \fill[fill=cyan] \agentthree;
    
    \draw \agentfour;
    \fill[fill=cyan] \agentfour;
    
    \draw \contone;
    \fill[fill=red] \contone;
    
    
    
    
    



    \begin{pgfonlayer}{back}

         \filldraw[fill=cyan!10!white, draw=cyan!50!black]
     (1.91,-2.31) arc (320.4147020709557:367.670448104312:3)
     arc (352.329551895688:399.5852979290443:3)
     arc (50.414702070955684:-50.414702070955684:3);
     
     \filldraw[fill=cyan!30!white, draw=cyan!50!black]
     (1.91,-2.31) arc (-50.41470207095567:-106.44337961485542:3)
     arc (-73.55662038514458:-59.75731932588542:3)
     arc (-85.98247895634071:-7.6704481043120065:3)
     arc (7.6704481043120065:-39.58529792904434:3);
     
     \filldraw[fill=cyan!30!white, draw=cyan!50!black]
     (2.57,0.00) arc (7.6704481043120065:85.98247895634071:3)
        arc (59.75731932588542:73.55662038514458:3)
        arc(106.44337961485542:50.41470207095567:3)
        arc(39.58529792904434:-7.6704481043120065:3);
        
        \filldraw[fill=cyan!50!white, draw=cyan!50!black]
     (-0.19, -2.59) arc (-59.75731932588542:59.75731932588542:3)
        arc (85.98247895634071:7.6704481043120065:3)
        arc(-7.6704481043120065:-85.98247895634071:3);
        
          \filldraw[fill=cyan!70!white, draw=cyan!50!black]
     (-0.85, -2.88) arc (253.55662038514458:106.44337961485542:3)
     arc(73.55662038514458:-73.55662038514458:3);
     
         \draw[fill=white,even odd rule]  (0,0) circle (0.2)
                                   (0,0) circle (1);
       
       \draw[fill=white,even odd rule]  (-1.7, 0) circle (0.2)
       (-1.7,0) circle (1);
       \draw[fill=white,even odd rule]  (-0.4, 0.4) circle (0.2)
       (-0.4,0.4) circle (1);
       \draw[fill=white,even odd rule]  (-0.4, -0.4) circle (0.2)
       (-0.4,-0.4) circle (1);

         \draw[color=cyan] (0,0)--(-1.7, 0);
    \draw[color=cyan] (-0.5,0.5)--(-1.7, 0);
    \draw[color=cyan] (-0.5,-0.5)--(-1.7, 0);
    
    \draw (2.85, 0)--(0, 0);
    \draw \smin;
    \draw \smax;
     
    \end{pgfonlayer}

\end{tikzpicture}\label{fig:a}}}
\hfil
\sidesubfloat[]{\scalebox{0.65}{ 
\def \agentone {(0,0) circle (0.2) node {$a_1$}}
\def \agenttwo {(-0.4,0.4) circle (0.2) node {$a_2$}}
\def \agentthree {(-0.4,-0.4) circle (0.2) node {$a_3$}}
\def \agentfour {(-1.7, 0) circle (0.2) node {$a_4$}}

\def \contone {(2.8, 0) circle (0.2) node {$c_1$}}
\def \conttwo {(2, 1.75) circle (0.2) node {$c_2$}}

\def \smin {[red] 
(0,0) circle (1)
(-0.4,0.4) circle (1)
(-0.4,-0.4) circle (1)
(-1.7, 0) circle (1)

(2.8, 0) circle (1)
(2, 1.75) circle (1)
}
\def \smax {[black!45!green] 
(0,0) circle (3) 
}

\begin{tikzpicture}
\tikzstyle{every node}=[font=\small]
    \draw \agentone;
    \fill[fill=cyan] \agentone;
    
    \draw \agenttwo;
    \fill[fill=cyan] \agenttwo;
    
    \draw \agentthree;
    \fill[fill=cyan] \agentthree;
    
    \draw \agentfour;
    \fill[fill=cyan] \agentfour;
    
    \draw \contone;
    \fill[fill=red] \contone;
    
    \draw \conttwo;
    \fill[fill=red] \conttwo;
    
    
    
    
    



    \begin{pgfonlayer}{back}

         \filldraw[fill=cyan!10!white, draw=cyan!50!black]
     (1.91,-2.31) arc (320.4147020709557:367.670448104312:3)
     arc (352.329551895688:399.5852979290443:3)
     arc (50.414702070955684:-50.414702070955684:3);
     
     \filldraw[fill=cyan!30!white, draw=cyan!50!black]
     (1.91,-2.31) arc (-50.41470207095567:-106.44337961485542:3)
     arc (-73.55662038514458:-59.75731932588542:3)
     arc (-85.98247895634071:-7.6704481043120065:3)
     arc (7.6704481043120065:-39.58529792904434:3);
     
     \filldraw[fill=cyan!30!white, draw=cyan!50!black]
     (2.57,0.00) arc (7.6704481043120065:85.98247895634071:3)
        arc (59.75731932588542:73.55662038514458:3)
        arc(106.44337961485542:50.41470207095567:3)
        arc(39.58529792904434:-7.6704481043120065:3);
        
        \filldraw[fill=cyan!50!white, draw=cyan!50!black]
     (-0.19, -2.59) arc (-59.75731932588542:59.75731932588542:3)
        arc (85.98247895634071:7.6704481043120065:3)
        arc(-7.6704481043120065:-85.98247895634071:3);
        
          \filldraw[fill=cyan!70!white, draw=cyan!50!black]
     (-0.85, -2.88) arc (253.55662038514458:106.44337961485542:3)
     arc(73.55662038514458:-73.55662038514458:3);
     
         \draw[fill=white,even odd rule]  (0,0) circle (0.2)
                                   (0,0) circle (1);
       
       \draw[fill=white,even odd rule]  (-1.7, 0) circle (0.2)
       (-1.7,0) circle (1);
       \draw[fill=white,even odd rule]  (-0.4, 0.4) circle (0.2)
       (-0.4,0.4) circle (1);
       \draw[fill=white,even odd rule]  (-0.4, -0.4) circle (0.2)
       (-0.4,-0.4) circle (1);

         \draw[color=cyan] (0,0)--(-1.7, 0);
    \draw[color=cyan] (-0.5,0.5)--(-1.7, 0);
    \draw[color=cyan] (-0.5,-0.5)--(-1.7, 0);
        \draw[color=red] (2.85, 0)--(2, 1.75);

    \draw (2.85, 0)--(0, 0);
    \draw (2, 1.75)--(0,0);
    \draw (2, 1.75)--(-0.4,0.4);
    
    \draw[color=red] (2, 1.75)--(2.85,0);
    \draw \smin;
    \draw \smax;
     
    \end{pgfonlayer}

\end{tikzpicture}\label{fig:b}}}

\medskip
\sidesubfloat[]{\scalebox{0.65}{ 
\def \agentone {(0,0) circle (0.2) node {$a_1$}}
\def \agenttwo {(-0.4,0.4) circle (0.2) node {$a_2$}}
\def \agentthree {(-0.4,-0.4) circle (0.2) node {$a_3$}}
\def \agentfour {(-1.7, 0) circle (0.2) node {$a_4$}}

\def \contone {(2.8, 0) circle (0.2) node {$c_1$}}
\def \conttwo {(2, 1.75) circle (0.2) node {$c_2$}}
\def \contthree {(1.5, 0.55) circle (0.2) node {$c_3$}}

\def \smin {[red] 
(0,0) circle (1)
(-0.4,0.4) circle (1)
(-0.4,-0.4) circle (1)
(-1.7, 0) circle (1)

(2.8, 0) circle (1)
(2, 1.75) circle (1)
(1.5, 0.55) circle (1)
}
\def \smax {[black!45!green] 
(0,0) circle (3) 
}

\begin{tikzpicture}
\tikzstyle{every node}=[font=\small]
    \draw \agentone;
    \fill[fill=cyan] \agentone;
    
    \draw \agenttwo;
    \fill[fill=cyan] \agenttwo;
    
    \draw \agentthree;
    \fill[fill=cyan] \agentthree;
    
    \draw \agentfour;
    \fill[fill=cyan] \agentfour;
    
    \draw \contone;
    \fill[fill=red] \contone;
    
    \draw \conttwo;
    \fill[fill=red] \conttwo;
    
    \draw \contthree;
    \fill[fill=red] \contthree;
    
    
    
    
    



    \begin{pgfonlayer}{back}

         \filldraw[fill=cyan!10!white, draw=cyan!50!black]
     (1.91,-2.31) arc (320.4147020709557:367.670448104312:3)
     arc (352.329551895688:399.5852979290443:3)
     arc (50.414702070955684:-50.414702070955684:3);
     
     \filldraw[fill=cyan!30!white, draw=cyan!50!black]
     (1.91,-2.31) arc (-50.41470207095567:-106.44337961485542:3)
     arc (-73.55662038514458:-59.75731932588542:3)
     arc (-85.98247895634071:-7.6704481043120065:3)
     arc (7.6704481043120065:-39.58529792904434:3);
     
     \filldraw[fill=cyan!30!white, draw=cyan!50!black]
     (2.57,0.00) arc (7.6704481043120065:85.98247895634071:3)
        arc (59.75731932588542:73.55662038514458:3)
        arc(106.44337961485542:50.41470207095567:3)
        arc(39.58529792904434:-7.6704481043120065:3);
        
        \filldraw[fill=cyan!50!white, draw=cyan!50!black]
     (-0.19, -2.59) arc (-59.75731932588542:59.75731932588542:3)
        arc (85.98247895634071:7.6704481043120065:3)
        arc(-7.6704481043120065:-85.98247895634071:3);
        
          \filldraw[fill=cyan!70!white, draw=cyan!50!black]
     (-0.85, -2.88) arc (253.55662038514458:106.44337961485542:3)
     arc(73.55662038514458:-73.55662038514458:3);
     
         \draw[fill=white,even odd rule]  (0,0) circle (0.2)
                                   (0,0) circle (1);
       
       \draw[fill=white,even odd rule]  (-1.7, 0) circle (0.2)
       (-1.7,0) circle (1);
       \draw[fill=white,even odd rule]  (-0.4, 0.4) circle (0.2)
       (-0.4,0.4) circle (1);
       \draw[fill=white,even odd rule]  (-0.4, -0.4) circle (0.2)
       (-0.4,-0.4) circle (1);

         \draw[color=cyan] (0,0)--(-1.7, 0);
    \draw[color=cyan] (-0.5,0.5)--(-1.7, 0);
    \draw[color=cyan] (-0.5,-0.5)--(-1.7, 0);
    
    \draw (2.85, 0)--(0, 0);
    \draw (2, 1.75)--(0,0);
    \draw (2, 1.75)--(-0.4,0.4);
    
    \draw (1.5, 0.55)--(0,0);
    \draw (1.5, 0.55)--(-0.4,0.4);
    \draw (1.5, 0.55)--(-0.4,-0.4);
        \draw[color=red] (2.85, 0)--(2, 1.75);

    \draw[color=red] (1.5, 0.55)--(2.85,0);
    \draw[color=red] (1.5, 0.55)--(2.85,0);
    \draw[color=red] (1.5, 0.55)--(2, 1.75);
    \draw \smin;
    \draw \smax;
     
    \end{pgfonlayer}

\end{tikzpicture}\label{fig:c}}}
\hfil
\sidesubfloat[]{\scalebox{0.65}{ 
\def \agentone {(0,0) circle (0.2) node {$a_1$}}
\def \agenttwo {(-0.4,0.4) circle (0.2) node {$a_2$}}
\def \agentthree {(-0.4,-0.4) circle (0.2) node {$a_3$}}
\def \agentfour {(-1.7, 0) circle (0.2) node {$a_4$}}

\def \contone {(2.8, 0) circle (0.2) node {$c_1$}}
\def \conttwo {(2, 1.75) circle (0.2) node {$c_2$}}
\def \contthree {(1.5, 0.55) circle (0.2) node {$c_3$}}
\def \contfour {(1.7, -0.7) circle (0.2) node {$c_4$}}

\def \smin {[red] 
(0,0) circle (1)
(-0.4,0.4) circle (1)
(-0.4,-0.4) circle (1)
(-1.7, 0) circle (1)

(2.8, 0) circle (1)
(2, 1.75) circle (1)
(1.5, 0.55) circle (1)
(1.7, -0.7) circle (1)
}
\def \smax {[black!45!green] 
(0,0) circle (3) 
}

\begin{tikzpicture}
\tikzstyle{every node}=[font=\small]
    \draw \agentone;
    \fill[fill=cyan] \agentone;
    
    \draw \agenttwo;
    \fill[fill=cyan] \agenttwo;
    
    \draw \agentthree;
    \fill[fill=cyan] \agentthree;
    
    \draw \agentfour;
    \fill[fill=cyan] \agentfour;
    
    \draw \contone;
    \fill[fill=red] \contone;
    
    \draw \conttwo;
    \fill[fill=red] \conttwo;
    
    \draw \contthree;
    \fill[fill=red] \contthree;
    
    \draw \contfour;
    \fill[fill=red] \contfour;
    
    
    
    
    



    \begin{pgfonlayer}{back}

         \filldraw[fill=cyan!10!white, draw=cyan!50!black]
     (1.91,-2.31) arc (320.4147020709557:367.670448104312:3)
     arc (352.329551895688:399.5852979290443:3)
     arc (50.414702070955684:-50.414702070955684:3);
     
     \filldraw[fill=cyan!30!white, draw=cyan!50!black]
     (1.91,-2.31) arc (-50.41470207095567:-106.44337961485542:3)
     arc (-73.55662038514458:-59.75731932588542:3)
     arc (-85.98247895634071:-7.6704481043120065:3)
     arc (7.6704481043120065:-39.58529792904434:3);
     
     \filldraw[fill=cyan!30!white, draw=cyan!50!black]
     (2.57,0.00) arc (7.6704481043120065:85.98247895634071:3)
        arc (59.75731932588542:73.55662038514458:3)
        arc(106.44337961485542:50.41470207095567:3)
        arc(39.58529792904434:-7.6704481043120065:3);
        
        \filldraw[fill=cyan!50!white, draw=cyan!50!black]
     (-0.19, -2.59) arc (-59.75731932588542:59.75731932588542:3)
        arc (85.98247895634071:7.6704481043120065:3)
        arc(-7.6704481043120065:-85.98247895634071:3);
        
          \filldraw[fill=cyan!70!white, draw=cyan!50!black]
     (-0.85, -2.88) arc (253.55662038514458:106.44337961485542:3)
     arc(73.55662038514458:-73.55662038514458:3);
     
         \draw[fill=white,even odd rule]  (0,0) circle (0.2)
                                   (0,0) circle (1);
       
       \draw[fill=white,even odd rule]  (-1.7, 0) circle (0.2)
       (-1.7,0) circle (1);
       \draw[fill=white,even odd rule]  (-0.4, 0.4) circle (0.2)
       (-0.4,0.4) circle (1);
       \draw[fill=white,even odd rule]  (-0.4, -0.4) circle (0.2)
       (-0.4,-0.4) circle (1);

         \draw[color=cyan] (0,0)--(-1.7, 0);
    \draw[color=cyan] (-0.5,0.5)--(-1.7, 0);
    \draw[color=cyan] (-0.5,-0.5)--(-1.7, 0);
    
    \draw (2.85, 0)--(0, 0);
    \draw (2, 1.75)--(0,0);
    \draw (2, 1.75)--(-0.4,0.4);
    
    \draw[color=red] (2.85, 0)--(2, 1.75);

    \draw (1.5, 0.55)--(0,0);
    \draw (1.5, 0.55)--(-0.4,0.4);
    \draw (1.5, 0.55)--(-0.4,-0.4);
    
    \draw[color=red] (1.5, 0.55)--(2.85,0);
    \draw[color=red] (1.5, 0.55)--(2.85,0);
    \draw[color=red] (1.5, 0.55)--(2, 1.75);
    
    \draw[color=red] (1.7, -0.7)--(2.85,0);
    \draw[color=red] (1.7, -0.7)--(2.85,0);
    \draw[color=red] (1.7, -0.7)--(2, 1.75);
    \draw[color=red] (1.7, -0.7)--(1.5, 0.55);
    
    \draw (1.7, -0.7)--(-0.4,0.4);
    \draw (1.7, -0.7)--(-0.4,-0.4);
    \draw (1.7, -0.7)--(0,0);

    \draw \smin;
    \draw \smax;
     
    \end{pgfonlayer}

\end{tikzpicture}\label{fig:d}}}
\caption{Conquering $a_1$ using four agents. (i) Agent $c_1$ is placed in the subarea of $O(a_1)$ that can only be observed by $a_1$. (ii) A second contaminated agent, $c_2$, is placed in the sub area observed by $a_1$ and $a_2$. (iii) In order to let $c_2$ preserve its state, we place another agent in the area observed by $a_1$, $a_2$ and $a_3$. (iv) In order to let $c_3$ preserve its state, we place another agent in the area observed by $a_1$, $a_2$ and $a_3$. At the end of the attack all the contaminated agents maintain their state and $a_1$ is conquered since it observes more contaminated agents than healthy ones.}
    \label{fig:possible_attack}
    \end{figure}
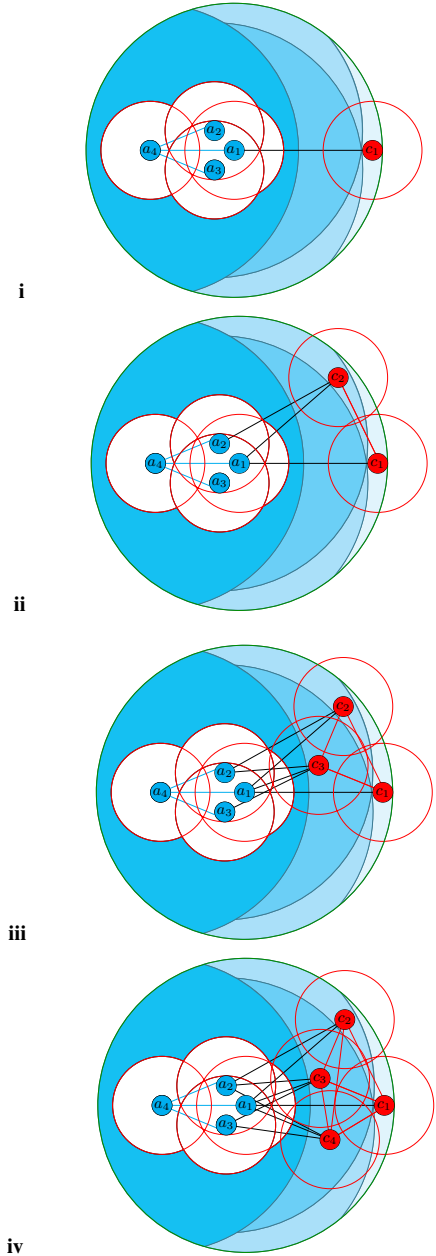

As an example, Figure \ref{fig:possible_attack} illustrates a possible attack which conquers $a_1$ using four contaminated agents.
In this paper we use the aforementioned lower bound  as an approximation of the bareness factor of each agent in the connected component. In other words, from now on when we refer to the bareness factor of an agent $a_i$ at time $t$ we refer to the value of the lower bound: $cf_{t}(a_i) + 1$.
From the attacker's perspective, there are several strategies that can be used when attacking a connected component of agents $C \in \ccnodes$. Initially, the attacker can only conquer agents which are on the fence of $C$ and he can do it by conquering one agent at a time or multiple at once. In most cases, it is infeasible to conquer all of the agents in $C$ at once. Therefore, an attacker must perform an iterative attack in which he conquers a subset of agents of $C$ at a time, this subset can even contain all of the agents in $C$ when we can conquer all of them at once. We refer to an algorithm that performs such an attack as an \textit{iterative conquering algorithm}, or iterative algorithm, in short.

There is a large variety of possible iterative algorithms that can be used to conquer a connected component of agents. Let $\mathbb{I}$ be the set of all the possible iterative conquering algorithms. We wish to find the iterative algorithm $A \in \mathbb{I}$ which conquers a connected component while allocating the minimal number of agents in the process.
The structure of each iterative algorithm is similar. Iterative algorithms mainly differ in the decision of which agents will be conquered at each iteration of the algorithm.

\begin{defn}
The \textit{decision rule} of an iterative conquering algorithm $A \in \mathbb{I}$, denoted by \\${\mathcal{D}^{A}:\ccnodes \times 2^{V_{t}(\mathcal{S})} \times \mathbb{Z}^{+} \mapsto 2^{V_{t}(\mathcal{S})}}$, is a function that receives a connected component of agents, the set of agents that were already conquered by algorithm $A$ and the current time of the game and returns the set of agents that should be conquered by algorithm $A$ at its next iteration.
\end{defn}
The decision rule embodies the core of the iterative algorithm as it is the function that decides which agents of the connected component will be conquered at each iteration.

\begin{algorithm}[!h]
    \SetKwComment{Comment}{$\triangleright$\ }{}
  \DontPrintSemicolon
\KwIn{Connected component of agents $C$.\\
    \qquad~~~~ Current time step $t$.\\
    \qquad~~~~ Decision rule $\mathcal{D}^{A}$.}
\KwOut{Number of required agents to conquer all the agents in $C$.}
        $\DataSty{r} \gets 0$  \Comment*[f]{Counts number of required agents to conquer $C$.}\\
        $\DataSty{c} \gets 0$ \Comment*[f]{Counts the total number of agents, includes agents of $C$.}\\
        $\mathcal{H} \gets \{\}$ \Comment*[f]{Set of all the agents of $C$ that $A$ conquered.}\\
        \While{$\mathcal{H} \neq C$}{
            $\Gamma \gets \mathcal{D}^{A}(C, \mathcal{H}, t)$\\
            $m \gets \max\limits_{a_{j} \in \Gamma}\{ \FuncSty{PBF}(a_{j},t,\mathcal{H})\}$\\
            $\Delta \gets m - c$\\
            \If{$\Delta > 0$}{
                $r \gets r + \Delta$\\
                $c \gets c + \Delta$
            }
            $c \gets c + \lvert \Gamma \rvert$\\
            $\mathcal{H} \gets \mathcal{H} \cup \Gamma$\\
        }
        \Return \DataSty{r}
        
        \SetKwFunction{PBF}{PBF}
  \SetKwProg{Fn}{Function}{:} {\KwRet \DataSty{barenessFactor}}
  \Fn{\PBF{$a_{j}$, $t$, $\mathcal{H}$}}
        {
        $\DataSty{removed} \gets 0$\\
        \ForEach{$a_{k} \in O(a_{j})$}{
            \If{$a_{k} \in \mathcal{H}$}{
                $\DataSty{removed} \gets \DataSty{removed} + 1$\\
            }
        }
        $\DataSty{barenessFactor} \gets cf_{t}(a_{j}) + 1 - \DataSty{removed}$\\
        }
    \caption{
    $\FuncSty{
    IterativeConqueringAlgorithm}$\newline($C, t, \mathcal{D}^{A}$)
    }
\end{algorithm}

Algorithm 1 describes the general form of an iterative conquering algorithm. The algorithm iteratively conquers agents of the connected component and stops when the whole connected component is conquered. At each iteration the algorithm uses its decision rule to decide which subset of agents must be conquered at the current iteration. The algorithm computes the maximal predicted bareness factor of all the agents in the chosen subset, the predictions are made using the \FuncSty{PBF} function which receives an agent and the set of all conquered agents and returns its predicted bareness factor. If the maximal bareness factor is larger than the total number of agents accumulated thus far, then it means that more agents are required to conquer the current subset of agents. Otherwise, we simply conquer the subset of agents without allocating any additional agents in the process. Finally, the algorithm returns the number of required agents to conquer all the agents in $C$.

The execution of an iterative algorithm $A \in \mathbb{I}$ on a connected component $C \in \ccnodes$ can be described by a function which maps each iteration to the set of agents conquered during it.
\begin{defn}
Given a connected component $C \in \ccnodes$ and iterative algorithm $A \in \mathbb{I}$, the \textit{attacking sequence} based on $A$, denoted by $\phi_{C}^{A}: \mathbb{Z} \mapsto 2^{C}$, maps each iteration of the attack against $C$ according to $A$ to the subset of agents of $C$ that will be conquered in this iteration. Further, let $\Phi(C)$ be the set of all the possible attacking sequences of $C$.
\end{defn}

In other words, when conquering connected component $C \in \ccnodes$ at time $t$ by following attacking sequence $\phi_{C}$ we conquer the agents in $\phi_{C}(i)$ at iteration $i$ of our iterative attack.

\begin{defn}
The \textit{length} of an attacking sequence ${\phi_{C} \in \Phi(C)}$ of connected component ${C \in \ccnodes}$, denoted by $\ell(\phi_{C})$, is defined to be the number of steps which are required to conquer $C$ according to $\phi_{C}$. Formally,
$$\ell(\phi_{C}) = \max\limits_{i \in \mathbb{Z}}\{i|~~\phi_{C}(i) \neq \varnothing\}$$
\end{defn}

Given a connected component $C \in \ccnodes$ and an attacking sequence $\phi_{C} \in \Phi(C)$, let $N(\phi_{C})$ be the total number of required agents to conquer $C$ based on attacking sequence $\phi_{C}$. It is equivalent to the output of the iterative algorithm that executes the attacking sequence $\phi_{C}$, i.e., given iterative algorithm ${A \in \mathbb{I}}$ and connected component ${C \in \ccnodes}$ we have that ${A(C) = N(\phi_{C}^{A})}$. 
Given a connected component ${C \in \ccnodes}$ and iterative algorithm ${A \in \mathbb{I}}$ (resp. attacking sequence ${\phi_{C} \in \Phi(C)}$), denote by $c_{i}^{A}(C)$ (resp. $c_{i}^{\phi_{C}}$) the value of $\DataSty{c}$ at the $\mathrm{i}_{\mathrm{th}}$ iteration of $A$ (resp. $\phi_{C}$) given input $C$. Similarly, denote by $r_{i}^{A}(C)$ (resp. $r_{i}^{\phi_{C}}$) the value of $\DataSty{r}$ at the $\mathrm{i}_{\mathrm{th}}$ iteration of $A$ (resp. $\phi_{C}$) given input $C$. The connectivity factor of agents changes throughout the execution of the iterative algorithm based on the agents that were conquered in previous iterations. Given an agent $a_j \in C \in \ccnodes$, let $cf_{t}^{A_{i}}(C, a_j)$ (resp. $cf_{t}^{\phi_{C_{i}}}(C, a_j)$) and $b_{t}^{A_{i}}(C, a_j)$ (resp. $b_{t}^{\phi_{C_{i}}}(C, a_j)$) be the connectivity and bareness factors of $a_{j}$ following $i$ iterations of iterative algorithm $A \in \mathbb{I}$ (resp. attacking sequence $\phi_{C} \in \Phi(C)$) given input $C$ at time $t$, respectively.

Similarly to previous notations, we denote the value of variables $m$, $\Delta$ and $\mathcal{H}$ at iteration $i$ of algorithm $A \in \mathbb{I}$ given input $C$ by $m_{i}^{A}(C)$,$\Delta_{i}^{A}(C)$ and $\mathcal{H}_{i}^{A}(C)$, respectively. Furthermore, given an agent $a_{j} \in C \in \ccnodes$ and an iterative algorithm $A \in \mathbb{I}$ (resp. attacking sequence $\phi_{C} \in \Phi(C)$), let $\Delta_{i}^{A}(C, a_{j})$ (resp. $\Delta_{i}^{\phi_{C}}(C, a_{j})$)  be the difference between the bareness factor of $a_{j}$ and the total number of agents accumulated by algorithm $A$ (resp. attacking sequence $\phi_{C}$) at its $\mathrm{i}_{\mathrm{th}}$ iteration given input $C$, i.e.,
\begin{align*}
    \Delta_{i}^{A}(C, a_{j}) &= cf_{t}^{A_{i}}(C, a_j) + 1 - c_{i}^{A}(C)\\
    \Delta_{i}^{\phi_{C}}(C, a_{j}) &= cf_{t}^{\phi_{C_{i}}}(C, a_j) + 1 - c_{i}^{\phi_{C}}(C)
\end{align*}
Following iteration $i$ of iterative algorithm $A \in \mathbb{I}$ given input $C \in \ccnodes$, the values of $c_{i+1}^{A}(C)$ and $r_{i+1}^{A}(C)$ are decided based on the difference between the maximal bareness factor,$m_{i}^{A}(C)$, and the current number of agents maintained by the algorithm $c_{i}^{A}(C)$.

The values of $c_{i+1}^{A}(C)$ and $r_{i+1}^{A}(C)$ are decided majorly by the value of $\Delta_{i+1}^{A}(C)$. If $\Delta_{i+1}^{A}(C) > 0$ then this means that more agents need to be allocated by the algorithm, whereas if $\Delta_{i+1}^{A}(C) \leq 0$ then it means that the attacker can conquer the set chosen by its decision rule without allocating any additional agents. Formally, when conquering connected component $C \in \ccnodes$ the values of $c_{i+1}^{A}(C)$ and $r_{i+1}^{A}(C)$ are updated based on the following update rules:
\begin{align*}\tag{$\star$}\label{eqn:update}
    \Delta_{i+1}^{A}(C) &= \max\limits_{a_{j} \in \mathcal{D}^{A}(C,\mathcal{H}_{i+1}^{A}(C), t)} \Delta_{i+1}^{A}(C, a_{j})\\
    c_{i+1}^{A}(C) &= c_{i}^{A}(C) + \lvert \mathcal{D}^{A}(C,\mathcal{H}_{i+1}^{A}(C), t) \rvert + \mathds{1}_{\Delta_{i+1}^{A}(C) > 0} * \Delta_{i+1}^{A}(C)\\
    r_{i+1}^{A}(C) &= r_{i}^{A}(C) + \mathds{1}_{\Delta_{i}^{A}(C) > 0} * \Delta_{i+1}^{A}(C)
\end{align*}

We can represent an attacking sequence in a simpler form, we define a singular attacking sequence as follows:
\begin{defn}
Given a connected component $C \in \ccnodes$, an attacking sequence $\phi_{C} \in \Phi(C)$ is called a \textit{singular attacking sequence} if it produces an iterative attack that conquers a single agent at each iteration. Formally, $\phi_{C}$ is singular if
$$ \lvert \phi_{C}(i) \rvert = 1 ~~\forall~0 \leq i \leq \ell(\phi_{C})$$
\end{defn}

It can be seen that every attacking sequence $\phi_{C} \in \Phi(C)$ can be transformed into a singular attacking sequence $\phi_{C}'$. Moreover, we can show that $N(\phi_{C}) \geq N(\phi_{C}')$.
\begin{lemma}\label{singularity}
Given a connected component $C \in \ccnodes$ at time $t$, each attacking sequence ${\phi_{C} \in \Phi(C)}$ which is not singular can be transformed into singular form ${\phi_{C}' \in \Phi(C)}$ such that ${N(\phi_{C}) \geq N(\phi_{C}')}$.
\end{lemma}
\begin{proof}
First, we describe a procedure that converts a given attacking sequence $\phi_{C} \in \Phi(C)$ into a singular attacking sequence $\phi_{C}' \in \Phi(C)$.

\begin{algorithm}[!b]
    \renewcommand{\algorithmcfname}{Procedure}
    \renewcommand{\thealgocf}{}
    \SetKwComment{Comment}{$\triangleright$\ }{}
    
     \DontPrintSemicolon
\KwIn{Attacking sequence $\phi_{C}$.\\
    \qquad~~~~ Current time step $t$.\\}
\KwOut{A singular attacking sequence $\phi_{C}'$ which is equivalent to $\phi_{C}$.}
    
        $\phi_{C}' \gets \{\}$\\
        $\mathcal{H} \gets \{\}$\\
        $\DataSty{iter} \gets 0$\\
        \For{$i \gets 0$ to $\ell(\phi_{C})$}{
            \eIf{$\lvert \phi_{C}(i) \rvert == 1$}{
                $\phi_{C}'(\DataSty{iter}) \gets \phi_{C}(i)$\\
            }{
                $\DataSty{factorMap} \gets \FuncSty{map}()$\\
                \ForEach{$a_{j} \in \phi_{C}(i)$}{
                    $\DataSty{factorMap}[a_{j}] \gets$  \FuncSty{PBF($a_{j}, t, \mathcal{H}$)}\\
                }
                $\DataSty{factorMap}$.\FuncSty{sortValues}()\\
                $\DataSty{numAgent} \gets 0$\\
                \ForEach{$a_{j}$ \KwSty{in} \DataSty{factorMap}.\FuncSty{keys}()}{
                    $\phi_{C}'(\DataSty{iter} + \DataSty{numAgent}) = \{a_{j}\}$\\
                    
                    $\DataSty{numAgent} \gets \DataSty{numAgent} + 1$\\
                }
            }
            
            $\mathcal{H} \gets \mathcal{H} \cup \phi_{C}(i)$\\
            $\DataSty{iter} \gets \DataSty{iter} + \lvert \phi_{C}(i) \rvert$\\}
        \Return $ \phi_{C}'$        
    \caption{\FuncSty{TransformSequence}($\phi_{C}$, $t$)}
\end{algorithm}

The \FuncSty{TransformSequence} procedure converts a given attacking sequence ${\phi_{C} \in \Phi(C)}$ to one of singular form ${\phi_{C}' \in \Phi(C)}$ by going over the iterations of the attacking sequence and decomposing each iteration ${0 \leq j \leq \ell(\phi_{C})}$ of $\phi_{C}$ that conquers several agents into $\lvert \phi_{C}(j) \rvert$ iterations in $\phi_{C}'$ while maintaining the iterations that conquer a single agent (lines 6-8). In lines 8-19 of the procedure we predict the bareness factor of each agent in $\phi_{C}(i)$ and separate them in the singular attacking sequence $\phi_{C}'$.

The attacking sequence that is returned by the procedure is indeed singular. Furthermore, the length of the transformed attacking sequence is necessarily larger than the length of $\phi_{C}$, i.e., $\ell(\phi_{C}') \geq \ell(\phi_{C})$.
It is left to be shown that $N(\phi_{C}') \leq N(\phi_{C})$.

Given attacking sequence $\phi_{C} \in \Phi(C)$, assume there is at least one iteration $0 \leq i \leq \ell(\phi_{C})$ where $\lvert \phi_{C}(i) \rvert > 1$ and let $0 \leq j \leq \ell(\phi_{C})$ be the first such iteration, meaning that $c_{j-1}^{\phi_{C}}(C) = c_{j-1}^{\phi_{C}'}(C)$.

Let ${\phi_{C}(j) = \{a_{i_{1}},\dots,a_{i_{n}}\}}$ be the agents in $\phi_{C}(j)$ ordered by their bareness factors, i.e., $$b_{t}^{\phi_{C_{j}}}(C, a_{i_{k}}) \leq b_{t}^{\phi_{C_{j}}}(C, a_{i_{k+1}}) ~~\forall ~0 \leq k < \lvert \phi_{C}(j) \rvert$$
For convenience, we will have that $\Delta_{i_{k}} \defeq \Delta_{j}^{\phi_{C}}(C, a_{i_{k}})$. Once we conquer the agent $a_{i_{k}}$ it effects the values of $\Delta_{i_{m}}$ for any ${k < m \leq \lvert \phi_{C}(j) \rvert}$. Let $\alpha_{i_{k}}$ be the value of $\Delta_{i_{k}}$ when all the agents $\{a_{i_{1}},\dots,a_{i_{k-1}}\}$ are conquered. We will get the following series:
\begin{align*}
    \alpha_{i_{1}} &= \Delta_{i_{1}}\\
    \alpha_{i_{2}} &= \Delta_{i_{2}} - \Delta_{i_{1}} - 1\\
    &\vdots \\
    \alpha_{i_{n}} &= \Delta_{i_{n}} - \sum\limits_{k=1}^{n-1}\alpha_{i_{k}} - (n-1)\\
\end{align*}

Since the agents are ordered based on their bareness factors we know that
$$r_{j}^{\phi_{C}}(C) = r_{j-1}^{\phi_{C}}(C) + \alpha_{i_{n}}$$
Whereas when using attacking sequence $\phi_{C}'$ and conquering the agents in $\phi_{C}(j)$ the required number of agents increases by $\sum\limits_{k=1}^{n}\alpha_{i_{k}}$ which is maximal when $\alpha_{i_{k}} > 0$ for every ${i_{1} \leq i_{k} \leq i_{n}}$. Therefore, we have that

\begin{gather*}
    \sum\limits_{k=1}^{n}\alpha_{i_{k}} = \sum\limits_{k=1}^{n}\Delta_{i_{k}} - \sum\limits_{m=1}^{k-1}\alpha_{i_{m}} - (k-1) = \\
    \Delta_{i_{n}} - \sum\limits_{k=1}^{n-1}\alpha_{i_{k}} - (n-1) + (\sum\limits_{k=1}^{n-1}\Delta_{i_{k}}) -\\ \Bigg(\sum\limits_{k=1}^{n-1}\bigg((k-1) + \sum\limits_{m=1}^{k-1}\alpha_{i_{m}}\bigg)\Bigg) = 
    \Delta_{i_{n}} - (n-1) +\\ \sum\limits_{k=1}^{n-1}(\Delta_{i_{k}} - \alpha_{i_{k}}) - \Bigg(\sum\limits_{k=1}^{n-1}\bigg((k-1) + \sum\limits_{m=1}^{k-1}\alpha_{i_{m}}\bigg)\Bigg)= \\
    \Delta_{i_{n}} - (n-1) + \sum\limits_{k=1}^{n-1}\bigg((k-1) + \sum_{m=1}^{k-1}\alpha_{i_{m}}\bigg)-\\ \Bigg(\sum\limits_{k=1}^{n-1}\bigg((k-1) + \sum\limits_{m=1}^{k-1}\alpha_{i_{m}}\bigg)\Bigg) = \Delta_{i_{n}} - (n-1) < \Delta_{i_{n}}
\end{gather*}

This means that the required number of agents reduces when we conquer the agents of $\phi_{C}(j)$ one by one. This applies to each iteration $0 \leq j \leq \ell(\phi_{C})$ where $\lvert \phi_{C}(j) \rvert > 1$. Hence, we get that $N(\phi_{C}) \geq N(\phi_{C}')$. \quad\quad\quad\quad\quad\quad\quad\quad\qedsymbol 
\end{proof}
We can infer from Lemma 1 that an iterative algorithm that returns the minimal number of required agents to conquer a connected component of agents must operate based on a singular attacking sequence. Hence, we can narrow down the possible algorithms that we inspect to those that conquer one agent at a time.

To come up with an {\em optimal} iterative algorithm, one must provide a decision rule which requires in total the minimal number of agents to conquer any component. 
Inspired by the idiom \textit{"a chain is only as strong as its weakest link"}, we define the following:
\def \tkzscl{0.85}
  \begin{figure}[!t]
    \centering
\sidesubfloat[]{\scalebox{0.72}{ 
\def \agentone {(0,0) circle (0.25) node {$a_1$}}
\def \agenttwo {(-0.5,0.5) circle (0.25) node {$a_2$}}
\def \agentthree {(0.5,0.5) circle (0.25) node {$a_3$}}
\def \agentfour {(0.75, 1.25) circle (0.25) node {$a_4$}}
\def \agentfive {(-0.75, 1.25) circle (0.25) node {$a_5$}}

\def \agentsix {(0, 2) circle (0.25) node {$a_6$}}

\def \agentseven {(-1.25, 2) circle (0.25) node {$a_7$}}

\def \agenteight {(-1.5, 0.75) circle (0.25) node {$a_8$}}

\def \smin {[red] (0,0) circle (0.5) (-0.5,0.5) circle (0.5) (0.5,0.5) circle (0.5) (0.75,1.25) circle (0.5) (-0.75,1.25) circle (0.5) (0, 2) circle (0.5) (-1.25, 2) circle (0.5) (-1.5, 0.75) circle (0.5)}
\def \smax {[black!45!green] (0,0) circle (1.5) (-0.5,0.5) circle (1.5) (0.5,0.5) circle (1.5) (0.75,1.25) circle (1.5) (-0.75,1.25) circle (1.5)
(0, 2) circle (1.5) (-1.25, 2) circle (1.5) (-1.5, 0.75) circle (1.5)}

\begin{tikzpicture}[scale=\tkzscl]
\tikzstyle{every node}=[font=\small]
    \draw (-0.75,1.25) -- (-0.5,0.5) -- (0,0) -- (0.5,0.5) -- (0.75,1.25) -- (-0.75, 1.25)--(0,2)--(0.75,1.25);
    
    \draw (-1.25,2)--(-0.75,1.25);
    \draw (-1.25,2)--(0,2);
    \draw (-1.5, 0.75)--(-0.5,0.5);
    \draw (-1.5, 0.75)--(-0.75, 1.25);
    \draw (-1.5, 0.75)--(-1.25, 2);
    
    \draw (0.5,0.5)--(-0.75, 1.25);
    \draw (0.5,0.5)--(-0.5,0.5);
    \draw (-0.5,0.5)--(0.75, 1.25);
    
    
    
    
    
    
    

    \draw \agentone;
    \fill[fill=yellow] \agentone;
    
    \draw \agenttwo;
    \fill[fill=cyan] \agenttwo;
    
    \draw \agentthree;
    \fill[fill=yellow] \agentthree;
    
    \draw \agentfour;
    \fill[fill=yellow] \agentfour;
    
    \draw \agentfive;
    \fill[fill=cyan] \agentfive;
    
    \draw (-4,-0.5) node 
                       {$\phi_{C}^{wpc}(0) = \{a_1\}$};
    \draw (-4,-1) node 
                       {$c_{0}^{wpc}(C) = 0 ~~r_{0}^{wpc}(C)= 0$};
    
    \draw \agentsix;
    \fill[fill=yellow] \agentsix;
    
    \draw \agentseven;
    \fill[fill=yellow] \agentseven;
    
    \draw \agenteight;
    \fill[fill=yellow] \agenteight;
    
    
    
    
    
    

    \begin{pgfonlayer}{back}
    \draw \smin;
    \draw \smax;
    \end{pgfonlayer}

\end{tikzpicture}\label{fig:a}}}
\hfil
\sidesubfloat[]{\scalebox{0.72}{ 
\def \agentone {(0,0) circle (0.25) node {$a_1$}}
\def \agenttwo {(-0.5,0.5) circle (0.25) node {$a_2$}}
\def \agentthree {(0.5,0.5) circle (0.25) node {$a_3$}}
\def \agentfour {(0.75, 1.25) circle (0.25) node {$a_4$}}
\def \agentfive {(-0.75, 1.25) circle (0.25) node {$a_5$}}

\def \agentsix {(0, 2) circle (0.25) node {$a_6$}}

\def \agentseven {(-1.25, 2) circle (0.25) node {$a_7$}}

\def \agenteight {(-1.5, 0.75) circle (0.25) node {$a_8$}}

\def \smin {[red] (-0.5,0.5) circle (0.5) (0.5,0.5) circle (0.5) (0.75,1.25) circle (0.5) (-0.75,1.25) circle (0.5) (0, 2) circle (0.5) (-1.25, 2) circle (0.5) (-1.5, 0.75) circle (0.5)}
\def \smax {[black!45!green] (-0.5,0.5) circle (1.5) (0.5,0.5) circle (1.5) (0.75,1.25) circle (1.5) (-0.75,1.25) circle (1.5)
(0, 2) circle (1.5) (-1.25, 2) circle (1.5) (-1.5, 0.75) circle (1.5)}

\begin{tikzpicture}[scale=\tkzscl]
\tikzstyle{every node}=[font=\small]
    \draw (-0.75,1.25) -- (-0.5,0.5) -- (0.5,0.5) -- (0.75,1.25) -- (-0.75, 1.25)--(0,2)--(0.75,1.25);
    
    \draw (-1.25,2)--(-0.75,1.25);
    \draw (-1.25,2)--(0,2);
    \draw (-1.5, 0.75)--(-0.5,0.5);
    \draw (-1.5, 0.75)--(-0.75, 1.25);
    \draw (-1.5, 0.75)--(-1.25, 2);
    
    \draw (0.5,0.5)--(-0.75, 1.25);
    \draw (0.5,0.5)--(-0.5,0.5);
    \draw (-0.5,0.5)--(0.75, 1.25);
    
    
    
    
    
    
    

    
    \draw \agenttwo;
    \fill[fill=yellow] \agenttwo;
    
    \draw \agentthree;
    \fill[fill=yellow] \agentthree;
    
    \draw \agentfour;
    \fill[fill=yellow] \agentfour;
    
    \draw \agentfive;
    \fill[fill=cyan] \agentfive;
    
    \draw (-4,-0.5) node 
                       {$\phi_{C}^{wpc}(1) = \{a_8\}$};
    \draw (-4,-1) node 
                       {$c_{1}^{wpc}(C) = 4 ~~r_{1}^{wpc}(C) = 3$};
    
    \draw \agentsix;
    \fill[fill=yellow] \agentsix;
    
    \draw \agentseven;
    \fill[fill=yellow] \agentseven;
    
    \draw \agenteight;
    \fill[fill=yellow] \agenteight;
    
    
    
    
    
    

    \begin{pgfonlayer}{back}
    \draw \smin;
    \draw \smax;
    \end{pgfonlayer}

\end{tikzpicture}\label{fig:b}}}

\medskip
\sidesubfloat[]{\scalebox{0.72}{ 
\def \agentone {(0,0) circle (0.25) node {$a_1$}}
\def \agenttwo {(-0.5,0.5) circle (0.25) node {$a_2$}}
\def \agentthree {(0.5,0.5) circle (0.25) node {$a_3$}}
\def \agentfour {(0.75, 1.25) circle (0.25) node {$a_4$}}
\def \agentfive {(-0.75, 1.25) circle (0.25) node {$a_5$}}

\def \agentsix {(0, 2) circle (0.25) node {$a_6$}}

\def \agentseven {(-1.25, 2) circle (0.25) node {$a_7$}}

\def \agenteight {(-1.5, 0.75) circle (0.25) node {$a_8$}}

\def \smin {[red] (-0.5,0.5) circle (0.5) (0.5,0.5) circle (0.5) (0.75,1.25) circle (0.5) (-0.75,1.25) circle (0.5) (0, 2) circle (0.5) (-1.25, 2) circle (0.5)}
\def \smax {[black!45!green] (-0.5,0.5) circle (1.5) (0.5,0.5) circle (1.5) (0.75,1.25) circle (1.5) (-0.75,1.25) circle (1.5)
(0, 2) circle (1.5) (-1.25, 2) circle (1.5)}

\begin{tikzpicture}[scale=\tkzscl]
\tikzstyle{every node}=[font=\small]
    \draw (-0.75,1.25) -- (-0.5,0.5) -- (0.5,0.5) -- (0.75,1.25) -- (-0.75, 1.25)--(0,2)--(0.75,1.25);
    
    \draw (-1.25,2)--(-0.75,1.25);
    \draw (-1.25,2)--(0,2);
    
    \draw (0.5,0.5)--(-0.75, 1.25);
    \draw (0.5,0.5)--(-0.5,0.5);
    \draw (-0.5,0.5)--(0.75, 1.25);
    
    
    
    
    
    
    

    
    \draw \agenttwo;
    \fill[fill=yellow] \agenttwo;
    
    \draw \agentthree;
    \fill[fill=yellow] \agentthree;
    
    \draw \agentfour;
    \fill[fill=yellow] \agentfour;
    
    \draw \agentfive;
    \fill[fill=yellow] \agentfive;
    
    \draw (-4,-0.5) node 
                       {$\phi_{C}^{wpc}(2) = \{a_7\}$};
    \draw (-4,-1) node 
                       {$c_{2}^{wpc}(C) = 5 ~~r_{2}^{wpc}(C) = 3$};
    
    \draw \agentsix;
    \fill[fill=yellow] \agentsix;
    
    \draw \agentseven;
    \fill[fill=yellow] \agentseven;
    
    
    
    
    
    
    

    \begin{pgfonlayer}{back}
    \draw \smin;
    \draw \smax;
    \end{pgfonlayer}

\end{tikzpicture}\label{fig:c}}}
\hfil
\sidesubfloat[]{\scalebox{0.72}{ 
\def \agentone {(0,0) circle (0.25) node {$a_1$}}
\def \agenttwo {(-0.5,0.5) circle (0.25) node {$a_2$}}
\def \agentthree {(0.5,0.5) circle (0.25) node {$a_3$}}
\def \agentfour {(0.75, 1.25) circle (0.25) node {$a_4$}}
\def \agentfive {(-0.75, 1.25) circle (0.25) node {$a_5$}}

\def \agentsix {(0, 2) circle (0.25) node {$a_6$}}

\def \agentseven {(-1.25, 2) circle (0.25) node {$a_7$}}

\def \agenteight {(-1.5, 0.75) circle (0.25) node {$a_8$}}

\def \smin {[red] (-0.5,0.5) circle (0.5) (0.5,0.5) circle (0.5) (0.75,1.25) circle (0.5) (-0.75,1.25) circle (0.5) (0, 2) circle (0.5)}
\def \smax {[black!45!green] (-0.5,0.5) circle (1.5) (0.5,0.5) circle (1.5) (0.75,1.25) circle (1.5) (-0.75,1.25) circle (1.5)
(0, 2) circle (1.5)}

\begin{tikzpicture}[scale=\tkzscl]
\tikzstyle{every node}=[font=\small]
    \draw (-0.75,1.25) -- (-0.5,0.5) -- (0.5,0.5) -- (0.75,1.25) -- (-0.75, 1.25)--(0,2)--(0.75,1.25);
    
    
    \draw (0.5,0.5)--(-0.75, 1.25);
    \draw (0.5,0.5)--(-0.5,0.5);
    \draw (-0.5,0.5)--(0.75, 1.25);
    
    
    
    
    
    
    

    
    \draw \agenttwo;
    \fill[fill=yellow] \agenttwo;
    
    \draw \agentthree;
    \fill[fill=yellow] \agentthree;
    
    \draw \agentfour;
    \fill[fill=yellow] \agentfour;
    
    \draw \agentfive;
    \fill[fill=yellow] \agentfive;
    \draw (-4,-0.5) node 
                       {$\phi_{C}^{wpc}(3) = \{a_6\}$};
    \draw (-4,-1) node 
                       {$c_{3}^{wpc}(C) = 6 ~~r_{3}^{wpc}(C) = 3$};
    
    \draw \agentsix;
    \fill[fill=yellow] \agentsix;
    
    
    
    
    
    
    
    

    \begin{pgfonlayer}{back}
    \draw \smin;
    \draw \smax;
    \end{pgfonlayer}

\end{tikzpicture}\label{fig:d}}}

\medskip
\sidesubfloat[]{\scalebox{0.72}{ 
\def \agentone {(0,0) circle (0.25) node {$a_1$}}
\def \agenttwo {(-0.5,0.5) circle (0.25) node {$a_2$}}
\def \agentthree {(0.5,0.5) circle (0.25) node {$a_3$}}
\def \agentfour {(0.75, 1.25) circle (0.25) node {$a_4$}}
\def \agentfive {(-0.75, 1.25) circle (0.25) node {$a_5$}}

\def \agentsix {(0, 2) circle (0.25) node {$a_6$}}

\def \agentseven {(-1.25, 2) circle (0.25) node {$a_7$}}

\def \agenteight {(-1.5, 0.75) circle (0.25) node {$a_8$}}

\def \smin {[red] (-0.5,0.5) circle (0.5) (0.5,0.5) circle (0.5) (0.75,1.25) circle (0.5) (-0.75,1.25) circle (0.5)}
\def \smax {[black!45!green] (-0.5,0.5) circle (1.5) (0.5,0.5) circle (1.5) (0.75,1.25) circle (1.5) (-0.75,1.25) circle (1.5)}

\begin{tikzpicture}[scale=\tkzscl]
\tikzstyle{every node}=[font=\small]
    \draw (-0.75,1.25) -- (-0.5,0.5) -- (0.5,0.5) -- (0.75,1.25) -- (-0.75, 1.25)--(0.75,1.25);
    
    
    \draw (0.5,0.5)--(-0.75, 1.25);
    \draw (0.5,0.5)--(-0.5,0.5);
    \draw (-0.5,0.5)--(0.75, 1.25);
    
    
    
    
    
    
    

    
    \draw \agenttwo;
    \fill[fill=yellow] \agenttwo;
    
    \draw \agentthree;
    \fill[fill=yellow] \agentthree;
    
    \draw \agentfour;
    \fill[fill=yellow] \agentfour;
    
    \draw \agentfive;
    \fill[fill=yellow] \agentfive;
    
    \draw (-4,-0.5) node 
                       {$\phi_{C}^{wpc}(4) = \{a_5\}$};
    \draw (-4,-1) node 
                       {$c_{4}^{wpc}(C) = 7 ~~r_{4}^{wpc}(C) = 3$};
    
    
    
    
    
    
    
    
    

    \begin{pgfonlayer}{back}
    \draw \smin;
    \draw \smax;
    \end{pgfonlayer}

\end{tikzpicture}\label{fig:e}}}
\hfil
\sidesubfloat[]{\scalebox{0.72}{ 
\def \agentone {(0,0) circle (0.25) node {$a_1$}}
\def \agenttwo {(-0.5,0.5) circle (0.25) node {$a_2$}}
\def \agentthree {(0.5,0.5) circle (0.25) node {$a_3$}}
\def \agentfour {(0.75, 1.25) circle (0.25) node {$a_4$}}
\def \agentfive {(-0.75, 1.25) circle (0.25) node {$a_5$}}

\def \agentsix {(0, 2) circle (0.25) node {$a_6$}}

\def \agentseven {(-1.25, 2) circle (0.25) node {$a_7$}}

\def \agenteight {(-1.5, 0.75) circle (0.25) node {$a_8$}}

\def \smin {[red] (-0.5,0.5) circle (0.5) (0.5,0.5) circle (0.5) (0.75,1.25) circle (0.5)}
\def \smax {[black!45!green] (-0.5,0.5) circle (1.5) (0.5,0.5) circle (1.5) (0.75,1.25) circle (1.5)}

\begin{tikzpicture}[scale=\tkzscl]
\tikzstyle{every node}=[font=\small]
    \draw (-0.5,0.5) -- (0.5,0.5)--(0.75,1.25);
    
    
    \draw (0.5,0.5)--(-0.5,0.5);
    \draw (-0.5,0.5)--(0.75, 1.25);
    
    
    
    
    
    
    

    
    \draw \agenttwo;
    \fill[fill=yellow] \agenttwo;
    
    \draw \agentthree;
    \fill[fill=yellow] \agentthree;
    
    \draw \agentfour;
    \fill[fill=yellow] \agentfour;
    
    
    \draw (-4,-0.5) node 
                       {$\phi_{C}^{wpc}(5) = \{a_4\}$};
    \draw (-4,-1) node 
                       {$c_{5}^{wpc}(C) = 8 ~~r_{5}^{wpc}(C) = 3$};
    
    
    
    
    
    
    
    
    

    \begin{pgfonlayer}{back}
    \draw \smin;
    \draw \smax;
    \end{pgfonlayer}

\end{tikzpicture}\label{fig:f}}}

\medskip
\sidesubfloat[]{\scalebox{0.72}{ 
\def \agentone {(0,0) circle (0.25) node {$a_1$}}
\def \agenttwo {(-0.5,0.5) circle (0.25) node {$a_2$}}
\def \agentthree {(0.5,0.5) circle (0.25) node {$a_3$}}
\def \agentfour {(0.75, 1.25) circle (0.25) node {$a_4$}}
\def \agentfive {(-0.75, 1.25) circle (0.25) node {$a_5$}}

\def \agentsix {(0, 2) circle (0.25) node {$a_6$}}

\def \agentseven {(-1.25, 2) circle (0.25) node {$a_7$}}

\def \agenteight {(-1.5, 0.75) circle (0.25) node {$a_8$}}

\def \smin {[red] (-0.5,0.5) circle (0.5) (0.5,0.5) circle (0.5)}
\def \smax {[black!45!green] (-0.5,0.5) circle (1.5) (0.5,0.5) circle (1.5)}

\begin{tikzpicture}[scale=\tkzscl]
\tikzstyle{every node}=[font=\small]
    \draw (-0.5,0.5) -- (0.5,0.5);
    
    
    \draw (0.5,0.5)--(-0.5,0.5);
    
    
    
    
    
    
    

    
    \draw \agenttwo;
    \fill[fill=yellow] \agenttwo;
    
    \draw \agentthree;
    \fill[fill=yellow] \agentthree;
    
    
    
    \draw (-4,-0.5) node 
                       {$\phi_{C}^{wpc}(6) = \{a_3\}$};
    \draw (-4,-1) node 
                       {$c_{6}^{wpc}(C) = 9 ~~r_{6}^{wpc}(C) = 3$};
    
    
    
    
    
    
    
    
    

    \begin{pgfonlayer}{back}
    \draw \smin;
    \draw \smax;
    \end{pgfonlayer}

\end{tikzpicture}\label{fig:g}}}
\hfil
\sidesubfloat[]{\scalebox{0.72}{ 
\def \agentone {(0,0) circle (0.25) node {$a_1$}}
\def \agenttwo {(-0.5,0.5) circle (0.25) node {$a_2$}}
\def \agentthree {(0.5,0.5) circle (0.25) node {$a_3$}}
\def \agentfour {(0.75, 1.25) circle (0.25) node {$a_4$}}
\def \agentfive {(-0.75, 1.25) circle (0.25) node {$a_5$}}

\def \agentsix {(0, 2) circle (0.25) node {$a_6$}}

\def \agentseven {(-1.25, 2) circle (0.25) node {$a_7$}}

\def \agenteight {(-1.5, 0.75) circle (0.25) node {$a_8$}}

\def \smin {[red] (-0.5,0.5) circle (0.5)}
\def \smax {[black!45!green] (-0.5,0.5) circle (1.5)}

\begin{tikzpicture}
\tikzstyle{every node}=[font=\small]
    \draw (-0.5,0.5);
    
    
    
    
    
    
    
    
    

    
    \draw \agenttwo;
    \fill[fill=yellow] \agenttwo;
    
    
    
    
    \draw (-4,-0.5) node 
                       {$\phi_{C}^{wpc}(7) = \{a_2\}$};
    \draw (-4,-1) node 
                       {$c_{7}^{wpc}(C) = 10 ~~r_{7}^{wpc}(C) = 3$};
    
    
    
    
    
    
    
    
    

    \begin{pgfonlayer}{back}
    \draw \smin;
    \draw \smax;
    \end{pgfonlayer}

\end{tikzpicture}\label{fig:h}}}
\caption{Example of the execution of the \wpc{} algorithm on a given connected component of healthy agents $C = \{a_1, a_2, \dots, a_8\}$. Healthy agents are colored in cyan whereas the bare ones are colored in yellow. At each iteration we conquer the bare agent with the minimal bareness factor where ties are broken arbitrarily.}
    \label{fig:wpc_example}
    \end{figure}
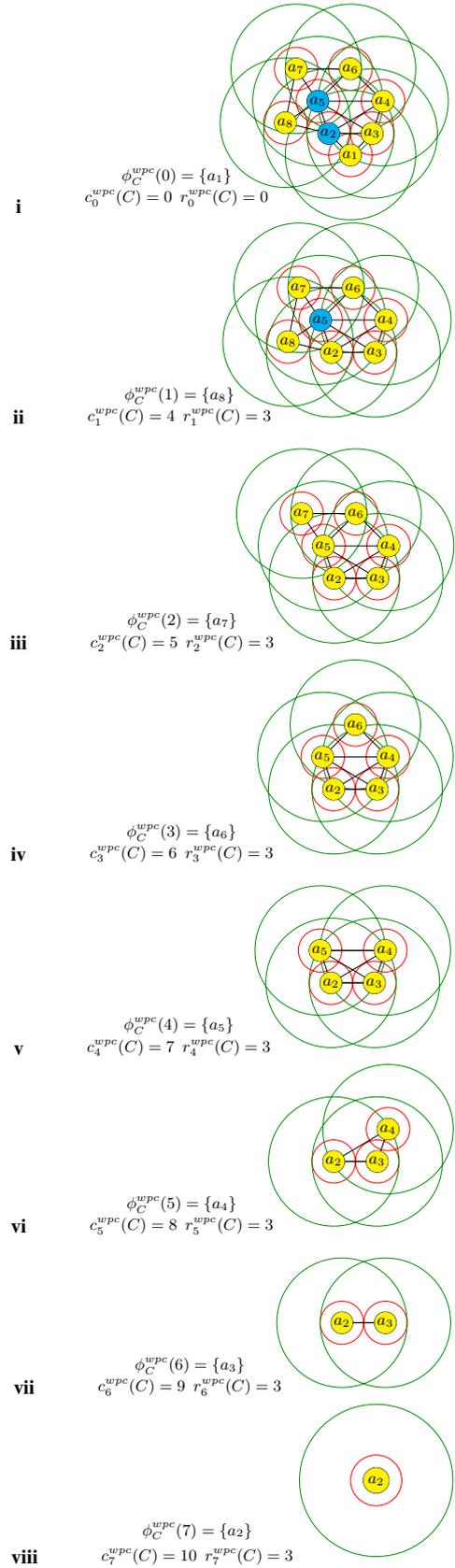

\begin{defn}
Given a connected component of agents ${C \in \ccnodes}$ at time $t$, the \textit{weak point} of $C$ is the bare agent with the minimal bareness factor, i.e. $$wp(C) = \argmin\limits_{a_i \in F(C)}\{b_{t}(a_i)\}$$
\end{defn}

Algorithm 2 represents an algorithm that iteratively conquers the weak points of a connected component of agents which we refer to as the Weak Point Conquer (WPC) algorithm. 

  \begin{algorithm}[!b]
  \renewcommand{\thealgocf}{2}
\KwIn{Connected component of agents $C$.\\
    \qquad~~~~ Current time step $t$.\\}
\KwOut{Number of required agents to conquer all the agents in $C$ by iteratively conquering its weak points.}
         \DataSty{wpcDr} $\gets {\{(C, \mathcal{H} ,t) \rightarrow \argmin\limits_{a_{j} \in F(C \setminus \mathcal{H})}\{\FuncSty{PBF}(a_{j},t,\mathcal{H})\}}\}$\\
    \Return \FuncSty{IterativeConqueringAlgorithm}($C$, $t$, \DataSty{wpcDr})\\
    \caption{\FuncSty{WPC}($C$, $t$)}
  \end{algorithm}

The WPC algorithm works based on the general iterative conquering algorithm presented in Algorithm 1. The algorithm iterates over the given connected component and conquers the weak point of the connected component as described in the $\DataSty{wpcDr}$ variable until the whole component is conquered.
Going forward, we refer to the output of the \wpc{} algorithm for the connected component of agents ${C \in \ccnodes}$, $wpc(C)$, as the \wpc{} value of $C$.

As an example, Figure \ref{fig:wpc_example} illustrates an execution of the \wpc{} algorithm on a connected component of eight healthy agents. At each iteration $i$ we conquer the weak point of the component and update $c_{i+1}^{wpc}(C)$ and $r_{i+1}^{wpc}(C)$ according to $\eqref{eqn:update}$.
We would like to keep track of the agents that are left of the component following $i$ iterations of an iterative algorithm $A$. Given a connected component $C \in \ccnodes$ and iterative algorithm $A \in \mathbb{I}$ denote by $C_{i}^{A} \subseteq C$ the set of agents that were not conquered after $i$ iterations of algorithm $A$, i.e., ${C_{i}^{A} = C \setminus \mathcal{H}_{i}^{A}}$.
To justify using the \wpc{} algorithm as a measure of the quality of a connected component of agents we have to prove the correctness of the output of the algorithm, i.e., we have to show that the \wpc{} algorithm returns the minimal number of agents which are required to conquer the given connected component as stated in the following lemma: 
\begin{theorem}
Given a connected component $C \in \ccnodes$ of $n$ agents, $wpc(C)$ returns the minimal number of agents which are required to conquer $C$.
\end{theorem}
\begin{proof}
Assume by contradiction that there is an iterative algorithm $A \in \mathbb{I}$ such that ${r_{\lvert C \rvert}^{A}(C) < r_{\lvert C \rvert}^{wpc}(C)}$. This means that there must be an iteration $t$ where \wpc{} and $A$ chose to conquer different agents. Denote the agent chosen by \wpc{} as $a_i$ and the agent chosen by $A$ as $a_j$. Since \wpc{} always chooses the weak point we know that $cf_{t}^{wpc}(C, a_i) \leq cf_{t}^{A}(C, a_j)$. This means that $\Delta_{t}^{wpc}(C, a_i) \leq \Delta_{t}^{A}(C, a_j)$ which implies that $r_{t+1}^{wpc}(C) \leq r_{t+1}^{A}(C)$.
We can now make the observation that the connectivity factor of all the agents in the connected component can only decrease as we iteratively conquer agents of the connected component, i.e., given iterative algorithm $A \in \mathbb{I}$ and connected component $C \in \ccnodes$ we have that ${cf_{t}^{A}(C, a_i) \geq cf_{t+1}^{A}(C, a_i) ~\forall a_i \in C}$. Consequently, this means that given iterative algorithm $A \in \mathbb{I}$ and connected component $C \in \ccnodes$ then $\Delta_{t}^{A}(C, a_i) \geq \Delta_{t+1}^{A}(C, a_i)$ for any agent $a_i \in S$.
Agent $a_j$ must be conquered by the \wpc{} algorithm at some iteration $t' > t$. 
At iteration $t'$ we have that $c_{t'}^{wpc}(C) > c_{t}^{wpc}(C) = c_{t}^{A}(C)$, i.e., \wpc{} has more agents at his disposal at iteration $t'$ in comparison to the number of agents held by algorithm $A$ at iteration $t$. Recall that \wpc{} iteratively conquers the agent with the minimal connectivity factor. From the update rule which is described in \eqref{eqn:update} we can claim that under the assumption that ${\Delta_{t}^{A}(C, a_j) > \Delta_{t}^{wpc}(C,a_j) > 0}$ we can say that
\begin{gather*}
    r_{t+1}^{A}(C) = r_{t}^{A}(C) + \Delta_{t}^{A}(C, a_j) =\\
    r_{t}^{wpc}(C) + \Delta_{t}^{wpc}(C, a_j) \geq  r_{t'+1}^{wpc}(C)
\end{gather*}

We can now compare the states of algorithms $A$ and \wpc{} following iterations $t$ and $t'$, we showed that the \wpc{} algorithm has more agents at his disposal and allocated less agents than the number that was allocated by algorithm $A$. Furthermore, we can obviously see that $C^{wpc}_{t'+1} \subseteq C^{A}_{t+1}$ and from the fact that the connectivity factor can only decrease we can say that 
$$\Delta_{t'+1}^{wpc}(C, a_k) \leq \Delta_{t+1}^{A}(C, a_k)~~\forall a_k \in C^{wpc}_{t'+1}$$
To conclude, we found that following iteration $t'$ the \wpc{} algorithm has allocated less agents than algorithm $A$ allocated in $t$ iterations while having more agents at his disposal. Furthermore, there are less agents left to be conquered and their connectivity factors are lower which means that we necessarily must have that ${r_{\lvert C \rvert}^{wpc}(C) \leq r_{\lvert C \rvert}^{A}(C)}$ in contradiction. \quad\quad\quad\quad\quad\quad\quad\quad\qedsymbol 
\end{proof}

Theorem 2 shows that the \wpc{} is the sole metric which can be used to quantify the number of agents that are required to conquer a connected component $C$, this means that $wpc(C)$ can be seen as a metric for the resiliency of $C$. It is important to note that no other metrics should be examined since we have shown that the \wpc{} metric computes the exact minimal number of agents that are required to conquer any component.
\subsection{Placing Agents in Connected Components}
We can conclude from the previous subsection that agents in the contamination game should aim to form connected components that have high \wpc{} values.
The \wpc{} value of a connected component of agents $C \in \ccnodes$ does not solely depend on the number of agents in $C$. The way the agents in $C$ are placed relative to the center of the connected component and to one another, or $C$'s \textit{placement}, plays a major role in determining the \wpc{} value of $C$. Formally, we define the placement of a component as follows: 

 \begin{defn}
Given a connected component $C \in \ccnodes$ of agents in $k$-dimensional space, a \textit{placement} $p: [1, n] \mapsto \mathbb{R}^k$ is a mapping between each agent's index in the component to its position relative to the center of the component. We denote the group of possible placements for connected component $C$ as $P(C)$. We denote the result of applying placement $p \in P(C)$ on the agents of $C$ by $C_{p}$, and the current placement of $C$ by $p_{C}$.
\end{defn}
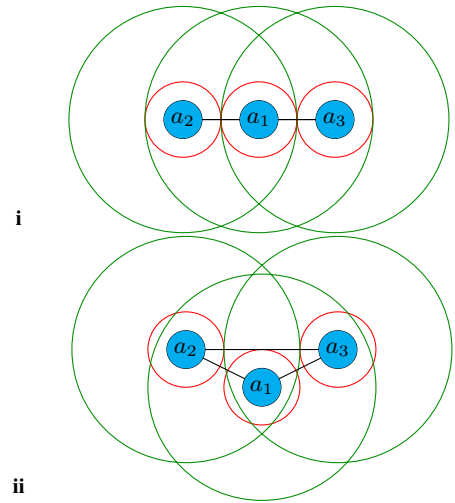
\begin{figure}[!b]
    \centering
\sidesubfloat[]{ 
\def \agentone {(0,0) circle (0.25) node {$a_1$}}
\def \agenttwo {(-1,0) circle (0.25) node {$a_2$}}
\def \agentthree {(1,0) circle (0.25) node {$a_3$}}

\def \smin {[red] (0,0) circle (0.5) (-1,0) circle (0.5) (1,0) circle (0.5)}
\def \smax {[black!45!green] 
(0,0) circle (1.5) 
(-1,0) circle (1.5)
(1,0) circle (1.5)}

\begin{tikzpicture}
\tikzstyle{every node}=[font=\small]

    \draw[-] (0,0)--(-1,0);
    \draw[-] (0,0)--(1,0);
    \draw[-] (-1,0)--(1,0);

    \draw \agentone;
    \fill[fill=cyan] \agentone;
    
    \draw \agenttwo;
    \fill[fill=cyan] \agenttwo;
    
    \draw \agentthree;
    \fill[fill=cyan] \agentthree;

    \begin{pgfonlayer}{back}
    \draw \smin;
    \draw \smax;
    \end{pgfonlayer}

\end{tikzpicture}\label{fig:a}}
\hfil
\sidesubfloat[]{ 
\def \agentone {(0,0) circle (0.25) node {$a_1$}}
\def \agenttwo {(-1,0.5) circle (0.25) node {$a_2$}}
\def \agentthree {(1,0.5) circle (0.25) node {$a_3$}}

\def \smin {[red] (0,0) circle (0.5) (-1,0.5) circle (0.5) (1,0.5) circle (0.5)}
\def \smax {[black!45!green] 
(0,0) circle (1.5) 
(-1,0.5) circle (1.5)
(1,0.5) circle (1.5)}

\begin{tikzpicture}
\tikzstyle{every node}=[font=\small]

    \draw[-] (0,0)--(-1,0.5);
    \draw[-] (0,0)--(1,0.5);
    \draw[-] (-1,0.5)--(1,0.5);

    \draw \agentone;
    \fill[fill=cyan] \agentone;
    
    \draw \agenttwo;
    \fill[fill=cyan] \agenttwo;
    
    \draw \agentthree;
    \fill[fill=cyan] \agentthree;

    \begin{pgfonlayer}{back}
    \draw \smin;
    \draw \smax;
    \end{pgfonlayer}

\end{tikzpicture}\label{fig:b}}
    \caption{Example of the effect of the placement of agents on $wpc(C)$, where $C=\{a_1,a_2,a_3\}$ is a connected component of three healthy agents.}
    \label{fig:placement_example}
\end{figure}
Figure \ref{fig:placement_example} represents two different placements of a connected component of agents, $C=\{a_1,a_2,a_3\}$. It illustrates the effect of the placement of agents on the value of $wpc(C)$. 
In the first subfigure the agents in $C$ are connected in a straight line such that the weak point of the connected component has a connectivity factor of $1$ which means that $wpc(C) = 1$. In the second subfigure all the agents in $C$ are connected to one another. Hence, we have that $wpc(C) = 2$ which is larger than the value of $wpc(C)$ in the first subfigure.

As previously mentioned, we search for a placement of agents that maximizes the \wpc{} value of connected components in our swarm, that is, an \textit{optimal placement}.
\begin{defn}
The \textit{optimal placement} of a connected component $C \in \ccnodes$, denoted by $p^{*}(C)$, is the placement which maximizes the output of the \wpc{} algorithm for the component, that is,
$$ p^*(C) = \argmax\limits_{p \in P(C)}wpc(C_{p})$$

We refer to the value of $wpc(C_{p^{*}(C)})$ as the \textit{maximal \wpc{}} of $C$.
\end{defn}

In previous work (\citet*{contaminationProblem}), it was shown that for each pair of values of $S_{min}$ and $S_{max}$ there is a maximal size of a clique of agents, which was referred to as a \textit{maximal stable cycle} (MSC), we denote this value by $c(S_{min}, S_{max})$, or $c$, in short. 
It can easily be seen that as long as the number of agents in a connected component is less than or equal to $c$ the optimal placement of the component would be a clique of agents whereas when the number of agents surpasses $c$ the optimal placement of the connected component becomes unclear. 
The \wpc{} value of a connected component $C \in \ccnodes$ is mainly effected by all the agents that required the algorithm to allocate additional agents to conquer it, or formally, all the iterations $i$ in which $\Delta_{i}^{wpc} > 0$. We refer to this subset of agents as the \textit{effective subset} of $C$ and denote it by 
$$ES(C) = \{\Delta_{i}^{wpc}(C, wp(C^{wpc}_{i})) > 0~|~~1 \leq i \leq \lvert C \rvert \} $$

\begin{figure}[!t]
    \centering
     \sidesubfloat[]{\scalebox{0.3}{ 
\def \agentone {(-3,0) circle (0.5) node {$a_1$}}
\def \agenttwo {(3,0) circle (0.5) node {$a_2$}}
\def \agentthree {(0,-5) circle (0.5) node {$a_3$}}

\def \agentfour {(0,-1.75) circle (0.5) node {$a_{4}$}}

\def \smin {[red] (-3,0) circle (2) (3,0) circle (2) (0,-5) circle (2) (0,-1.75) circle (2)
}
\def \smax {[black!45!green] 
(-3,0) circle (6) 
(3,0) circle (6)
(0,-5) circle (6)
(0,-1.75) circle (6)
}

\begin{tikzpicture}
\tikzstyle{every node}=[font=\Large]

    \draw[-] (-3,0)--(3,0);
    \draw[-] (3,0)--(0,-5);
    \draw[-] (-3,0)--(0,-5);
    
    \draw[-] (-3,0)--(0,-1.75);
    \draw[-] (3,0)--(0,-1.75);
    \draw[-] (0,-5)--(0,-1.75);

    \draw \agentone;
    \fill[fill=cyan] \agentone;
    
    \draw \agenttwo;
    \fill[fill=cyan] \agenttwo;
    
    \draw \agentthree;
    \fill[fill=cyan] \agentthree;
    
    \draw \agentfour;
    \fill[fill=cyan] \agentfour;

    \begin{pgfonlayer}{back}
    \draw \smin;
    \draw \smax;
    
    \end{pgfonlayer}

\end{tikzpicture}\label{fig:b}}}\\
    \sidesubfloat[]{\scalebox{0.3}{ 
\def \agentone {(-2.5,0) circle (0.5) node {$a_1$}}
\def \agenttwo {(3,0) circle (0.5) node {$a_2$}}
\def \agentthree {(-2.5,-5) circle (0.5) node {$a_3$}}

\def \agentfour {(3,-5) circle (0.5) node {$a_{4}$}}

\def \agentfive {(0.3,-1) circle (0.5) node {$a_{5}$}}

\def \agentsix {(0.3,-2.5)circle (0.5) node {$a_{6}$}}

\def \agentseven {(0.3,-4) circle (0.5) node {$a_{7}$}}

\def \smin {[red] (-2.5,0) circle (2) (3,0) circle (2) (-2.5,-5) circle (2) (3,-5) circle (2)
(0.3,-1) circle (2)
(0.3,-4) circle (2)
(0.3,-2.5) circle (2)
}
\def \smax {[black!45!green] 
(-2.5,0) circle (6) 
(3,0) circle (6)
(-2.5,-5) circle (6)
(3,-5) circle (6)
(0.3,-1) circle (6)
(0.3,-4) circle (6)
(0.3,-2.5) circle (6)
}

\begin{tikzpicture}
\tikzstyle{every node}=[font=\Large]
    
    \draw[-] (-2.5,0)--(-2.5,-5);
    \draw[-] (-2.5,0)--(3,0);
    
    \draw[-] (-2.5,-5)--(3,-5);
    \draw[-] (3,-5)--(3,0);
    
    \draw[-] (-2.5,0)--(0.3,-1);
    \draw[-] (-2.5,-5)--(0.3,-1);
    \draw[-] (3,0)--(0.3,-1);
    \draw[-] (3,-5)--(0.3,-1);
    
    \draw[-] (-2.5,0)--(0.3,-4);
    \draw[-] (-2.5,-5)--(0.3,-4);
    \draw[-] (3,0)--(0.3,-4);
    \draw[-] (3,-5)--(0.3,-4);
    
    \draw[-] (-2.5,0)--(0.3,-2.5);
    \draw[-] (-2.5,-5)--(0.3,-2.5);
    \draw[-] (3,0)--(0.3,-2.5);
    \draw[-] (3,-5)--(0.3,-2.5);
    
    

    \draw \agentone;
    \fill[fill=cyan] \agentone;
    
    \draw \agenttwo;
    \fill[fill=cyan] \agenttwo;
    
    \draw \agentthree;
    \fill[fill=cyan] \agentthree;
    
    \draw \agentfour;
    \fill[fill=cyan] \agentfour;
    
    \draw \agentfive;
    \fill[fill=cyan] \agentfive;
    
    \draw \agentsix;
    \fill[fill=cyan] \agentsix;
    
    \draw \agentseven;
    \fill[fill=cyan] \agentseven;

    \begin{pgfonlayer}{back}
    \draw \smin;
    \draw \smax;
    
    \end{pgfonlayer}

\end{tikzpicture}}}
   
    \caption{Two examples of monotonic connected components. The components in both subfigures have an effective set which is entirely comprised of agents in the fence of the component.}
    \label{fig:monotonic_example}
    
\end{figure}
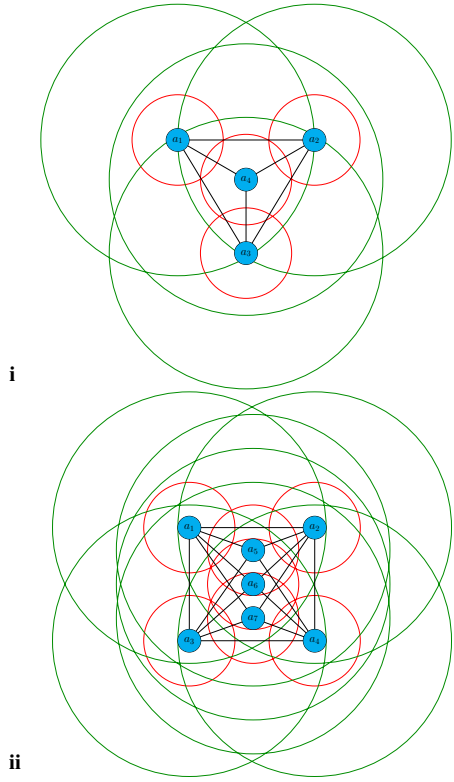

Since there is no way of predicting the probable effective subset of a general connected component, we define the following class of components:
\begin{defn}
A connected component of agents ${C \in \ccnodes}$ is called a \textit{monotonic connected component} if the effective subset of agents of $C$ is a subset of its fence, that is, $ES(C) \subseteq F(C)$.
\end{defn}
Figure \ref{fig:monotonic_example} illustrates several examples of monotonic connected components. It can be seen that the agents on the fence of monotonic components have high connectivity to the rest of the agents in the component in comparison to the ones which are enclosed inside the component.
The analysis of the problem of finding an optimal placement of $n$ agents becomes clearer once we choose to focus on the subset of monotonic connected components. Therefore, we analyse the complexity of the problem of finding an optimal placement of a monotonic component and deduce from it the complexity of the original problem.
We can separate the problem of finding an optimal placement for a monotonic connected component of $n$ agents into two separate ones. First, we must find the number of agents that we must allocate for the fence of the component. Second, we need to place the remaining agents inside the computed fence such that the component remains monotonic. Focusing first on the latter problem, we examine the problem of finding the placement of $n$ agents in a given area that maximizes the \wpc{} of a fixed fence as the Fixed Maximum WPC (FMWPC) problem. Since we must handle concealments of agents while placing agents in a given area, the effect of placing an agent in a given location $\ell$ in area $A$ can be described by a utility function that receives the observing agent from the fence, locations of agents placed in $A$ and the location in $A$ in which we wish to place an agent. Formally, given a set of agents $C$ and set of locations in two-dimensional space $A$, a utility function for the FMWPC problem, denoted by ${u: C \times \mathbb{R}^{2} \times 2^{A} \mapsto \mathbb{R}}$, computes the effect of placing an agent in a given location in our two-dimensional space given the locations of the other placed agents in area $A$ and the observing agent from $C$.
Formally, the problem is defined as follows:

{\small
\begin{quote}
{\bf Fixed Maximum WPC Problem (FMWPC):}\\
{\em Instance:} A fixed fence of $m$ agents $C=\{a_{1}, \dots, a_{m}\}$, a set of locations in which we can place agents $A$, a number $n$ and a utility function $u$.\\
{\em Objective:} Find the subset of at most $n$ locations  from $A$ that maximizes the minimal utility of any agent in $C$, that is,
 $$\argmax\limits_{A' \subseteq A~s.t.~\lvert A' \rvert \leq n} \min_{a_{i} \in C}\sum\limits_{\ell \in A'} u(a_{i}, \ell, A')$$
\end{quote}
}
The FMWPC problem is NP-hard, by reduction from the geometric version of the Minimum Membership Set Cover problem (MMSC) (\citet*{erlebach2008approximating}). In this problem a set of points must be covered using a given set of circles such that the maximal number of circles covering the same point is minimal.

\begin{theorem}
The FMWPC problem is NP-hard.
\end{theorem}
\begin{proof}
We prove this theorem by reduction from the geometric version of the Minimum Membership Set Cover problem. Let $\mathcal{P}$ and $\mathcal{R}$ be the set of points and circles in an instance of the MMSC problem where the circles have a radius of $r > 0$. Given the instance ($\mathcal{P}$, $\mathcal{R}$) a reduction can be made to an instance of the FMWPC problem using the following set of steps.

First, let $\mathcal{P}$ be the locations of the set of agents in $C$. Secondly, let the set of centers of the circles in $\mathcal{R}$ be the set of available locations $A$ and set $n$ to be equal to $\lvert \mathcal{R}\rvert$. Finally, define our utility function to be the following:
$$ u(a_{i}, \ell, A') = \left\{\begin{array}{lr}
        -1 & \lVert a_{i} - \ell \rVert_{k} \leq r\\
        0 & \text{Otherwise} 
        \end{array}\right\} $$
        
That is, the utility of an agent is the negation of the number of agents which are at a distance smaller than $r$ from it. The set of aforementioned transformations results in an instance of the FMWPC problem. It remains to be shown that a solution to the FMWPC problem is equivalent to a solution of the initial instance of the MMSC problem.

In the MMSC problem the \textit{membership} of a point $p \in \mathcal{P}$ in a subset of circles $\mathcal{R}' \subseteq \mathcal{R}$, denoted by $mem_{\mathcal{R}'}(p)$ is defined to be the number of circles in $\mathcal{R}'$ containing $p$. That is, if the radii of all the circles in $\mathcal{R}$ are equal to $r$ the membership of a point $p \in \mathcal{P}$ in a given subset of circles $\mathcal{R}' \subseteq \mathcal{R}$ is the number of circles in $\mathcal{R}'$ whose centers are closer than $r$ to $p$. The objective of the MMSC problem is to find the subset of circles which results in the minimal maximal membership of a point in $\mathcal{P}$ and covers all the points in $\mathcal{P}$, or formally find the following subset of circles:
$$\mathcal{R}^{*} = \argmin_{\{\mathcal{R}' \subseteq \mathcal{R}~|~mem_{\mathcal{R}'}(p) \geq 1 ~~\forall p \in \mathcal{P}\}} \max_{p \in \mathcal{P}} mem_{\mathcal{R}'}(p)$$

In the case of the aforementioned instance of the FMWPC problem, the objective is to find the subset of locations in $A$ of size at most $\lvert \mathcal{R} \rvert$ such that the minimal utility of any agent in $C$ is maximal. The utility of an agent in $C$ in the reduced instance of the FMWPC problem is equal to the negation of the membership of its matching point in $\mathcal{P}$. Therefore, given a subset of circles $\mathcal{R}'$ and its matching set of locations in $A$ (the centers of these circles), the point with the maximal membership is the location of the agent in $C$ with the minimal utility. Hence, the solution to the reduced instance of the FMWPC is the subset of locations in $A$ which results in the maximal minimal utility, or in other words, the set of circles in $\mathcal{R}$ that result in the minimal maximal membership in the initial instance of the MMSC problem.

This shows that the proposed polynomial reduction results in an instance of the FMWPC problem such that a solution to the reduced instance corresponds to a solution for the initial instance in the MMSC problem which proves our claim. \quad\qedsymbol
\end{proof}

The hardness of the FMWPC problem directly implies that the problem of finding the optimal placement of $n$ agents in a monotonic connected component is NP-hard. Altogether, Theorem 3 leads to the following corollary:
\begin{corollary}
The problem of finding the optimal placement of $n$ agents in a general connected component is NP-hard.
\end{corollary}
Since finding the optimal placement among monotonic connected components is NP-hard we can directly conclude that finding the optimal placement among general connected components is NP-hard.

\section{Bounding the \wpc{} Value}
\label{sec:4boundwpc}
We have shown that finding an optimal placement is NP-hard. This means that constructing an algorithm that finds an optimal placement for a given number of agents is improbable. Nevertheless, finding an upper bound for the \wpc{} value of a connected component can help us come up with an optimal strategy for players in the contamination game.

\begin{figure}[!t]
    \centering
     \sidesubfloat[]{\scalebox{0.4}{ 
\def \agentone {(0,0) circle (0.2) node {$a_i$}}
\def \agenttwo {(1,0) circle (0.2) node {$a_j$}}
\def \agentthree {(2,0) circle (0.2) node {$a_k$}}

\def \smin {[red] 
(0,0) circle (0.5)

}
\def \smax {[black!45!green] 
(0,0) circle (2.89) 
}

\begin{tikzpicture}
\tikzstyle{every node}=[font=\small]
    \draw \agentone;
    \fill[fill=cyan] \agentone;

    \draw \agenttwo;
    \fill[fill=cyan] \agenttwo;

    
    
    
    
    
    



    \begin{pgfonlayer}{back}
    

    \draw (0,0)--(2.8213,0.5642);
    \draw (0,0)--(2.8213,-0.5642);
    \draw \smin;
    \draw \smax;
     
    \end{pgfonlayer}

\end{tikzpicture}\label{fig:b}}}
    \sidesubfloat[]{\scalebox{0.4}{ 
\def \agentone {(0,0) circle (0.2) node {$a_i$}}
\def \agenttwo {(1.75,0) circle (0.2) node {$a_j$}}

\def \smin {[red] 
(0,0) circle (0.5)

}
\def \smax {[black!45!green] 
(0,0) circle (2.86) 
}

\begin{tikzpicture}
\tikzstyle{every node}=[font=\small]
    \draw \agentone;
    \fill[fill=cyan] \agentone;

    \draw \agenttwo;
    \fill[fill=cyan] \agenttwo;

    
    
    
    
    
    



    \begin{pgfonlayer}{back}
    

    \draw (0,0)--(2.82611,0.322742);
    \draw (0,0)--(2.82611,-0.322742);
    \draw \smin;
    \draw \smax;
     
    \end{pgfonlayer}

\end{tikzpicture}}}
    \sidesubfloat[]{\scalebox{0.4}{ 
\def \agentone {(0,0) circle (0.2) node {$a_i$}}
\def \agenttwo {(2.5,0) circle (0.2) node {$a_j$}}

\def \smin {[red] 
(0,0) circle (0.5)

}
\def \smax {[black!45!green] 
(0,0) circle (2.86) 
}

\begin{tikzpicture}
\tikzstyle{every node}=[font=\small]
    \draw \agentone;
    \fill[fill=cyan] \agentone;

    \draw \agenttwo;
    \fill[fill=cyan] \agenttwo;

    
    
    
    
    
    



    \begin{pgfonlayer}{back}
    

    \draw (0,0)--(2.827295, 0.22618);
    \draw (0,0)--(2.827295, -0.22618);
    \draw \smin;
    \draw \smax;
     
    \end{pgfonlayer}

\end{tikzpicture}}}
   
    \caption{When an agent $a_{i}$ observes another agent $a_{j}$. The further they are from one another the smaller the area that each one of them conceals from the other. In each subfigure the distance between $a_i$ and $a_j$ increases and accordingly the size of the concealed area decreases.}
    \label{fig:observation_dist}
\end{figure}
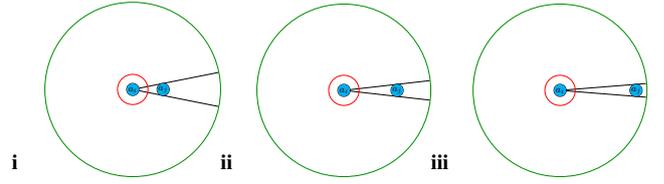

In this section, we perform a set of steps to compute the upper bound of the \wpc{} value of a connected component of agents.
First, we present a tight bound on the number of agents that can be observed by a single agent in the contamination game.
\begin{lemma}
Given an agent $a_{i}$ that observes a sector of its observation area. The maximal connectivity factor of $a_{i}$ is achieved by placing agents densely along the longest convex arc of the sector it observes.
\end{lemma}
\begin{proof}
Given a pair of agents $a_i$ and $a_j$ it can easily be seen that the closest $a_j$ is to $a_i$ the larger is the observation area it conceals from $a_i$. Figure \ref{fig:observation_dist} displays the effect of the distance between two agents on the area that one of them conceals from the other.
To maximize the connectivity factor of an agent we must place agents in its observation area while minimizing the portion of its observation area which is concealed from it. Therefore, we can conclude from our earlier observation that to maximize the connectivity factor of an agent we must place agents as furthest as possible from  it, that is, we must place agents at a distance of $S_{max}$ from it along the longest convex arc of the sector.\quad\quad\quad\quad\quad\quad\quad\qedsymbol
\end{proof}

\begin{lemma}
The maximal connectivity factor of an agent in the contamination game is $\floor{\frac{2\pi}{\arccos{(1 - \frac{2D_{r}^2}{S_{max}^2})}}}$
\end{lemma}
\begin{proof}
We know from Lemma 4 that to maximize the connectivity factor in a sector of the observation area of an agent we must place agents densely along the longest convex arc of the sector. This means that to maximize the connectivity factor of an agent we must place agents densely along the circumference of its observation area, that is, we need to compute the number of agents that can be placed at a distance of $S_{max}$ from it.
Each agent that is placed along the circumference of the circle can be represented as a sector of the circle whose arc can be computed using the cosine rule as depicted in Figure \ref{fig:agent_circumference}. Using the cosine rule we get that the size of the arc is 
$$ \alpha = \arccos{}(1 - \frac{2 D_{r}^2}{S_{max}^2})$$
Hence, the maximal number of agents that can be placed along the circumference of the circle is
\pushQED{\qed} 

$$\floor{\frac{2\pi}{\alpha}} = \floor{\frac{2\pi}{\arccos{}(1 - \frac{2D_{r}^2}{S_{max}^2})}} \qedhere
\popQED$$
\end{proof}
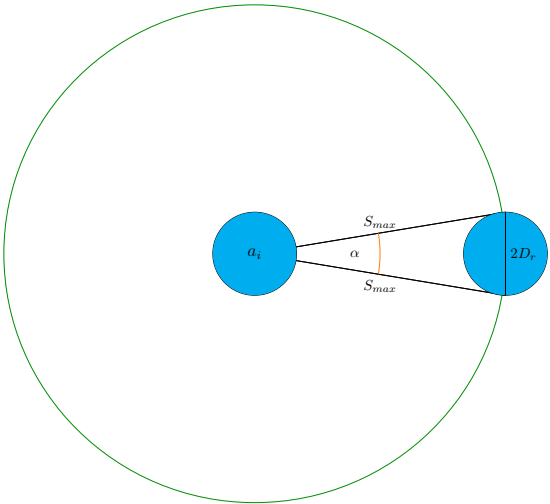
\begin{figure}[!b]
    \centering
     {\scalebox{0.55}{ 
\def \agentone {(0,0) circle (1) node {$a_i$}}
\def \agenttwo {(6,0) circle (1) node {}}
\def \agentthree {(2,0) circle (0.2) node {$a_k$}}

\def \smin {[red] 

}
\def \smax {[black!45!green] 
(0,0) circle (6) 
}

\begin{tikzpicture}
\tikzstyle{every node}=[font=\large]
    \draw \agentone;
    \fill[fill=cyan] \agentone;
    
    \draw \agenttwo;
    \fill[fill=cyan] \agenttwo;
    
    \draw (6,1)--(6,-1) node[midway, right] {\normalsize$2D_{r}$};
    
    \begin{pgfonlayer}{back}
    \draw (0,0)--(6,1) node [midway, above] {\normalsize $S_{max}$};
    \draw (0,0)--(6,-1) node [midway, below] {\normalsize $S_{max}$};
    
    \draw
    (6,1) coordinate (a) node[right] {}
    -- (0,0) coordinate (b) node[left] {}
    -- (6,-1) coordinate (c) node[above right] {}
    pic["\normalsize $\alpha$", draw=orange,angle radius = 3cm, angle eccentricity = 0.8]
    {angle=c--b--a};

    \draw \smin;
    \draw \smax;
     
    \end{pgfonlayer}

\end{tikzpicture}}}
    \caption{Each agent that is placed along the circumference of the $S_{max}$ circle can be seen as a sector of the circle with an angle which we denote by $\alpha$}
    \label{fig:agent_circumference}
\end{figure}
Since the \wpc{} value of a component is bounded by the maximal connectivity factor of any weak point of a component of agents, the aforementioned bound can be used as a non tight bound for the \wpc{} value of a component of any number of agents.

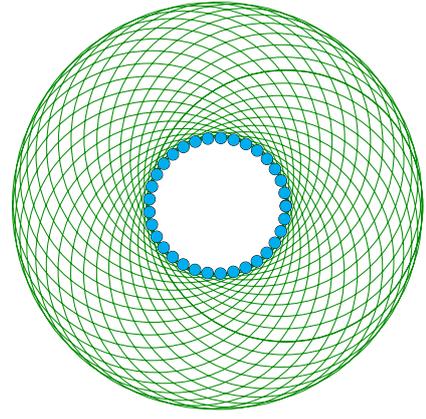
\begin{figure}[!t]
     \scalebox{0.3}{\def \A {(3.0, 0.0) circle (0.25) node {}}
\def \B {(2.9458744353310813, 0.5672951711964273) circle (0.25) node {}}
\def \C {(2.7854507924914773, 1.1141202280762161) circle (0.25) node {}}
\def \D {(2.524517751651148, 1.6207436939868782) circle (0.25) node {}}
\def \E {(2.1724907449276003, 2.068884714817121) circle (0.25) node {}}
\def \F {(1.7420722126659143, 2.4423727000311164) circle (0.25) node {}}
\def \G {(1.2487932522678453, 2.7277308175643538) circle (0.25) node {}}
\def \H {(0.7104531985806208, 2.9146622879206032) circle (0.25) node {}}
\def \I {(0.1464773578640526, 2.9964219301749155) circle (0.25) node {}}
\def \J {(-0.4227839293229178, 2.9700595531245284) circle (0.25) node {}}
\def \K {(-0.9767896039048566, 2.8365264091320204) circle (0.25) node {}}
\def \L {(-1.495549085904075, 2.6006408694111314) circle (0.25) node {}}
\def \M {(-1.9603436087268649, 2.2709145593182787) circle (0.25) node {}}
\def \N {(-2.35440166170469, 1.8592452273334454) circle (0.25) node {}}
\def \O {(-2.6635041684177096, 1.380487430157002) circle (0.25) node {}}
\def \P {(-2.876497563721645, 0.8519165251968304) circle (0.25) node {}}
\def \Q {(-2.9856962557554443, 0.29260531158528424) circle (0.25) node {}}
\def \R {(-2.9871599506074835, -0.27726418716938084) circle (0.25) node {}}
\def \S {(-2.880835832737516, -0.8371289654620383) circle (0.25) node {}}
\def \T {(-2.6705604707573, -1.3667870251171352) circle (0.25) node {}}
\def \U {(-2.363921379802262, -1.8471263384278211) circle (0.25) node {}}
\def \V {(-1.9719832358374063, -2.2608144810170145) circle (0.25) node {}}
\def \W {(-1.5088886211009904, -2.5929240500083974) circle (0.25) node {}}
\def \X {(-0.9913477072715087, -2.8314713000995653) circle (0.25) node {}}
\def \Y {(-0.438035290482488, -2.967848561549582) circle (0.25) node {}}
\def \Z {(0.13108306460138508, -2.997134836835792) circle (0.25) node {}}
\def \a {(0.6954714564252032, -2.9182733685002518) circle (0.25) node {}}
\def \b {(1.2347646580556018, -2.734109770879149) circle (0.25) node {}}
\def \c {(1.729502970119011, -2.4512893497809514) circle (0.25) node {}}
\def \d {(2.1618343989462447, -2.0800173152002195) circle (0.25) node {}}
\def \e {(2.5161588227310387, -1.6336905394819277) circle (0.25) node {}}
\def \f {(2.7796909017978337, -1.128414148467816) circle (0.25) node {}}
\def \g {(2.9429214210880525, -0.5824203888095588) circle (0.25) node {}}
\def \h {(2.999960417916506, -0.015410740871962588) circle (0.25) node {}}
\def \smax {[black!45!green] (3.0, 0.0) circle (6) (2.9458744353310813, 0.5672951711964273) circle (6) (2.7854507924914773, 1.1141202280762161) circle (6) (2.524517751651148, 1.6207436939868782) circle (6) (2.1724907449276003, 2.068884714817121) circle (6) (1.7420722126659143, 2.4423727000311164) circle (6) (1.2487932522678453, 2.7277308175643538) circle (6) (0.7104531985806208, 2.9146622879206032) circle (6) (0.1464773578640526, 2.9964219301749155) circle (6) (-0.4227839293229178, 2.9700595531245284) circle (6) (-0.9767896039048566, 2.8365264091320204) circle (6) (-1.495549085904075, 2.6006408694111314) circle (6) (-1.9603436087268649, 2.2709145593182787) circle (6) (-2.35440166170469, 1.8592452273334454) circle (6) (-2.6635041684177096, 1.380487430157002) circle (6) (-2.876497563721645, 0.8519165251968304) circle (6) (-2.9856962557554443, 0.29260531158528424) circle (6) (-2.9871599506074835, -0.27726418716938084) circle (6) (-2.880835832737516, -0.8371289654620383) circle (6) (-2.6705604707573, -1.3667870251171352) circle (6) (-2.363921379802262, -1.8471263384278211) circle (6) (-1.9719832358374063, -2.2608144810170145) circle (6) (-1.5088886211009904, -2.5929240500083974) circle (6) (-0.9913477072715087, -2.8314713000995653) circle (6) (-0.438035290482488, -2.967848561549582) circle (6) (0.13108306460138508, -2.997134836835792) circle (6) (0.6954714564252032, -2.9182733685002518) circle (6) (1.2347646580556018, -2.734109770879149) circle (6) (1.729502970119011, -2.4512893497809514) circle (6) (2.1618343989462447, -2.0800173152002195) circle (6) (2.5161588227310387, -1.6336905394819277) circle (6) (2.7796909017978337, -1.128414148467816) circle (6) (2.9429214210880525, -0.5824203888095588) circle (6) (2.999960417916506, -0.015410740871962588) circle (6) }
\begin{tikzpicture}
\tikzstyle{every node}=[font=\Large]
\draw \A;
\fill[fill=cyan] \A;
\draw \B;
\fill[fill=cyan] \B;
\draw \C;
\fill[fill=cyan] \C;
\draw \D;
\fill[fill=cyan] \D;
\draw \E;
\fill[fill=cyan] \E;
\draw \F;
\fill[fill=cyan] \F;
\draw \G;
\fill[fill=cyan] \G;
\draw \H;
\fill[fill=cyan] \H;
\draw \I;
\fill[fill=cyan] \I;
\draw \J;
\fill[fill=cyan] \J;
\draw \K;
\fill[fill=cyan] \K;
\draw \L;
\fill[fill=cyan] \L;
\draw \M;
\fill[fill=cyan] \M;
\draw \N;
\fill[fill=cyan] \N;
\draw \O;
\fill[fill=cyan] \O;
\draw \P;
\fill[fill=cyan] \P;
\draw \Q;
\fill[fill=cyan] \Q;
\draw \R;
\fill[fill=cyan] \R;
\draw \S;
\fill[fill=cyan] \S;
\draw \T;
\fill[fill=cyan] \T;
\draw \U;
\fill[fill=cyan] \U;
\draw \V;
\fill[fill=cyan] \V;
\draw \W;
\fill[fill=cyan] \W;
\draw \X;
\fill[fill=cyan] \X;
\draw \Y;
\fill[fill=cyan] \Y;
\draw \Z;
\fill[fill=cyan] \Z;
\draw \a;
\fill[fill=cyan] \a;
\draw \b;
\fill[fill=cyan] \b;
\draw \c;
\fill[fill=cyan] \c;
\draw \d;
\fill[fill=cyan] \d;
\draw \e;
\fill[fill=cyan] \e;
\draw \f;
\fill[fill=cyan] \f;
\draw \g;
\fill[fill=cyan] \g;
\draw \h;
\fill[fill=cyan] \h;
\begin{pgfonlayer}{back}
\draw \smax;
\end{pgfonlayer}
\end{tikzpicture}}
    \caption{Example of a dense circle when $S_{max} = 4$ and $D_{R} = 0.25$.}
    \label{fig:dense_example}
\end{figure}

To achieve a tighter bound we must analyze the properties of the possible observation areas of agents that are weak points of connected components of agents.
\begin{lemma}
The maximal connectivity factor of a weak point of a connected component in the contamination game is bounded by $\floor{\frac{\pi}{\arccos{}(1 - \frac{2D_{r}^2}{S_{max}^2})}}$
\end{lemma}
\begin{proof}
The fence of a connected component of agents always forms a closed polygon. We showed in Lemma 5 that placing agents along the circumference of the $S_{max}$ circle maximizes the connectivity factor of an agent. In this case, to maximize the connectivity factor of a bare agent we must place agents along the circumference of the sector of the circle whose angle is the internal angle of the polygon.

Hence, to bound the connectivity factor of the weak point we must have an upper bound for the minimal internal angle in the polygon. Since the polygon is closed it can be concluded that the minimal internal angle cannot exceed $\pi$.
Therefore, the connectivity factor of the weak point is bounded by the number of agents that fit in the circumference of the sector of the circle whose angle is $\pi$.  Using similar algebra as the one used in Lemma 5 we have that the connectivity factor of the weak point is bounded by 
${\floor{\frac{\pi}{\arccos{}(1 - \frac{2D_{r}^2}{S_{max}^2})}}}\pushQED{\qed}  \qedhere
\popQED$.
\end{proof}
Even though the aforementioned lemma presents a tighter bound for the \wpc{} value than the one presented in Lemma 5 it is still not a tight bound.
To tighten the bound we must delve into the properties of observations of bare agents in the contamination game.
Assume that there is a component $C \in \ccnodes$ that has a maximal \wpc{} value for any number of agents, that is, for any $C' \in \ccnodes$ of any size we have that $wpc(C') \leq wpc(C)$. We show that this component is monotonic by proving the following lemma:
\begin{lemma}
If there exists an optimal component $C \in \ccnodes$ it must be monotonic, and furthermore it must be that $\lvert ES(C) \rvert = 1$
\end{lemma}
\begin{proof}
Assume by contradiction there is an optimal component $C \in \ccnodes$ such that $ES(C) > 1$. That means that in addition to $wp(C)$ there is an additional agent $a_{i} \in C \setminus \{wp(C)\}$ that forced the \wpc{} algorithm to allocate agents to conquer it.
As previously stated, the number of agents allocated by the \wpc{} algorithm can only increase as we iteratively conquer the weak points of the component. Therefore, if $a_{i}$ forced the \wpc{} algorithm to allocate agents following the loss of $wp(C)$ it means that the weakpoint of $C \setminus \{wp(C)\}$ has a higher connectivity factor than that of $wp(C)$. 
Hence, we can conclude that ${wpc(C) < wpc(C \setminus \{wp(C)\})}$ which contradicts the optimality of $C$. Furthermore, we can conclude that $ES(C) = \{wp(C)\}$ which means that $\lvert ES(C) \rvert = 1$. \quad\quad\quad\quad\quad\quad\quad\quad\quad\quad\quad\quad\quad\quad\quad\quad\quad\qedsymbol
\end{proof}
We can conclude from Lemma 7 the following corollary:
\begin{corollary}
Given an optimal component $C \in \ccnodes$, the maximal difference in connectivity factors of two bare agents is bounded by 2, that is,
$$\max\limits_{a_{i}, a_{j} \in F(C)} \lvert cf(a_{i}) - cf(a_{j}) \rvert \leq 2$$
\end{corollary}
\begin{proof}
According to Lemma 7, if there is an optimal component $C \in \ccnodes$ we have that $\lvert ES(C) \rvert = 1$ . This means that besides the first iteration there is no iteration of the \wpc{} algorithm that allocates additional agents.
After conquering the first weak point of the component we have $cf(wp(C)) + 2$ agents at our disposal. Therefore, we can combine this with the fact that $\lvert ES(C) \rvert = 1$ to get that 
$$ cf(a_{i}) < cf(wp(C)) + 2~~\forall a_{i} \in C \setminus \{wp(C)\}$$
The connectivity factor of an agent can decrease by at most one following a single iteration of the \wpc{} algorithm. Therefore, we can conclude that
\pushQED{\qed} 
$$ cf(a_{i}) \leq cf(wp(C)) + 2~~\forall a_{i} \in C \qedhere \popQED$$
\end{proof}
The aforementioned corollary brings us closer to finding the optimal component which represents a tighter bound for the \wpc{} value of a connected component. We can conclude from it that in an optimal component $C \in \ccnodes$ the \wpc{} value is directly determined based on the minimal connectivity factor of an agent on the fence of the component, that is, 
$$wpc(C) = \min\limits_{a_{i} \in F(C)} cf(a_{i}) + 1$$
To ease our effort, we wish to focus on a specific subset of components which is defined as follows:
\begin{defn}
A component $C \in \ccnodes$ will be referred to as a \textit{bare component} if all of its agents are bare, that is, $F(C) = C$.
\end{defn}
Any bare component can be represented as a simple polygon where the locations of the agents represent the points of the polygon. The polygon is simple since if we would assume that there is a hole in the polygon or that it intersects itself it would mean that there is necessarily an agent which is not bare which contradicts the assumption that the component is bare.
\begin{defn}
A component $C \in \ccnodes$ is called a \textit{symmetric} component if all of its bare agents observe the exact same image (only change is orientation). 
\end{defn}
The convex hull of a symmetrical component has the shape of a regular polygon, that is, a simple polygon where all sides and angles are congruent. This notion helps us prove the following lemma:
\begin{lemma}
The optimal component $C^{*} \in \ccnodes$ must be symmetric.
\end{lemma}
\begin{proof}
The sum of angles of a regular polygon with $n$ sides is $(n-2) * \pi$. In a component $C \in \ccnodes$ each agent in the convex hull $a_{i} \in C$ has an internal angle $\alpha_{i}$ with a size that is directly correlated to the size of the area that $a_{i}$ observes inside $C$ (see Figure \ref{fig:angle_circle}).
\begin{figure}[!t]
     \scalebox{0.9}{\def \A {(1.5, -1.0) circle (0.35) node {$a_{i+1}$}}
\def \B {(0, 0) circle (0.35) node {$a_{i}$}}
\def \C {(-1.5, -1) circle (0.35) node {$a_{i-1}$}}
 
\def \smax {[black!45!green] (0, 0) circle (3cm) }
\begin{tikzpicture}
\tikzstyle{every node}=[font=\small]

\draw \A;
\fill[fill=cyan] \A;
\draw \B;
\fill[fill=cyan] \B;
\draw \C;
\fill[fill=cyan] \C;

\begin{pgfonlayer}{back}
\draw \smax;
\draw[very thick,fill=green!30] (0,0) --  (213.6905:3) arc(213.6900675:326.3099325:3) -- cycle;
\draw[latex-latex]  (213.6905:1.1) arc(213.6905:326.3093:1.1) node[midway,above]{$\alpha_{i}$};

\end{pgfonlayer}
\end{tikzpicture}}
    \caption{Each agent $a_{i}$ observes an area inside the polygon whose size is directly determined based on its internal angle $\alpha_{i}$ with its direct bare neighbors $a_{i-1}$ and $a_{i+1}$.}
    \label{fig:angle_circle}
\end{figure}
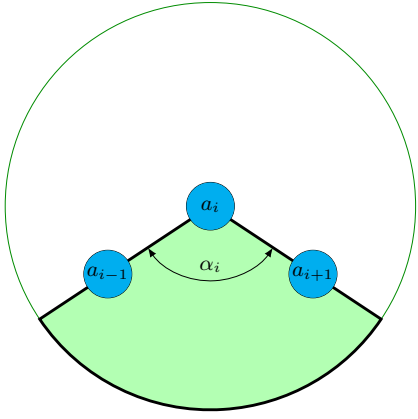
In a symmetrical component all the agents have the same internal angle ${\alpha^{*} = \frac{n-2}{n} \pi}$, that is, they observe an equal-sized area in the component. We know from Lemma 7 that the \wpc{} value of the optimal component is determined by the lowest connectivity factor of an agent on the fence of the component. Since we are looking at an optimal component we have that
$$wpc(C^{*}) = \min_{a_{i} \in F(C^{*})} cf(a_{i}) + 1$$
Assume by contradiction that there is a component ${C \in \ccnodes}$ which is not regular and has a higher \wpc{} value than $C^{*}$. Since $C$ is not regular there is an agent $a_{i} \in C$ such that $\alpha_{i} < \alpha^{*}$.  This means that
$$cf(a_{i}) < cf(a_{j})~~\forall~a_{j} \in C^{*}$$
which contradicts the assumption that ${wpc(C) > wpc(C^{*}) ~}$.\qedsymbol
\end{proof}
Since we view the agents in a two dimensional space, any symmetric component must have a convex hull that has the shape of a regular polygon (we refer to this simply as a {\em circle}). In the case of bare components it means that all the agents are necessarily placed along the circumference of a circle.
To maximize the connectivity factor of agents along the circle we intuitively need to place agents densely along the circumference of the circle as defined as follows:
\begin{defn}
A \textit{dense circle} of radius $r$, denoted by ${DC_{r} \in \ccnodes}$, is a component of agents which are placed densely along the circumference of a circle of radius $r$.
\end{defn}
Figure \ref{fig:dense_example} illustrates an example of a dense circle that has a radius of $\frac{S_{max}}{2}$. In this component all agents are bare and have the same connectivity factor. Meaning that any agent can be the weak point of the component.
\begin{lemma}
Any dense circle of any radius has a \wpc{} value that is bounded by that of a dense circle with a radius of $\frac{S_{max}}{2}$, that is,
$$wpc(DC_{r}) \leq wpc(DC_{\frac{S_{max}}{2}})~~\forall r \geq 0$$
\end{lemma}
\begin{proof}
In the formation of a dense circle each agent only observes a portion of the circle while there is another portion that is concealed from it.
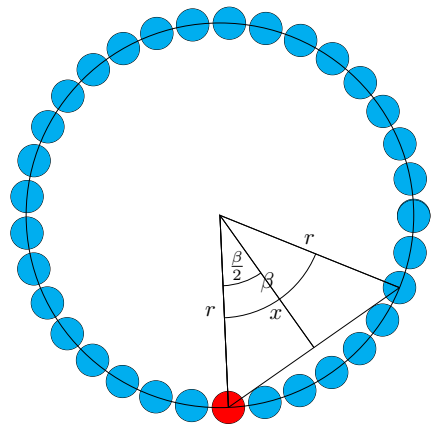
\begin{figure}[!b]
    \leftskip -1cm
     \scalebox{0.85}{\def \A {(3.0, 0.0) circle (0.25) node {}}
\def \B {(2.9458744353310813, 0.5672951711964273) circle (0.25) node {}}
\def \C {(2.7854507924914773, 1.1141202280762161) circle (0.25) node {}}
\def \D {(2.524517751651148, 1.6207436939868782) circle (0.25) node {}}
\def \E {(2.1724907449276003, 2.068884714817121) circle (0.25) node {}}
\def \F {(1.7420722126659143, 2.4423727000311164) circle (0.25) node {}}
\def \G {(1.2487932522678453, 2.7277308175643538) circle (0.25) node {}}
\def \H {(0.7104531985806208, 2.9146622879206032) circle (0.25) node {}}
\def \I {(0.1464773578640526, 2.9964219301749155) circle (0.25) node {}}
\def \J {(-0.4227839293229178, 2.9700595531245284) circle (0.25) node {}}
\def \K {(-0.9767896039048566, 2.8365264091320204) circle (0.25) node {}}
\def \L {(-1.495549085904075, 2.6006408694111314) circle (0.25) node {}}
\def \M {(-1.9603436087268649, 2.2709145593182787) circle (0.25) node {}}
\def \N {(-2.35440166170469, 1.8592452273334454) circle (0.25) node {}}
\def \O {(-2.6635041684177096, 1.380487430157002) circle (0.25) node {}}
\def \P {(-2.876497563721645, 0.8519165251968304) circle (0.25) node {}}
\def \Q {(-2.9856962557554443, 0.29260531158528424) circle (0.25) node {}}
\def \R {(-2.9871599506074835, -0.27726418716938084) circle (0.25) node {}}
\def \S {(-2.880835832737516, -0.8371289654620383) circle (0.25) node {}}
\def \T {(-2.6705604707573, -1.3667870251171352) circle (0.25) node {}}
\def \U {(-2.363921379802262, -1.8471263384278211) circle (0.25) node {}}
\def \V {(-1.9719832358374063, -2.2608144810170145) circle (0.25) node {}}
\def \W {(-1.5088886211009904, -2.5929240500083974) circle (0.25) node {}}
\def \X {(-0.9913477072715087, -2.8314713000995653) circle (0.25) node {}}
\def \Y {(-0.438035290482488, -2.967848561549582) circle (0.25) node {}}
\def \Z {(0.13108306460138508, -2.997134836835792) circle (0.25) node {}}
\def \a {(0.6954714564252032, -2.9182733685002518) circle (0.25) node {}}
\def \b {(1.2347646580556018, -2.734109770879149) circle (0.25) node {}}
\def \c {(1.729502970119011, -2.4512893497809514) circle (0.25) node {}}
\def \d {(2.1618343989462447, -2.0800173152002195) circle (0.25) node {}}
\def \e {(2.5161588227310387, -1.6336905394819277) circle (0.25) node {}}
\def \f {(2.7796909017978337, -1.128414148467816) circle (0.25) node {}}
\def \g {(2.9429214210880525, -0.5824203888095588) circle (0.25) node {}}
\def \h {(2.999960417916506, -0.015410740871962588) circle (0.25) node {}}

\def \full {(0,0) circle (3) node {}}
\def \smax {[black!45!green] (3.0, 0.0) circle (6) (2.9458744353310813, 0.5672951711964273) circle (6) (2.7854507924914773, 1.1141202280762161) circle (6) (2.524517751651148, 1.6207436939868782) circle (6) (2.1724907449276003, 2.068884714817121) circle (6) (1.7420722126659143, 2.4423727000311164) circle (6) (1.2487932522678453, 2.7277308175643538) circle (6) (0.7104531985806208, 2.9146622879206032) circle (6) (0.1464773578640526, 2.9964219301749155) circle (6) (-0.4227839293229178, 2.9700595531245284) circle (6) (-0.9767896039048566, 2.8365264091320204) circle (6) (-1.495549085904075, 2.6006408694111314) circle (6) (-1.9603436087268649, 2.2709145593182787) circle (6) (-2.35440166170469, 1.8592452273334454) circle (6) (-2.6635041684177096, 1.380487430157002) circle (6) (-2.876497563721645, 0.8519165251968304) circle (6) (-2.9856962557554443, 0.29260531158528424) circle (6) (-2.9871599506074835, -0.27726418716938084) circle (6) (-2.880835832737516, -0.8371289654620383) circle (6) (-2.6705604707573, -1.3667870251171352) circle (6) (-2.363921379802262, -1.8471263384278211) circle (6) (-1.9719832358374063, -2.2608144810170145) circle (6) (-1.5088886211009904, -2.5929240500083974) circle (6) (-0.9913477072715087, -2.8314713000995653) circle (6) (-0.438035290482488, -2.967848561549582) circle (6) (0.13108306460138508, -2.997134836835792) circle (6) (0.6954714564252032, -2.9182733685002518) circle (6) (1.2347646580556018, -2.734109770879149) circle (6) (1.729502970119011, -2.4512893497809514) circle (6) (2.1618343989462447, -2.0800173152002195) circle (6) (2.5161588227310387, -1.6336905394819277) circle (6) (2.7796909017978337, -1.128414148467816) circle (6) (2.9429214210880525, -0.5824203888095588) circle (6) (2.999960417916506, -0.015410740871962588) circle (6) }
\begin{tikzpicture}
\tikzstyle{every node}=[font=\Large]
\draw \full;
\draw[-] (2.7796909017978337, -1.128414148467816)--(0.13108306460138508, -2.997134836835792);
\draw[-](0.13108306460138508, -2.997134836835792)--(0,0) node [midway, left] {\normalsize $r$};

\draw[-](2.7796909017978337, -1.128414148467816)--(0,0) node [midway, above] {\normalsize $r$};

\draw
    (1.4553865,-2.062774) coordinate (a) node[right] {}
    -- (0,0) coordinate (b) node[left] {}
    -- (0.13108306460138508, -2.997134836835792) coordinate (c) node[above right] {}
    pic["\normalsize $\frac{\beta}{2}$", draw=black,angle radius = 1.1cm, angle eccentricity = 0.75]
    {angle=c--b--a};
    
    (2.7796909017978337, -1.128414148467816)--(0.13108306460138508, -2.997134836835792);
\draw
    (2.7796909017978337, -1.128414148467816) coordinate (d) node[right] {}
    -- (0,0) coordinate (e) node[left] {}
    -- (0.13108306460138508, -2.997134836835792) coordinate (f) node[above right] {}
    pic["\normalsize $\beta$", draw=black,angle radius = 1.6cm, angle eccentricity = 0.8]
    {angle=f--e--d};

\begin{pgfonlayer}{back}
\draw[-](1.4553865,-2.062774)--(0,0) node [near start, left] {\normalsize $x$};
\draw \A;
\fill[fill=cyan] \A;
\draw \B;
\fill[fill=cyan] \B;
\draw \C;
\fill[fill=cyan] \C;
\draw \D;
\fill[fill=cyan] \D;
\draw \E;
\fill[fill=cyan] \E;
\draw \F;
\fill[fill=cyan] \F;
\draw \G;
\fill[fill=cyan] \G;
\draw \H;
\fill[fill=cyan] \H;
\draw \I;
\fill[fill=cyan] \I;
\draw \J;
\fill[fill=cyan] \J;
\draw \K;
\fill[fill=cyan] \K;
\draw \L;
\fill[fill=cyan] \L;
\draw \M;
\fill[fill=cyan] \M;
\draw \N;
\fill[fill=cyan] \N;
\draw \O;
\fill[fill=cyan] \O;
\draw \P;
\fill[fill=cyan] \P;
\draw \Q;
\fill[fill=cyan] \Q;
\draw \R;
\fill[fill=cyan] \R;
\draw \S;
\fill[fill=cyan] \S;
\draw \T;
\fill[fill=cyan] \T;
\draw \U;
\fill[fill=cyan] \U;
\draw \V;
\fill[fill=cyan] \V;
\draw \W;
\fill[fill=cyan] \W;
\draw \X;
\fill[fill=cyan] \X;
\draw \Y;
\fill[fill=cyan] \Y;
\draw \Z;
\fill[fill=red] \Z;
\draw \a;
\fill[fill=cyan] \a;
\draw \b;
\fill[fill=cyan] \b;
\draw \c;
\fill[fill=cyan] \c;
\draw \d;
\fill[fill=cyan] \d;
\draw \e;
\fill[fill=cyan] \e;
\draw \f;
\fill[fill=cyan] \f;
\draw \g;
\fill[fill=cyan] \g;
\draw \h;
\fill[fill=cyan] \h;
\end{pgfonlayer}
\end{tikzpicture}}
    \caption{The closest agent that the red agent observes is the one for which the chord between their centers is at a distance smaller than $r - D_{r}$ from the center of the circle.}
    \label{fig:closest_computation}
\end{figure}
We can show that as the radius of the dense circle increases the number of agents which are concealed from the component increases.
Given an agent $a_{i}$ on the circle, the closest agent it can observe is the one for which the chord between their centers is at a distance smaller than $r-D_{r}$ from the center of the circle. Figure \ref{fig:closest_computation} illustrates a dense circle and the computation of the closest agent observed by an agent in the circle. As seen in the figure we can compute the angle of the circle from which we can observe agents relative to a given agent (in the case of the figure it is the one marked in red). We have that
\begin{gather*}
    \frac{r-D_{r}}{r} = 1 - \frac{D_{r}}{r} > \frac{x}{r} = \cos{}(\frac{\beta}{2})\\
    \beta = 2\arccos{}(\frac{x}{r}) > 2\arccos{}(1 - \frac{D_{r}}{r})
\end{gather*}

Hence, two sectors (both clockwise and counterclockwise) of the circle of angles $\beta = 2\arccos{}(1 - \frac{D_{r}}{r})$ cannot be observed by each agent in the dense circle. According to a previous computation the number of agents that can fit in a sector of a circle that has an angle $\beta$ is 
\begin{align}\tag{$\star$}\label{eqn:BETA}
    \floor{\frac{\beta}{\arccos{}(1 - \frac{2D_{r}^2}{r^2})}} = \floor{\frac{2\arccos{}(1 - \frac{D_{r}}{r})}{\arccos{}(1 - \frac{2D_{r}^2}{r^2})}}
\end{align}
Since $D_{r} << r$ the term $\frac{D_{r}}{r} - \frac{2D_{r}^2}{r^2} > 0$ is always positive and increases as long as $r$ increases. Therefore, the number of agents that are concealed from each agent on the dense circle (described in $\eqref{eqn:BETA}$) increases as $r$ increases.
There are two distinct cases for the radius of a dense circle. First, there is the case where $r \leq \frac{S_{max}}{2}$, that is, the whole circle is contained in the observation area of each agent. In this case the each agent that is contained in the concealed sector of the circle has a matching agent on the other side of the circle that can be observed (see Figure \ref{fig:symmetrical_dense}). Therefore, we must have that
$$wpc(DC_{r}) \leq wpc(DC_{\frac{S_{max}}{2}})~~\forall~r \leq \frac{S_{max}}{2}$$
Secondly, there is the case where $r > \frac{S_{max}}{2}$. In this case each agent cannot observe the whole component and the number of closer agents that are concealed keeps increasing. Consequently, we must have that
$$ wpc(DC_{r}) < wpc(DC_{\frac{S_{max}}{2}})~~\forall~r > \frac{S_{max}}{2}$$
Overall, we have that
\pushQED{\qed}
$$wpc(DC_{r}) < wpc(DC_{\frac{S_{max}}{2}})~~\forall~r \geq 0 \qedhere \popQED$$
\begin{figure}[!t]
    \leftskip -1cm
     \scalebox{0.85}{\def \A {(3.0, 0.0) circle (0.25) node {}}
\def \B {(2.9458744353310813, 0.5672951711964273) circle (0.25) node {}}
\def \C {(2.7854507924914773, 1.1141202280762161) circle (0.25) node {}}
\def \D {(2.524517751651148, 1.6207436939868782) circle (0.25) node {}}
\def \E {(2.1724907449276003, 2.068884714817121) circle (0.25) node {}}
\def \F {(1.7420722126659143, 2.4423727000311164) circle (0.25) node {}}
\def \G {(1.2487932522678453, 2.7277308175643538) circle (0.25) node {}}
\def \H {(0.7104531985806208, 2.9146622879206032) circle (0.25) node {}}
\def \I {(0.1464773578640526, 2.9964219301749155) circle (0.25) node {}}
\def \J {(-0.4227839293229178, 2.9700595531245284) circle (0.25) node {}}
\def \K {(-0.9767896039048566, 2.8365264091320204) circle (0.25) node {}}
\def \L {(-1.495549085904075, 2.6006408694111314) circle (0.25) node {}}
\def \M {(-1.9603436087268649, 2.2709145593182787) circle (0.25) node {}}
\def \N {(-2.35440166170469, 1.8592452273334454) circle (0.25) node {}}
\def \O {(-2.6635041684177096, 1.380487430157002) circle (0.25) node {}}
\def \P {(-2.876497563721645, 0.8519165251968304) circle (0.25) node {}}
\def \Q {(-2.9856962557554443, 0.29260531158528424) circle (0.25) node {}}
\def \R {(-2.9871599506074835, -0.27726418716938084) circle (0.25) node {}}
\def \S {(-2.880835832737516, -0.8371289654620383) circle (0.25) node {}}
\def \T {(-2.6705604707573, -1.3667870251171352) circle (0.25) node {}}
\def \U {(-2.363921379802262, -1.8471263384278211) circle (0.25) node {}}
\def \V {(-1.9719832358374063, -2.2608144810170145) circle (0.25) node {}}
\def \W {(-1.5088886211009904, -2.5929240500083974) circle (0.25) node {}}
\def \X {(-0.9913477072715087, -2.8314713000995653) circle (0.25) node {}}
\def \Y {(-0.438035290482488, -2.967848561549582) circle (0.25) node {}}
\def \Z {(0.13108306460138508, -2.997134836835792) circle (0.25) node {}}
\def \a {(0.6954714564252032, -2.9182733685002518) circle (0.25) node {}}
\def \b {(1.2347646580556018, -2.734109770879149) circle (0.25) node {}}
\def \c {(1.729502970119011, -2.4512893497809514) circle (0.25) node {}}
\def \d {(2.1618343989462447, -2.0800173152002195) circle (0.25) node {}}
\def \e {(2.5161588227310387, -1.6336905394819277) circle (0.25) node {}}
\def \f {(2.7796909017978337, -1.128414148467816) circle (0.25) node {}}
\def \g {(2.9429214210880525, -0.5824203888095588) circle (0.25) node {}}
\def \h {(2.999960417916506, -0.015410740871962588) circle (0.25) node {}}

\def \full {(0,0) circle (3) node {}}
\def \smax {[black!45!green] (3.0, 0.0) circle (6) (2.9458744353310813, 0.5672951711964273) circle (6) (2.7854507924914773, 1.1141202280762161) circle (6) (2.524517751651148, 1.6207436939868782) circle (6) (2.1724907449276003, 2.068884714817121) circle (6) (1.7420722126659143, 2.4423727000311164) circle (6) (1.2487932522678453, 2.7277308175643538) circle (6) (0.7104531985806208, 2.9146622879206032) circle (6) (0.1464773578640526, 2.9964219301749155) circle (6) (-0.4227839293229178, 2.9700595531245284) circle (6) (-0.9767896039048566, 2.8365264091320204) circle (6) (-1.495549085904075, 2.6006408694111314) circle (6) (-1.9603436087268649, 2.2709145593182787) circle (6) (-2.35440166170469, 1.8592452273334454) circle (6) (-2.6635041684177096, 1.380487430157002) circle (6) (-2.876497563721645, 0.8519165251968304) circle (6) (-2.9856962557554443, 0.29260531158528424) circle (6) (-2.9871599506074835, -0.27726418716938084) circle (6) (-2.880835832737516, -0.8371289654620383) circle (6) (-2.6705604707573, -1.3667870251171352) circle (6) (-2.363921379802262, -1.8471263384278211) circle (6) (-1.9719832358374063, -2.2608144810170145) circle (6) (-1.5088886211009904, -2.5929240500083974) circle (6) (-0.9913477072715087, -2.8314713000995653) circle (6) (-0.438035290482488, -2.967848561549582) circle (6) (0.13108306460138508, -2.997134836835792) circle (6) (0.6954714564252032, -2.9182733685002518) circle (6) (1.2347646580556018, -2.734109770879149) circle (6) (1.729502970119011, -2.4512893497809514) circle (6) (2.1618343989462447, -2.0800173152002195) circle (6) (2.5161588227310387, -1.6336905394819277) circle (6) (2.7796909017978337, -1.128414148467816) circle (6) (2.9429214210880525, -0.5824203888095588) circle (6) (2.999960417916506, -0.015410740871962588) circle (6) }
\begin{tikzpicture}
\tikzstyle{every node}=[font=\Large]
\draw \full;
\draw[-] (2.7796909017978337, -1.128414148467816)--(0.13108306460138508, -2.997134836835792);

\draw[-](2.9429214210880525, -0.5824203888095588)--(0.13108306460138508, -2.997134836835792);
\draw[-](2.999960417916506, -0.015410740871962588)--(0.13108306460138508, -2.997134836835792);
\draw[-](3.0, 0.0)--(0.13108306460138508, -2.997134836835792);
\draw[-](2.9458744353310813, 0.5672951711964273)--(0.13108306460138508, -2.997134836835792);
\draw[-](2.7854507924914773, 1.1141202280762161)--(0.13108306460138508, -2.997134836835792);
\draw[-](2.524517751651148, 1.6207436939868782)--(0.13108306460138508, -2.997134836835792);
\draw[-](2.1724907449276003, 2.068884714817121)--(0.13108306460138508, -2.997134836835792);
\draw[-](1.7420722126659143, 2.4423727000311164)--(0.13108306460138508, -2.997134836835792);
\draw[-](1.2487932522678453, 2.7277308175643538)--(0.13108306460138508, -2.997134836835792);
\draw[-](0.7104531985806208, 2.9146622879206032)--(0.13108306460138508, -2.997134836835792);
\draw[-](0.1464773578640526, 2.9964219301749155)--(0.13108306460138508, -2.997134836835792);

\draw[-](-0.4227839293229178, 2.9700595531245284)--(0.13108306460138508, -2.997134836835792);
\draw[-](-0.9767896039048566, 2.8365264091320204)--(0.13108306460138508, -2.997134836835792);
\draw[-](-1.495549085904075, 2.6006408694111314)--(0.13108306460138508, -2.997134836835792);
\draw[-](-1.9603436087268649, 2.2709145593182787)--(0.13108306460138508, -2.997134836835792);
\draw[-](-2.35440166170469, 1.8592452273334454)--(0.13108306460138508, -2.997134836835792);
\draw[-](-2.6635041684177096, 1.380487430157002)--(0.13108306460138508, -2.997134836835792);
\draw[-](-2.876497563721645, 0.8519165251968304)--(0.13108306460138508, -2.997134836835792);
\draw[-](-2.9856962557554443, 0.29260531158528424)--(0.13108306460138508, -2.997134836835792);
\draw[-](-2.9871599506074835, -0.27726418716938084)--(0.13108306460138508, -2.997134836835792);
\draw[-](-2.880835832737516, -0.8371289654620383)--(0.13108306460138508, -2.997134836835792);
\draw[-](-2.6705604707573, -1.3667870251171352)--(0.13108306460138508, -2.997134836835792);

\begin{pgfonlayer}{back}
\draw \A;
\fill[fill=cyan] \A;
\draw \B;
\fill[fill=cyan] \B;
\draw \C;
\fill[fill=cyan] \C;
\draw \D;
\fill[fill=cyan] \D;
\draw \E;
\fill[fill=cyan] \E;
\draw \F;
\fill[fill=cyan] \F;
\draw \G;
\fill[fill=cyan] \G;
\draw \H;
\fill[fill=cyan] \H;
\draw \I;
\fill[fill=cyan] \I;
\draw \J;
\fill[fill=cyan] \J;
\draw \K;
\fill[fill=cyan] \K;
\draw \L;
\fill[fill=cyan] \L;
\draw \M;
\fill[fill=cyan] \M;
\draw \N;
\fill[fill=cyan] \N;
\draw \O;
\fill[fill=cyan] \O;
\draw \P;
\fill[fill=cyan] \P;
\draw \Q;
\fill[fill=cyan] \Q;
\draw \R;
\fill[fill=cyan] \R;
\draw \S;
\fill[fill=cyan] \S;
\draw \T;
\fill[fill=cyan] \T;
\draw \U;
\fill[fill=black] \U;
\draw \V;
\fill[fill=black] \V;
\draw \W;
\fill[fill=black] \W;
\draw \X;
\fill[fill=black] \X;
\draw \Y;
\fill[fill=black] \Y;
\draw \Z;
\fill[fill=red] \Z;
\draw \a;
\fill[fill=black] \a;
\draw \b;
\fill[fill=black] \b;
\draw \c;
\fill[fill=black] \c;
\draw \d;
\fill[fill=black] \d;
\draw \e;
\fill[fill=black] \e;
\draw \f;
\fill[fill=cyan] \f;
\draw \g;
\fill[fill=cyan] \g;
\draw \h;
\fill[fill=cyan] \h;
\end{pgfonlayer}
\end{tikzpicture}}
    \caption{For each black agent that is concealed from the red agent there is a symmetrically placed cyan agent on the other side of the circle.}
    \label{fig:symmetrical_dense}
\end{figure}
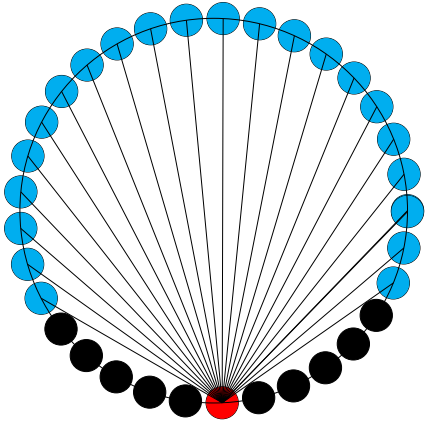
\end{proof}

Lemma 9 shows that a dense circle with a radius of $\frac{S_{max}}{2}$ is the dense circle with the maximal \wpc{} value. Therefore, we refer to it as an \textit{optimal dense circle} (ODC).
\begin{corollary}
The \wpc{} value of any bare component is bounded by that of the ODC.
\end{corollary}

Up to this point we have shown that the ODC serves as a tight bound for the \wpc{} value of a bare connected component of agents. The main question that is left unanswered is what happens in the case of components which are not bare?
To answer this question we prove the following lemma:

\begin{lemma}
If there is an optimal non-bare component $C \in \ccnodes$, its \wpc{} value is bounded by that of the ODC, that is,
$$wpc(C) < wpc(DC_{\frac{S_{max}}{2}})$$
\end{lemma}
\begin{proof}
Given a non-bare component $C \in \ccnodes$ which is optimal, there are two distinct cases depending on the diameter of $C$, that is, the largest distance between a pair of agents in $C$.
First, there is the case where the diameter of $C$ is less than $S_{max}$. In this case we can see that the minimal connectivity factor of an agent on the fence in $C$ cannot surpass the minimal connectivity factor of an agent in $DC_{\frac{S_{max}}{2}}$ since we have shown in Lemma 8 that any bare component with a diameter lower than $S_{max}$ cannot have a higher \wpc{} value than that of $DC_{\frac{S_{max}}{2}}$. Furthermore, the non-bare agents that are contained in $C$ can only decrease the connectivity factor of the bare ones. From Lemma 7 we know that the \wpc{} value of an optimal component is determined by the minimal connectivity factor of an agent in its fence which means that $wpc(C) < wpc(DC_{\frac{S_{max}}{2}})$.
Secondly, there is the case in which the diameter of $C$ exceeds $S_{max}$.
Assume by contradiction that there is an optimal non-bare component with a diameter larger than $S_{max}$ that has a higher \wpc{} value than the ODC.
We have shown in Lemma 9 that 
$$wpc(DC_{r}) <  wpc(DC_{\frac{S_{max}}{2}})~~\forall~r > \frac{S_{max}}{2}$$
Let $\Delta_{r}$ be the difference between the \wpc{} values of the ODC and a dense circle of a given radius $r > 0$, that is, 
$$\Delta_{r} = wpc(DC_{\frac{S_{max}}{2}}) - wpc(DC_{r}) $$
Therefore, we must have that each agent on the fence of $C$ must observe at least ${\Delta_{r}+ 1}$ non-bare agents.
We know from Lemma 8 that $C$ must be symmetric. Since we observe components in two dimensions it means that rotating $C$ such that one agent on the fence takes the place of another will result in the exact same image of the component that we had previously. This means that besides the bare agents having the shape of a circle, the non-bare ones must have the shape of a circle as well. 
Due to the symmetry each agent on the fence of the component observes a convex arc of a circle whereas in $DC_{\frac{S_{max}}{2}}$ each bare agent observes a convex arc along its $S_{max}$ circle. We know from Lemma 5 that the maximal connectivity factor of an agent who observes a sector of a circle is achieved by placing agents densely along the longest convex arc of the circle which makes it impossible that each agent in $C$ observes more than $\Delta_{r}$ agents.
\end{proof}
The following theorem directly follows from Lemmas 9 and 10:
\begin{theorem}
The \wpc{} value of any component is bounded by that of the ODC, or in other words, the ODC represents a tight bound for the \wpc{} value of a connected component of agents.
\end{theorem}

\section{Distributed Strategy for the Contamination Game}
\label{sec:6distributed}
We can conclude from the results of the previous sections that the problem of reaching a consensus in the contamination game can be reduced to the problem of forming dense circles while facing adversaries. 
In this section, we discuss previous attempts towards solving the problem of distributed pattern formation. We then show that forming dense circles in a distributed manner is impossible due to the challenging characteristics of swarm members. Finally, we perform several relaxations to the settings of the contamination problem in order to be able to assess the performance of a strategy that gathers agents in dense circles in simulation.
Our problem is an extension to the \textit{pattern formation problem}(\citet*{fujinaga2015pattern}) in which a group of robots is required to form a predefined geometric pattern. In our case, a group of agents is required to place themselves on the vertices of a regular polygon to form a dense circle. In the literature, this problem is known as the \textit{uniform circle formation} problem~(\citet*{dieudonne2008squaring,dieudonne2008circle,flocchini2006self,flocchini2017distributed,datta2013circle,jiang2017uniform}).
The uniform circle formation problem plays a crucial role in the domain of coordination problems due to the observation by \citet*{suzuki1999distributed} that uniform circles and points are the only patterns formable from any initial configuration under a fully synchronous model.
The problem is challenging in the distributed domain due to the fact that the robots have to make independent decisions by their own, while avoiding collisions.
Previous attempts at solving the uniform circle formation problem had sets of assumptions which are unfeasible in our work. \citet*{flocchini2017distributed} provided a constructive proof that the uniform circle formation problem is solvable for any initial configuration. In their work, it was assumed that all robots can observe one another which is obviously not possible under our problem's settings.
\citet*{datta2013circle} proposed a distributed algorithm for circle formation by a set of oblivious mobile robots where each robot has a unit disk size. Even though it proposes a fully distributed algorithm, this work assumes that robots cannot conceal one another and that the resulting circle is not uniform.
\citet*{jiang2017uniform} proposes a new approach towards solving the uniform circle formation that constitutes three main phases: consensus on the circle, circle formation and uniform transformation. While the proposed approach showed promising simulation results, it assumes that robots can send messages back and forth between one another which is not possible in our settings.
The existence of adversaries in our problem adds another layer of difficulty to the already challenging problem of distributed pattern formation. \citet*{pattanayak2020distributed} presented the problem of distributed pattern formation where robots are susceptible to crash faults, i.e., they stop moving after the crash and never recover. In our work, the robots faults are caused by them switching their own state and adopting an adversarial behavior which is similar to the definition of byzantine faults (\citet*{agmon2006fault}) in the fault-tolerant algorithms literature. To our knowledge, no study has been conducted on the problem of distributed pattern formation where the robots are susceptible to byzantine faults.
\subsection{Impossibility of Coordinated Movement In Dense Circles}
As previously mentioned, recent literature which proposed solutions to the distributed uniform circle formation problem assumed that once the robots reach the desired formation they stop moving. One might wonder whether converging to a stationary uniform circle might be a satisfactory solution to the contamination game. We answer this question by proposing the following example: assume there is a contamination game played between two swarms of agents where the healthy swarm is gathered in stationary dense circles while the contaminated agents are spread randomly and are stationary as well. Since there is no movement of any agent in the game it is impossible that either the healthy or the contaminated group will win the game, resulting in a definite tie between the two groups of agents. This simple example shows the importance of moving agents in their constructed formations. Therefore, we aim to develop strategies that do not only gather agents in dense circles but also move them while maintaining their formations.
Unfortunately, the following theorem proves that there is no deterministic distributed algorithm that moves robots in a formation of a uniform circle under the swarm settings examined so far.
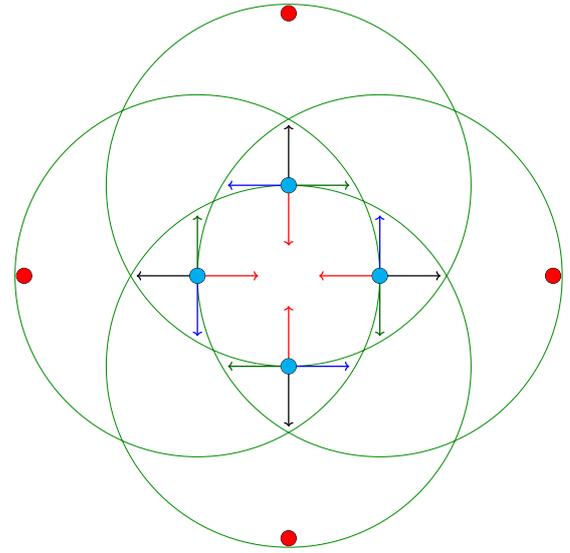
\begin{figure}[!t]
     \scalebox{0.4}{\def \A {(3.0, 0.0) circle (0.25) node {}}
\def \B {(-3.0, 0.0) circle (0.25) node {}}

\def \C {(0.0, 3.0) circle (0.25) node {}}
\def \D {(0.0, -3.0) circle (0.25) node {}}

\def \E {(0.0, -8.7) circle (0.25) node {}}
\def \F {(0.0, 8.7) circle (0.25) node {}}
\def \G {(8.7, 0.0) circle (0.25) node {}}
\def \H {(-8.7, 0.0) circle (0.25) node {}}
\def \smax {[black!45!green] (3.0, 0.0) circle (6)
(-3.0, 0.0) circle (6)
(0.0, 3.0) circle (6)
(0.0, -3.0) circle (6)}
\begin{tikzpicture}
\tikzstyle{every node}=[font=\Large]
\draw \A;
\fill[fill=cyan] \A;

\draw \B;
\fill[fill=cyan] \B;

\draw \C;
\fill[fill=cyan] \C;

\draw \D;
\fill[fill=cyan] \D;

\draw \E;
\fill[fill=red] \E;

\draw \F;
\fill[fill=red] \F;

\draw \G;
\fill[fill=red] \G;

\draw \H;
\fill[fill=red] \H;

\begin{pgfonlayer}{back}
\draw \smax;

\draw[->, very thick] (0,-3)->(0,-5);

\draw[->, very thick] (3,0)->(5,0);

\draw[->, very thick] (0,3)->(0,5);

\draw[->, very thick] (-3,0)->(-5,0);

\draw[->, very thick, blue] (0,-3)->(2,-3);

\draw[->, very thick, blue] (3,0)->(3,2);

\draw[->, very thick, blue] (0,3)->(-2,3);

\draw[->, very thick, blue] (-3,0)->(-3,-2);

\draw[->, very thick, red] (0,-3)->(0,-1);

\draw[->, very thick, red] (3,0)->(1,0);

\draw[->, very thick, red] (0,3)->(0,1);

\draw[->, very thick, red] (-3,0)->(-1,0);

\draw[->, very thick, black!60!green] (0,-3)->(-2,-3);

\draw[->, very thick, black!60!green] (3,0)->(3,-2);

\draw[->, very thick, black!60!green] (0,3)->(2,3);

\draw[->, very thick, black!60!green] (-3,0)->(-3,2);
\end{pgfonlayer}
\end{tikzpicture}}
    \caption{A uniform circle of four healthy agents that observe the exact same image which contains the uniform circle's members and a contaminated agent. Each colored arrow represents a possible direction of movement which is computed by a deterministic algorithm. Each colored arrow leads to the same outcome which is the separation of the circle.}
    \label{fig:full_symmetric_image}
\end{figure}
\begin{theorem}
Given a swarm of anonymous, oblivious robots that do not share the same coordinate system and use no explicit communication, there is no deterministic algorithm that moves robots in a uniform circle where the robots' time cycles are fully synchronous.
\end{theorem}
\begin{proof}
Assume by contradiction that there is a deterministic algorithm $\mathcal{A}$ that moves the robots in a uniform circle. Take as an example the scenario presented in Figure \ref{fig:full_symmetric_image} in which each agent observes the exact same image. Under the assumption that each agent operates based on $\mathcal{A}$, each agent in the dense circle will be ordered to move in the same direction relative to its position in the circle. If this implies a movement inwards or outwards of the circle, this will break the formation. If this is a movement on the circumference of the circle, this does not move the formation, which concludes the proof. \qquad\qquad\qquad\qquad\qquad\qquad\qquad\qquad\qquad\quad \qedsymbol
\end{proof} 
Theorem 12 showed that under the settings of the contamination game we cannot propose a distributed strategy that moves agents in a uniform circle even when they are fully synchronous. To overcome this, we perform several relaxations over the initial settings of the contamination game.
\subsection{Distributed Strategy for the Simplified Contamination Game}
Following Theorem 12, we transform the settings of the contamination game by allowing the following set of relaxations:
\begin{itemize}
    \item Agents operate in a random scheduled order in their \textit{Look-Compute-Move} cycles.
    \item Agents are no longer anonymous, that is, each agent has its own unique identifier.
    \item Each agent can internally save its own state throughout the game.
    \item Agents share the same coordinate system.
    \item Each agent can communicate with each agent it observes.
\end{itemize}
We propose a distributed strategy that utilizes each of these relaxations to place agents in uniform circles.
We showed in previous subsections that the lack of communication in the initial settings of the contamination game made the development of a distributed strategy for moving agents in uniform circles impossible. Therefore, we propose a distributed strategy that utilizes the assumption that the agents are fully synchronous and the (now possible) communication channels to form uniform  circles of agents. Additionally, the proposed communication protocols rely on the fact that each agent has its own unique identifier, that is, there is no anonymity.
As a part of our proposed solution, each agent will hold an additional state to its pre-existing healthy or contaminated state which will indicate whether it is a part of a uniform circle or not. We refer to this state as the \textit{formation-state} of an agent. There are three distinct formation-states for agents in the contamination game. First, there is the \textit{single} state which is the state of any agent that is not a part of a uniform circle or a group that aims to form a uniform circle. Second, there is the \textit{converging} state which is the state of an agent which is part of a group of agents that moves together in order to form a uniform circle. Third, there is the \textit{circle} state which is the state of an agent which is a part of an existing uniform circle.
In our proposed solution, each agent adopts a different behavior for each possible formation-state.
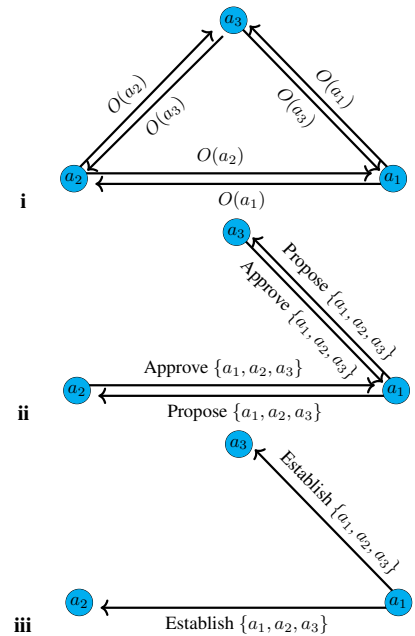
\begin{figure}[!b]
    \centering
     \sidesubfloat[]{\scalebox{0.7}{\def \A {(3.0, 0.0) circle (0.25) node {$a_{1}$}}
\def \B {(-3.0, 0.0) circle (0.25) node {$a_{2}$}}

\def \C {(0.0, 3.0) circle (0.25) node {$a_{3}$}}
\def \smax {[black!45!green] (3.0, 0.0) circle (6)
(-3.0, 0.0) circle (6)
(0.0, 3.0) circle (6)}
\begin{tikzpicture}
\draw \A;
\fill[fill=cyan] \A;

\draw \B;
\fill[fill=cyan] \B;

\draw \C;
\fill[fill=cyan] \C;

\begin{pgfonlayer}{back}

\draw [->, very thick]
    (-3.0,0.1) -- (2.6,0.1)
    node [above,align=center,midway]
    {
      $O(a_{2})$
    };
    
\draw [->, very thick]
(3.0,-0.1) -- (-2.6,-0.1)
node [below,align=center,midway]
{
  $O(a_{1})$
};

\draw [->, very thick]
    (-3.1,0) -- (-0.4,2.8)
    node [above,align=center,midway, sloped]
    {
      $O(a_{2})$
    };
    
\draw [->, very thick]
    (-0.2,2.7) -- (-2.75, 0.15)
    node [below,align=center,midway, sloped]
    {
      $O(a_{3})$
    };

\draw [->, very thick]
    (3.1,0) -- (0.3,2.9)
    node [above,align=center,midway, sloped]
    {
      $O(a_{1})$
    };
    
\draw [->, very thick]
    (0.1,2.9) -- (2.75, 0.15)
    node [below,align=center,midway, sloped]
    {
      $O(a_{3})$
    };

\end{pgfonlayer}
\end{tikzpicture}\label{fig:b}}}\\
    \sidesubfloat[]{\scalebox{0.7}{\def \A {(3.0, 0.0) circle (0.25) node {$a_{1}$}}
\def \B {(-3.0, 0.0) circle (0.25) node {$a_{2}$}}

\def \C {(0.0, 3.0) circle (0.25) node {$a_{3}$}}
\def \smax {[black!45!green] (3.0, 0.0) circle (6)
(-3.0, 0.0) circle (6)
(0.0, 3.0) circle (6)}
\begin{tikzpicture}
\draw \A;
\fill[fill=cyan] \A;

\draw \B;
\fill[fill=cyan] \B;

\draw \C;
\fill[fill=cyan] \C;

\begin{pgfonlayer}{back}

\draw [->, very thick]
    (-3.0,0.1) -- (2.6,0.1)
    node [above,align=center,midway]
    {
      Approve $\{a_{1},a_{2}, a_{3}\}$
    };
    
\draw [->, very thick]
(3.0,-0.1) -- (-2.6,-0.1)
node [below,align=center,midway]
{
  Propose $\{a_{1}, a_{2}, a_{3}\}$
};

\draw [->, very thick]
    (3.1,0) -- (0.3,2.9)
    node [above,align=center,midway, sloped] {Propose $\{a_{1}, a_{2}, a_{3}\}$};
    
\draw [->, very thick]
    (0.1,2.9) -- (2.75, 0.15)
    node [below,sloped,align=center,midway]
    {
      Approve $\{a_{1},a_{2}, a_{3}\}$
    };

\end{pgfonlayer}
\end{tikzpicture}}}\\
    \sidesubfloat[]{\scalebox{0.7}{\def \A {(3.0, 0.0) circle (0.25) node {$a_{1}$}}
\def \B {(-3.0, 0.0) circle (0.25) node {$a_{2}$}}

\def \C {(0.0, 3.0) circle (0.25) node {$a_{3}$}}
\def \smax {[black!45!green] (3.0, 0.0) circle (6)
(-3.0, 0.0) circle (6)
(0.0, 3.0) circle (6)}
\begin{tikzpicture}
\draw \A;
\fill[fill=cyan] \A;

\draw \B;
\fill[fill=cyan] \B;

\draw \C;
\fill[fill=cyan] \C;

\begin{pgfonlayer}{back}

    
\draw [->, very thick]
(3.0,-0.1) -- (-2.6,-0.1)
node [below,align=center,midway]
{
  Establish $\{a_{1}, a_{2}, a_{3}\}$
};

\draw [->, very thick]
    (3.1,0) -- (0.3,2.9)
    node [above,align=center,midway, sloped]
    {
       Establish $\{a_{1}, a_{2}, a_{3}\}$
    };

\end{pgfonlayer}
\end{tikzpicture}}}
   
    \caption{The three step communication protocol to gather agents in uniform circles. Assume that the agents are scheduled in the order of their indices, that is, $a_1$ executes before $a_{2}$ which is followed by $a_{3}$. (i) Each agent shares its observation with its neighbors. (ii) $a_{1}$ identifies $\{a_{1}, a_{2}, a_{3}\}$ as a possible uniform circle and proposes it to $a_{2}$ and $a_{3}$. $a_{2}$ and $a_{3}$ will approve this circle since they do not have better options. (iii) $a_{1}$ sends an establishment message which notifies that there is an approval of all the members of the circle, agents move to converging formation-state.}
    \label{fig:communication_proto}
\end{figure}
\subsubsection{Single Formation-State}
In the \textit{single} formation-state agents aim to form an initial uniform circle of any size. To do so we propose an algorithm that is based on the idea of converging to the self enclosing circle (SEC) of a group of agents as proposed in \citet*{flocchini2017distributed}.
The strategy of a single formation-state agent is described in broad terms in Algorithm 3.1. We avoid delving into the intricacies of the strategy to make the idea behind it more clear.
\footnote{Full implementation at \url{https://github.com/LiorMoshe/Contamination-Simulator}}.
In each cycle the agent is given the observation that its sensors gathered in its \textit{Look} phase along with messages that were sent to it from other agents that can observe it with the goal of gathering in a uniform circle as fast as possible. Each message can have a different type where the type differs based on the relationship between the sender and the receiver of the message.

\begin{algorithm}[!b]
    \SetKwInOut{Input}{Input}
    \SetKwInOut{Output}{Output}
    \SetKwComment{Comment}{$\triangleright$\ }{}
    \renewcommand{\algorithmcfname}{Algorithm}
    \renewcommand{\thealgocf}{3.1}
    \Input{Current observation of the agent \DataSty{obs}.\\
    List of messages that were sent to the agent \DataSty{messages}}
        $\DataSty{proposal} \gets \{\}$  \Comment*[f]{The proposal which contains the largest circle.}\\
        \ForEach{$\DataSty{m} \in \DataSty{messages}$}{%
      \uIf{\DataSty{m} \KwSty{is} \CommentSty{an observation}}{
                \CommentSty{Save agent's observation.}\\
              }
              \uElseIf{\DataSty{m} \KwSty{is} \CommentSty{a circle proposal}}{
                \CommentSty{Update \DataSty{proposal} if the given proposal contains a larger circle.}\\
              }
              \uElseIf{\DataSty{m} \KwSty{is} \CommentSty{a circle approval}}{
                \CommentSty{Save the id of the approving agent.}\\
              }
              \uElseIf{\DataSty{m} \KwSty{is} \CommentSty{a circle establishment message}}{
                \CommentSty{Change formation-state to CONVERGING.}\\
                \CommentSty{Move toward target location in the circle.}\\
              }
              \ElseIf{\DataSty{m} \KwSty{is} \CommentSty{an approval from an agent of formation-state circle}}{
                \CommentSty{Change formation-state to CONVERGING.}\\
                \CommentSty{Move toward target location in the circle.}\\
              }
    }
    \If{\CommentSty{formation-state \KwSty{is} CONVERGING}} {
            \Return
    }
    $\DataSty{clique} \gets$ \CommentSty{Largest clique based on observations.}\\
        \If{\CommentSty{Current agent proposed a circle in the previous time-step}}{
            \uIf{\CommentSty{All members of the proposed circle sent approval}} {
                \CommentSty{Send an establishment message to all the approvers.}
            } \ElseIf {\CommentSty{Some members of the proposed circle sent approval}} {
                \CommentSty{Send a message proposing a new circle which contains the approvers.}
            }
        }
    \If {$\lvert \DataSty{proposal} \rvert \geq \lvert \DataSty{clique} \rvert~~\KwSty{and}~~\lvert \DataSty{proposal} \rvert > 1$} {
            \CommentSty{Send approval message to the sender of \DataSty{proposal}}.\\
            \Return
        }
        \caption{\FuncSty{SingleStrategy}(\DataSty{obs}, \DataSty{messages})}
\end{algorithm}
\begin{algorithm}[!t]
    \algsetup{linenosize=\tiny}
    \SetKwInOut{Input}{Input}
    \SetKwInOut{Output}{Output}
    \SetKwComment{Comment}{$\triangleright$\ }{}
    \renewcommand{\algorithmcfname}{Algorithm}
    \renewcommand{\thealgocf}{3.1}
        \setcounter{AlgoLine}{31}
        \eIf {$\lvert \DataSty{clique} \rvert > 1$} {
            \CommentSty{Send a circle proposal to the members of the clique.}
        } {
            \eIf{\CommentSty{No agents are observed}}{
                \CommentSty{Move randomly.}\\
            } {
                \CommentSty{Find the closest agent that has a circle formation-state.}\\
                \CommentSty{Send a message proposing that you join the circle.}\\
            }
        }
    \caption{\FuncSty{SingleStrategy}(\DataSty{obs}, \DataSty{messages})}
\end{algorithm}
There are three distinct cases for an agent with a single formation-state. First, there is the simple case in which the agent cannot observe any other agent and we have nothing to do besides a simple random walk (lines 35-37). Second, there is the case where an agent can observe a set of single formation-state agents and they form together a clique of agents. In this case we use a communication protocol that is made of three time-steps whose purpose is to gather the agents in the clique together in a uniform circle as presented in Figure \ref{fig:communication_proto}. First, each agent shares its observation with its neighboring single formation-state agents (lines 3-5). Then the first agent that is scheduled in the next time step can compute the maximal sized clique that it has among its neighbors (line 20) and send a proposal to the members of this clique to gather together in a dense circle (lines 32-34). In the following time step the members of the proposed circle will send an approval message to the proposing agent if they do not have a larger clique within their set of neighbors (lines 28-31). Finally, the proposing agent will check that it got the approval of all the agents in the potential circle. If it did it will send a message to all the members of the clique to say that the circle is established. If only a subset of the members of the potential circle approved the message then the agent will compose a new proposal containing only the approving subset (lines 21-26).
The third and final case of a single formation-state agent is the case in which it observes only circle formation-state agents. In this case the agent will send a proposal to the member of the existing circle to merge together (lines 37-40). The members of the existing circle decide whether to accept or reject the proposal using a communication protocol which we cover in the upcoming subsection about the circle formation-state. If they approve the proposal both the single formation-state agent and the agents of the existing circle move to converging formation-state and gather in a new uniform circle with a diameter of $S_{max}$.
\begin{algorithm}[!t]
    \SetKwInOut{Input}{Input}
    \SetKwInOut{Output}{Output}
    \SetKwComment{Comment}{$\triangleright$\ }{}
    \renewcommand{\algorithmcfname}{Algorithm}
    \renewcommand{\thealgocf}{3.2}
    
    \Input{Current observation of the agent \DataSty{obs}.\\
    List of messages that were sent to the agent \DataSty{messages}.}
        $\DataSty{C} \gets \FuncSty{set}()$\\
      \ForEach{$\DataSty{m} \in \DataSty{messages}$}{
            \If{\DataSty{m} \KwSty{is} \CommentSty{of type CONVERGENCE STATE}~~\KwSty{and}~~\CommentSty{\DataSty{m}.\FuncSty{publisher}() is in our circle}}{
                \CommentSty{Update \DataSty{C} based on the content of the message.}   
            }
      }
      \eIf{\CommentSty{Current agent converged to its target}} {
            \CommentSty{Add current agent to \DataSty{C}.}\\
      }{
        \CommentSty{Move toward the agents target location in the circle.}\\
      }
       
      \CommentSty{Send the converged set of agents \DataSty{C} in a CONVERGENCE STATE message.}\\
      \If{\CommentSty{$\lvert \DataSty{C} \rvert$ is equal to the size of our circle}}{
            \CommentSty{Switch to CIRCLE formation-state.}\\
      }
    \caption{\FuncSty{ConvergingStrategy}(\DataSty{obs}, \DataSty{messages})}
\end{algorithm}
\subsubsection{Converging Formation-State}
Agents in \textit{converging} formation-state interact only with other converging agents that aim to form a uniform circle with them. The strategy for the converging formation-state relies on the simple idea that if each agent will relay information regarding the convergence of the agents it can observe we will get to a point in which each agent knows whether all the members of the uniform circle converged. Notice that this works since all the agents share the same coordinate system with one another.
Similarly to the single formation-state, we describe the flow of the converging strategy in broad terms in Algorithm 3.2. Each agent goes over the messages it receives and checks whether there are any messages that contain the convergence state of other agents in its circle (lines  2-5). Then, the agent checks whether it converged to its own target position in the circle in order to determine its action in the current cycle (lines 7-11). Finally, the agent sends to the agents it observes all the information it gathered about which members of the circle converged to their required locations and changes its state to \textit{circle} once all the members of the circle converged to their required locations (lines 12-15).
\begin{algorithm}[!b]
    \SetKwInOut{Input}{Input}
    \SetKwInOut{Output}{Output}
    \SetKwComment{Comment}{$\triangleright$\ }{}
    \renewcommand{\algorithmcfname}{Algorithm}
    \renewcommand{\thealgocf}{3.3}
    
    \Input{Current observation of the agent \DataSty{obs}.\\
    List of messages that were sent to the agent \DataSty{messages}\\
    Optimal dense circle size \DataSty{threshold}.}
        $\DataSty{mode} \gets \CommentSty{Current circle mode.}$\\
        \ForEach{$\DataSty{m} \in \DataSty{messages}$}{
            \uIf{\DataSty{m} \KwSty{is} \CommentSty{a merge proposal}}{
                \CommentSty{Save the given proposal in our state.}\\
                \If{\CommentSty{\DataSty{m}'s circle was previously proposed to by the current agent}} {
                    \CommentSty{Send approval message to all the agents we observe.}\\
                    \CommentSty{Move to CONVERGING formation-state and update target location.}\\
    
                }
              }
              \uElseIf{\DataSty{m} \CommentSty{contains exterior info}}{
                \CommentSty{Save the given information in $\DataSty{m}$ in our state.}\\
              } \uElseIf{\DataSty{m} \KwSty{is} \CommentSty{a circle publication}}{
                \CommentSty{Save the information about the circle in $\DataSty{m}$ in our state.}\\
              }
              \uElseIf{\DataSty{m} \KwSty{is} \CommentSty{approval of a merge proposal}}{
                \CommentSty{Send \DataSty{m} to all the agents we observe.}\\
                \CommentSty{Move to CONVERGING formation-state and update target location.}\\
                \Return
              }
              \ElseIf{\DataSty{m} \KwSty{is} \CommentSty{a random direction}~\KwSty{and}~\DataSty{mode}~\KwSty{is}~\CommentSty{MOVE}}{
                \CommentSty{Send $\DataSty{m}$ to all the agents we observe.}\\
                \CommentSty{Move in the given random direction.}\\
                \CommentSty{Switch the circle mode.}\\
                \Return
              }
      }
      \uIf{$\DataSty{mode}~\KwSty{is}~\CommentSty{PUBLICIZE}$}{
            \CommentSty{Send information about our current circle to all observed agents.}\\
      }
      \uElseIf{$\DataSty{mode}~\KwSty{is}~\CommentSty{DISCOVERY}$}{
            \CommentSty{Send exterior info gathered to all the agents we observe.}\\
        } 
    \caption{\FuncSty{CircleStrategy}(\DataSty{obs}, \DataSty{messages})}
\end{algorithm}
\begin{algorithm}[!t]
    \SetKwInOut{Input}{Input}
    \SetKwInOut{Output}{Output}
    \SetKwComment{Comment}{$\triangleright$\ }{}
    \renewcommand{\algorithmcfname}{Algorithm}
    \renewcommand{\thealgocf}{3.3}
    \setcounter{AlgoLine}{27}
    \uElseIf{$\DataSty{mode}~\KwSty{is}~\CommentSty{COORDINATE}$}{
            $\DataSty{F} \gets$ \CommentSty{Maximal sized neighbor 
            circle.}\\
            $\DataSty{P} \gets$ \CommentSty{Maximal sized circle merge proposal.}\\
            
            \uIf{$\lvert \DataSty{P} \rvert > \lvert \DataSty{F} \rvert$~\KwSty{and}~\CommentSty{the agent that proposed \DataSty{P} is observed}} {
                \CommentSty{Send approval to the proposal.}\\
                \CommentSty{Switch to CONVERGING formation-state and update target location.}\\
                \Return
            }\ElseIf {$\lvert \DataSty{F} \rvert > 0$~\KwSty{and}~\CommentSty{size of proposed circle does not excess \DataSty{threshold}}}{
                \CommentSty{Send merge proposal with \DataSty{F}.}\\
            }
        } 
        \Else {
            \CommentSty{Send randomly computed angle to all the agents we observe.}\\
            \CommentSty{Move in the direction of the randomly computed angle.}\\
        }
        \CommentSty{Switch the circle mode.}\\
    \caption{\FuncSty{CircleStrategy}(\DataSty{obs}, \DataSty{messages})}
\end{algorithm}
\subsubsection{Circle Formation-State}
Agents in the \textit{circle} formation-state aim to merge together with other uniform circles to strengthen themselves. We describe the strategy for an agent in a circle formation-state broadly in Algorithm 3.3. The strategy of the circle formation-state relies on a similar idea to the one used in the converging formation-state strategy. Each agent passes its knowledge regarding the surroundings of the circle to the portion of the circle it observes such that after a single cycle of messages all the members of the circle have full knowledge of the surroundings of the circle.
Each member of a uniform circle operates according to a four stepped communication protocol whose goal is to ensure that the agents of the circle make coordinated decisions while maintaining the uniform circle formation. We refer to the current step of the communication protocol that the agent executes as its \textit{circle mode}.
  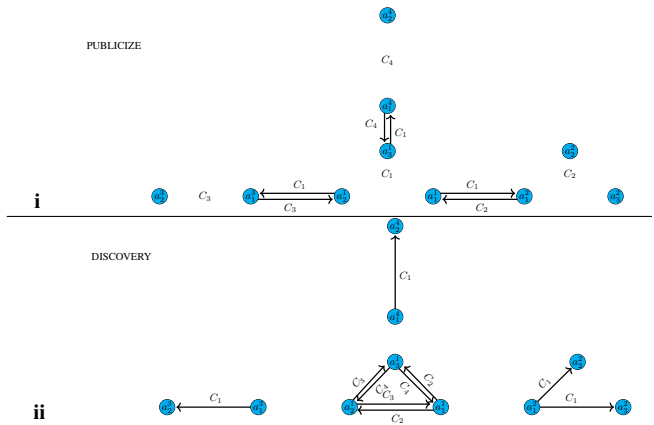
\begin{figure}[!b]
    \centering
\sidesubfloat[]{\scalebox{0.4}{\def \A {(1.5, 0.0) circle (0.25) node {$a_{1}^{1}$}}
\def \B {(-1.5, 0.0) circle (0.25) node {$a_{2}^{1}$}}

\def \C {(0.0, 1.5) circle (0.25) node {$a_{3}^{1}$}}

\def \D {(0,3.0)  circle (0.25) node {$a_{1}^{4}$}}
\def \d {(0,6)  circle (0.25) node {$a_{2}^{4}$}}

\def \E {(-4.5,0.0)  circle (0.25) node {$a_{1}^{3}$}}
\def \F {(-7.5,0.0)  circle (0.25) node {$a_{2}^{3}$}}

\def \G {(4.5,0.0)  circle (0.25) node {$a_{1}^{2}$}}
\def \H {(6.0,1.5)  circle (0.25) node {$a_{2}^{2}$}}
\def \I {(7.5,0.0)  circle (0.25) node {$a_{3}^{2}$}}
\def \smax {[black!45!green] (3.0, 0.0) circle (3)
(-3.0, 0.0) circle (6)
(0.0, 3.0) circle (6)}
\begin{tikzpicture}
\draw \A;
\fill[fill=cyan] \A;

\draw \B;
\fill[fill=cyan] \B;

\draw \C;
\fill[fill=cyan] \C;

\draw \D;
\fill[fill=cyan] \D;

\draw \d;
\fill[fill=cyan] \d;

\draw \E;
\fill[fill=cyan] \E;

\draw \F;
\fill[fill=cyan] \F;

\draw \G;
\fill[fill=cyan] \G;

\draw \H;
\fill[fill=cyan] \H;

\draw \I;
\fill[fill=cyan] \I;

\draw (-9,5.0) node 
                       {PUBLICIZE};
                       
\draw (0,0.75) node 
                       {$C_{1}$};
                       
\draw (6,0.75) node 
{$C_{2}$};

\draw (-6,0) node 
{$C_{3}$};

\draw (0,4.5) node 
{$C_{4}$};

\begin{pgfonlayer}{back}

\draw [->, very thick]
    (-1.5,0.1) -- (-4.2, 0.1)
    node [above,align=center,midway, sloped]
    {
      $C_{1}$
    };
    
\draw [->, very thick]
(-4.5,-0.1) -- (-1.8, -0.1)
node [below,align=center,midway, sloped]
{
  $C_{3}$
};

\draw [->, very thick]
(0.1, 1.5) -- (0.1, 2.7)
node [right,align=center,midway]
{
  $C_{1}$
};

\draw [->, very thick]
(-0.1, 3) -- (-0.1, 1.8)
node [left,align=center,midway]
{
  $C_{4}$
};

\draw [->, very thick]
(1.5, 0.1) -- (4.2, 0.1)
node [above,align=center,midway]
{
  $C_{1}$
};

\draw [->, very thick]
(4.5, -0.1) -- (1.8, -0.1)
node [below,align=center,midway]
{
  $C_{2}$
};

\end{pgfonlayer}
\end{tikzpicture}\label{fig:a}}}
\rulesep
\medskip
\sidesubfloat[]{\scalebox{0.4}{\def \A {(1.5, 0.0) circle (0.25) node {$a_{1}^{1}$}}
\def \B {(-1.5, 0.0) circle (0.25) node {$a_{2}^{1}$}}

\def \C {(0.0, 1.5) circle (0.25) node {$a_{3}^{1}$}}

\def \D {(0,3.0)  circle (0.25) node {$a_{1}^{4}$}}
\def \d {(0,6)  circle (0.25) node {$a_{2}^{4}$}}

\def \E {(-4.5,0.0)  circle (0.25) node {$a_{1}^{3}$}}
\def \F {(-7.5,0.0)  circle (0.25) node {$a_{2}^{3}$}}

\def \G {(4.5,0.0)  circle (0.25) node {$a_{1}^{2}$}}
\def \H {(6.0,1.5)  circle (0.25) node {$a_{2}^{2}$}}
\def \I {(7.5,0.0)  circle (0.25) node {$a_{3}^{2}$}}
\def \smax {[black!45!green] (3.0, 0.0) circle (3)
(-3.0, 0.0) circle (6)
(0.0, 3.0) circle (6)}
\begin{tikzpicture}
\draw \A;
\fill[fill=cyan] \A;

\draw \B;
\fill[fill=cyan] \B;

\draw \C;
\fill[fill=cyan] \C;

\draw \D;
\fill[fill=cyan] \D;

\draw \d;
\fill[fill=cyan] \d;

\draw \E;
\fill[fill=cyan] \E;

\draw \F;
\fill[fill=cyan] \F;

\draw \G;
\fill[fill=cyan] \G;

\draw \H;
\fill[fill=cyan] \H;

\draw \I;
\fill[fill=cyan] \I;

\draw (-9,5.0) node 
                       {DISCOVERY};

\begin{pgfonlayer}{back}

\draw [->, very thick]
    (-1.6, 0.0) -- (-0.35, 1.375)
    node [above,align=center,midway, sloped]
    {
      $C_{3}$
    };
    
\draw [->, very thick]
    (0.0, 1.5) -- (-1.25, 0.2)
    node [below,align=center,midway, sloped]
    {
      $C_{4}$
    };
    
\draw [->, very thick]
    (-1.5, 0.1) -- (1.15, 0.1)
    node [above,align=center,midway, sloped]
    {
      $C_{3}$
    };

\draw [->, very thick]
    (1.5, -0.1) -- (-1.25, -0.1)
    node [below,align=center,midway, sloped]
    {
      $C_{2}$
    };

\draw [->, very thick]
    (1.6, 0) -- (0.3, 1.35)
    node [above,align=center,midway, sloped]
    {
      $C_{2}$
    };
    
\draw [->, very thick]
    (-0.1, 1.5) -- (1.25, 0.15)
    node [below,align=center,midway, sloped]
    {
      $C_{4}$
    };

\draw [->, very thick]
    (-4.5, 0) -- (-7.2, 0.0)
    node [above,align=center,midway, sloped]
    {
      $C_{1}$
    };
    
\draw [->, very thick]
    (0, 3) -- (0, 5.7)
    node [right,align=center,midway]
    {
      $C_{1}$
    };

\draw [->, very thick]
    (4.5,0.0) -- (5.8,1.3)
    node [above,align=center,midway, sloped]
    {
      $C_{1}$
    };

\draw [->, very thick]
    (4.5,0.0) -- (7.2,0.0)
    node [above,align=center,midway, sloped]
    {
      $C_{1}$
    };

\end{pgfonlayer}
\end{tikzpicture}\label{fig:c}}}
\caption{Usage of PUBLICIZE and DISCOVERY circle modes to help the circle members gain full knowledge regarding the surroundings of the circle in the case of four neighboring dense circles. (i) Each member of each circle sends to its neighbors from other circles information regarding its own circle. (ii) Each agent shares the information it discovered to its circle members.}
    \label{fig:pub_disc}
    \end{figure}
The communication protocol of dense circles operates according to the following four modes. First, there is the \DataSty{PUBLICIZE} mode which is the mode in which each agent sends to the agent it can observe information about its own circle (lines 24-26). Second, there is the \DataSty{DISCOVERY} mode which is the mode of operation where each agent shares the information it gathered thus far to the agents it observes (lines 26-28). Third, there is the \DataSty{COORDINATE} mode which is the mode of operation where the agents decide which circle they should merge with (lines 28-38). Then, there is \DataSty{MOVE} mode of operation which is the mode where the circle moves towards a random location (lines 39-41).
Figure \ref{fig:pub_disc} serves as an example for the way in which the PUBLICIZE and DISCOVERY circle modes aid the members in the circle gain full knowledge regarding the surroundings of the circle.
Similarly to the strategies in other formation-states, the strategy of an agent in a circle formation-state relies on having messages of several types passed between agents in order to coordinate the dense circles and allow them to merge with one another. Once an agent receives any information regarding the surroundings of its circle it immediately save the received information in its own state (lines 3-13). In the case in which an agent receives an agreement for a merge proposal or a direction of movement it forwards this information to all the agents it observes in order to fully synchronize the circle (lines 13-21). Notice that each merge of two circles is done only following the agreement of all the members of both circles since we require full approval of all the circle members to send an approval message. Furthermore, we restrict the size of each circle by forbidding each possible merge of a pair of circles whose combined size exceeds that of the ODC (line 35).
\subsection{Simulation Results}
For the experimental results in this section, we consider agents that move in a 2D world synchronously.
Throughout this work we assume that the $S_{min}$ and $S_{max}$ values are equal among all the agents in the game, that is, all agents face the same physical limitations. The major contribution of these configuration settings are the inflicted maximal clique size and maximal dense circle size which could be constructed by the agents in the game. The results of any game in which the ratio $\frac{S_{min}}{S_{max}}$ has an extremely small (resp. large) values are of little interest since they result in cases in which we might simply have extremely large (resp. small) components. Therefore, our results focused on a configuration in which had an average value for the ratio $\frac{S_{min}}{S_{max}}$. Throughout our experimentation process we observed that different configurations which are not extremely small or large did not effect our results in any significant way. Hence, we avoid going through several possible configurations.

\begin{figure}[!t]
    \begin{center}
      \scalebox{0.45}{\includegraphics{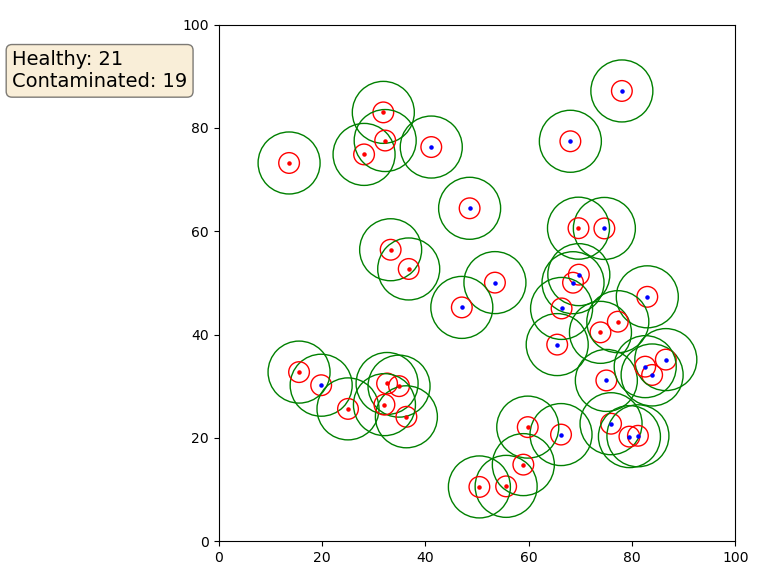}} 
    \end{center}
    \caption{The display of the simulation environment. Blue and red filled circles represent the healthy and contaminated agents, respectively. Each one of the agents is covered by red and green circles which represent the $S_{min}$ and $S_{max}$ observation radii, respectively. On the top left, the simulator presents the number of agents in each group in the game.}
    \label{fig:sim_environment}
\end{figure}

\begin{figure}[!t]
    \centering
    \scalebox{0.45}{\includegraphics{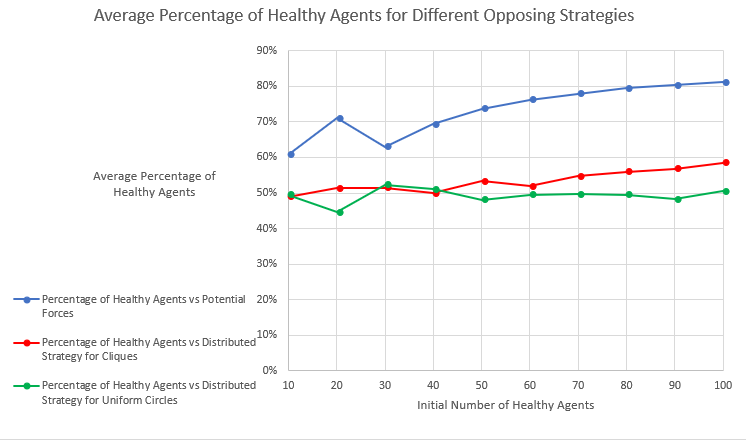}}
    \caption{The average final percentage of healthy agents achieved by the distributed strategy versus different opposing strategies. The percentage is averaged over a 100 simulated games for each number of agents. Each game begins with an equal number of healthy and contaminated agents, that is, if a game has 50 healthy agents initially it has a total of a 100 agents.}
    \label{fig:sim_results}
\end{figure}

We assume that each agent observes agents based on the observation radii $S_{min} = 2$ and $S_{max}=6$ with a diameter of $D_r=0.25$. We can conclude from these settings that a clique can contain up to 9 agents and a dense uniform circle can contain up to 37 agents.
In each time-step, each agent sets its state based on the majority transition rule and moves in the 2D world according to a computed two dimensional action vector.
The simulation ends either when all agents share the same state, when the time bound of $T=1024$ time-steps has passed or when there were at least 200 time-steps in which there was no change in the number of healthy and contaminated agents.
The simulation environment was implemented in python and uses the matplotlib library to display the game as presented in Figure \ref{fig:sim_environment}.
Each game started with an equal number of healthy and contaminated agents that are spread randomly in the game's area. For each number of agents we simulated a series of 100 games where each game had a length of 1024 time-steps.
We tested our distributed strategy against three main strategies. First, there is the strategy that is based on potential forces which was proposed in \citet*{contaminationProblem}. The main disadvantage of this strategy is that it can cause agents to remain stationary instead of actively exploring the space of the game. We could see in our simulations that small connected components of agents remained stationary which gave an advantage to the opposing swarm that actively explored the game's area.  Secondly, we evaluated our strategy against a directly similar strategy which has the minor difference of converging to cliques instead of gathering in uniform circles, that is, the \DataSty{threshold} that is used by Algorithm 3.3 is the size of the maximal clique in the game. The purpose of this evaluation is to test whether longer convergence time to global optimum hurts the swarm members in comparison to the option of converging faster to a local optimum. Finally, we decided to test our strategy against itself to see the effect of having the opposing strategy forming uniform circles instead of cliques.
All the games in our simulations did not end with an absolute winner as the outcome of the game is directly correlated to the random exploration done by agents. Hence, we use the average final percentage of healthy agents as our evaluation metric for the performance of our strategy. We strive to achieve consistently a percentage higher than 50\% since that means that more than half of the agents are healthy at the end of the game.
Figure \ref{fig:sim_results} displays the results of our simulations against both of the aforementioned strategies. Note that for each game the initial percentage of healthy agents was 50\%, that is, if a game began with 50 healthy agents it means there are a total of a 100 agents.
It can be seen that the distributed strategy outperformed the potential forces strategy consistently for each number of  agents. Further inspection showed that these results are statistically significant (p-value $<$ 0.05) for each number of agents. In the case of the distributed strategy which aims to form cliques, we can see that forming uniform circles has an advantage once we exceed 50 agents. It can be seen that the gap in performance gets larger as the number of agents increases. Furthermore, it can be seen that once we exceed 50 agents the strategy performs better against the distributed strategy that forms cliques in comparison to its performance against itself. Further analysis showed that these results are statistically significant once we exceed 50 agents (p-value $<$ 0.05). When there are less than 60 agents it is harder to form large uniform circles faster than it is to form cliques. Hence, the performance of a strategy that forms uniform circles is nearly identical to the performance of the strategy that forms cliques. Once we exceed 50 agents it is easier to form uniform circles faster which explains why the strategy that form uniform circles consistently outperforms the strategy that forms cliques.
\section{Conclusions and Future Work}
\label{sec:7conclude}
In this work we tackled  the robotic swarm contamination game by taking a top-down approach, that is, inspecting theoretical properties of the game from a centralized point of view and using these properties to develop an efficient distributed strategy. A series of theoretical discoveries led us to finding the globally optimal behavior of swarm members in the contamination game which is to gather in dense uniform circles of agents. We have shown that under the current settings of the problem it is impossible to implement this behavior in a distributed manner. Moreover, we covered the performance of an implementation of the aforementioned behavior in a simplified version of the contamination game. This means that an efficient solution to the contamination game under the  current settings can only achieve a local optima such as formation of agents in cliques as presented in \citet*{contaminationProblem}.
To conclude, our findings show that the contamination game can be directly reduced to the problem of distributed formation of non-stationary uniform circles, that is, a solution to the problem of forming non-stationary uniform circles can be used to design an optimal solution to the contamination game.
Going forward, we believe that future work should focus on development of broader frameworks for analyzing games in large populations of players such as in the case of games involving swarms since we can observe that realistic strategies in those games can only achieve a local optimum. Hence, future research directions should focus on deconstructing the large space of local optimum solutions for games involving swarms in a way that would help the research community develop solutions that achieve a local optimum which is guaranteed to be an efficient one.
\newpage
\bibliographystyle{SageH}
\bibliography{paper.bib}

\end{document}